\documentclass{aa}  
\usepackage{natbib,epsfig,graphicx,color,psfrag,pdflscape}
\usepackage{amsmath}
\usepackage{txfonts}
\bibpunct{(}{)}{,}{a}{}{,}

\newcommand{\rsun}{\ensuremath{R_{\odot}}}

\newcommand{\Teff}{\ensuremath{T_{\rm eff}}}
\newcommand{\vinf}{\ensuremath{v_{\infty}}}
\newcommand{\mdot}{\ensuremath{\dot{M}}}

\newcommand{\msunyr}{\ensuremath{M_{\odot} {\rm yr}^{-1}}}

\newcommand{\mdu}{\ensuremath{10^{-6}\,M_{\odot} {\rm yr}^{-1}}}

\newcommand{\beq}{\begin{equation}}
\newcommand{\eeq}{\end{equation}}
\newcommand{\beqa}{\begin{eqnarray}}
\newcommand{\eeqa}{\end{eqnarray}}

\newcommand{\nbeq}{\begin{equation*}}
\newcommand{\neeq}{\end{equation*}}

\newcommand{\kms}{\ensuremath{{\rm km}\,{\rm s}^{-1}}}
\newcommand{\cms}{\ensuremath{{\rm cm}\,{\rm s}^{-2}}}

\newcommand{\dd}{{\rm d}}

\newcommand{\HeI} {He\,{\sc i}}
\newcommand{\HeII}{He\,{\sc ii}}
\newcommand{\HeIII}{He\,{\sc iii}}

\newcommand{\CII}{C\,{\sc ii}}
\newcommand{\CIII}{C\,{\sc iii}}
\newcommand{\CIV}{C\,{\sc iv}}
\newcommand{\CV}{C\,{\sc v}}

\newcommand\NIII{N\,{\sc iii}}
\newcommand\NIV{N\,{\sc iv}}
\newcommand\NV{N\,{\sc v}}
\newcommand\NVI{N\,{\sc vi}}

\newcommand\OIII{O\,{\sc iii}}
\newcommand\OIV{O\,{\sc iv}}
\newcommand\OV{O\,{\sc v}}
\newcommand\OVI{O\,{\sc vi}}
\newcommand\OVII{O\,{\sc vii}}

\newcommand\SiIII{Si\,{\sc iii}}
\newcommand\SiIV{Si\,{\sc iv}}
\newcommand\SiV{Si\,{\sc v}}
\newcommand\SiVI{Si\,{\sc vi}}

\newcommand{\PIV}{P\,{\sc iv}}
\newcommand{\PV}{P\,{\sc v}}
\newcommand{\PVI}{P\,{\sc vi}}
\newcommand{\PVII}{P\,{\sc vii}}

\newcommand{\Rstar}{\ensuremath{R_{\ast}}}

\newcommand{\Rmin}{\ensuremath{R_{\rm min}}}

\newcommand{\logg}{\ensuremath{\log g}}

\newcommand{\Tshock}{\ensuremath{T_{\rm s}}}
\newcommand{\Tshockmax}{\ensuremath{T_{\rm s}^{\infty}}}

\newcommand{\vsini}{\ensuremath{v{\thinspace}\sin{\thinspace}i}}
\newcommand{\vturb}{\ensuremath{v_{\rm turb}}}

\newcommand{\mx}{\ensuremath{m_{\rm x}}}

\newcommand{\adiabsound}{\ensuremath{a_{\rm s}}}

\newcommand{\fcl}{\ensuremath{f_{\rm cl}}}
\newcommand{\fv}{\ensuremath{f_{\rm X}}}

\newcommand{\taur}{\ensuremath{\tau_{\rm Ross}}}

\newcommand{\elecden}{\ensuremath{n_{\rm e}}}
\newcommand{\protden}{\ensuremath{n_{\rm p}}}

\newcommand{\lxray}{\ensuremath{L_{\rm x}}}
\newcommand{\lbol}{\ensuremath{L_{\rm bol}}}

\begin{document}
\title{Atmospheric NLTE-Models for the Spectroscopic Analysis of\\ 
Blue Stars with Winds}
\subtitle{III. X-ray emission from wind-embedded
shocks\thanks{Appendices A, 
B, and C are only available in electronic form 
at http://www.edpsciences.org}
}

\author{Luiz P. Carneiro\inst{1}, J. Puls\inst{1}, J. O. Sundqvist\inst{2,3}, \and T. L. Hoffmann\inst{1}
}

\institute{LMU Munich, Universit\"atssternwarte, Scheinerstr. 1, 81679 M\"unchen, 
           Germany, \email{luiz@usm.uni-muenchen.de} 
           \and
           Centro de Astrobiolog{\'i}a, CSIC-INTA,\ Ctra. Torrej{\'o}n a 
	   Ajalvir km.4, 28850 Madrid, Spain
	   \and
           Instituut voor Sterrenkunde, KU Leuven, Celestijnenlaan 200D, 
	   3001 Leuven, Belgium
}

\date{Received; Accepted}

\abstract 
%
%{context}
%
%{aims}
%
%{methods}
%
%{results}
%
%{conclusions}
%
{X-rays/EUV radiation emitted from wind-embedded shocks in hot,
massive stars can affect the ionization balance in their outer
atmospheres, and can be the mechanism responsible for the production
of highly ionized atomic species detected in stellar wind UV spectra.}
{To allow for these processes in the context of spectral analysis, we
have implemented the emission from wind-embedded shocks and related
physics into our unified, NLTE model atmosphere/spectrum synthesis
code {\sc FASTWIND}.}
{The shock structure and corresponding emission is calculated as a
function of user-supplied parameters (volume filling factor, radial
stratification of shock strength, and radial onset of emission). We
account for a temperature and density stratification inside the
post-shock cooling zones, calculated for radiative and adiabatic
cooling in the inner and outer wind, respectively. The high-energy
absorption of the cool wind is considered by adding important K-shell
opacities, and corresponding Auger ionization rates have been
included into the NLTE network. To test our implementation and
to check the resulting effects, we calculated a comprehensive
model grid with a variety of X-ray emission parameters.}
{We tested and verified our implementation carefully against
corresponding results from various alternative model atmosphere codes,
and studied the effects from shock emission for important ions from
He, C, N, O, Si, and P. Surprisingly, dielectronic recombination
turned out to play an essential role for the ionization balance of
\OIV/\OV\ in stars (particularly dwarfs) with \Teff\ $\sim$ 45,000~K. 
Finally, we investigated the frequency dependence and 
{\textit{radial}} behavior of the mass absorption coefficient, $\kappa_\nu(r)$, 
important in the context of X-ray line formation in massive star winds.}
{In almost all considered cases, {\textit{direct}} ionization is of major
influence (because of the enhanced EUV radiation field), and Auger
ionization significantly affects only \NVI\ and \OVI. The
approximation of a radially constant $\kappa_\nu$ is justified for $r
\gtrsim 1.2 \Rstar$ and $\lambda \lesssim 18$~\AA, and also for many
models at longer wavelengths. To estimate the actual {\textit{value}} of
this quantity, however, the \HeII\ opacities need to be calculated
from detailed NLTE modeling, at least for wavelengths longer than 18
to 20~\AA, and information on the individual CNO abundances has to be
present.}


\titlerunning{{\sc FASTWIND} -- X-ray emission from wind-embedded shocks}
\authorrunning{L.P. Carneiro et al.}

\maketitle
%
%________________________________________________________________
%
\section{Introduction} 
\label{Introduction}

Most of our knowledge about the physical parameters of hot stars
%(e.g., effective temperatures, gravities, wind-properties, chemical
%composition of the outer layers) 
has been inferred by means of quantitative spectroscopy, i.e., the
analysis of stellar spectra based on atmospheric models. The
computation of such models is quite challenging, mostly because of the
intense radiation fields of hot stars leading to various effects that
are absent in the atmospheres of cooler ones, such as the requirement
for a kinetic equilibrium description (also simply called NLTE =
non-LTE) and the presence of strong, radiation-driven winds.

In recent decades, a number of numerical codes have been developed
which enable the calculation of synthetic profiles/spectral energy
distributions (SEDs) from such hot stars. Apart from plane-parallel,
hydrostatic codes that can be used to analyze those atmospheres which
are less affected by the wind (e.g., {\sc tlusty}, \citealt{hubeny98};
{\sc Detail/Surface}, \citealt{Giddings81, ButlerGiddings85}), all of
these codes apply the concept of {\textit{unified}} (or {\textit{global}}) model
atmospheres \citep{Gabler89} which aims at a consistent treatment of
both photosphere and wind, i.e., including (steady-state) mass loss
and velocity fields. Examples of such codes are {\sc CMFGEN}
\citep{hilliermiller98}, {\sc PHOENIX} \citep{haus92}, {\sc PoWR}
\citep{Graf02}, WM-{\sc basic} \citep{pauldrach01} and {\sc FASTWIND}
\citep{Puls05, rivero12}.\footnote{The multi-component code developed
by \citet{KK01} that will be referred to later on has been designed to
calculate the wind properties, and has not been used for diagnostic
purposes so far.} A brief comparison of these different codes can be
found in \citet{puls09}.

In the present paper, we report on recent progress to improve the
capabilities of {\sc FASTWIND}, which is widely used to analyze the {\textit{
optical}} spectra of hot massive stars (e.g., in the context of the
VLT-{\sc flames} survey of massive stars, \citealt{evans08}, and the
VLT-{\sc flames} Tarantula Survey, \citealt{evans11}). One of the most
challenging aspects of these surveys was the analysis of the
atmospheric nitrogen content (processed in the stellar core by the
CNO-cycle and transported to the outer layers by rotational mixing),
in order to derive stringent constraints for up-to-date evolutionary
calculations. Though the optical nitrogen analysis of B-stars (dwarfs
and supergiants with not too dense winds) could still be performed by
a hydrostatic code (in this case {\sc TLUSTY}, e.g.,
\citealt{hunter07, hunter08}), a similar analysis of hotter stars with
denser winds required the application of unified model atmospheres,
due to the wind impact onto the strategic nitrogen lines
\citep{rivero11, rivero12, martins12}. Moreover, because of the
complexity of the involved processes, the precision of the derived
nitrogen abundances\footnote{which, for early-type O-stars, suggest 
very efficient mixing processes already at quite early stages
\citep{rivero122}} is still questionable. To independently check this
precision and to obtain further constraints, a parallel investigation
of the carbon (and oxygen) abundances is urgently needed, since at
least the N/C abundance {\textit{ratio}} as a function of N/O might be
predicted almost independent from the specific evolutionary scenario
\citep{przybilla10}, and thus allows individually derived 
spectroscopic abundances to be tested (see also
\citealt{martins15a}).

As shown by \citet{MartinsHillier12}, however, the optical diagnostics
of carbon in O-stars is even more complex than the nitrogen analysis,
since specific, important levels are pumped by a variety of UV
resonance lines. Thus, an adequate treatment of UV lines (both for the
optical diagnostics, but also to constrain the results by an
additional analysis of carbon lines located in the UV) is inevitable.
If at least part of these lines are formed in the wind, the inclusion
of X-ray and EUV emission from wind-embedded shocks turns out to be
essential (see below); this is the main reason (though
not the only one) for our current update of {\sc FASTWIND}. Other codes such as {\sc
CMFGEN}, {\sc PoWR}, and WM-{\sc basic} already include these
processes, thus enabling the modeling of the UV (e.g.,
\citealt{pauldrach01, Crowther02, HamannOskinova12}) and the analysis
of carbon (plus nitrogen and oxygen, e.g., \citealt{bouret12,
martins15a, martins15b} for the case of Galactic O-stars). 

X-ray emission from hot stars has been measured at soft (0.1 to $\ga$
2~keV) and harder energies, either at low resolution in the form of a
quasi-continuum, or at high resolution allowing the investigation of 
individual lines (e.g., \citealt{Oskinova06, OwockiCohen06, Herve13,
leutenegger13, Cohen14, rauw15}). Already the first X-ray satellite
observatory, EINSTEIN, revealed that O-stars are soft X-ray sources
\citep{harnden79, seward79}, and \citet{cassinelli83} 
were the first to show that the observed X-ray emission is due to thermal
emission, dominated by lines.
%interpreted the 
%emission near 2Kev in the spectrum of $\zeta$ Ori, as line emission 
%of Si xiii and S xv from a very hot component (~1.5$\cdot$10$^7$~K).}
Follow-up investigations, particularly by
ROSAT, have subsequently allowed us to quantify X-ray properties for
many OB-stars (see \citealt{KP00} and references therein). Accounting
also for more recent work based on CHANDRA and XMM-{\sc Newton}, it
was found that the intrinsic X-ray emission of `normal' O-stars is
highly constant w.r.t. time (e.g., \citealt{naze13}), and that the
level of X-ray emission is quite strictly related to basic stellar and
wind parameters, e.g., $L_x/L_{\rm bol} \approx 10^{-7}$ for O-stars
\citep{Chlebo89, Sana06, naze11}. 

Such X-ray emission is widely believed to originate from wind-embedded
shocks, and to be related to the line-driven instability (LDI, e.g.,
\citealt{LS70, ORI, OCR88, Owocki94b, Feldmeier95}). In terms of a
stationary description, a simple model (e.g., \citealt{Hillier93,
Cassinelli94}) assumes randomly distributed shocks above a minimum
radius, \Rmin\ $\approx$ 1.5 \Rstar\ (consistent with X-ray line
diagnostics, e.g., \citealt{leutenegger13}, but see also
\citealt{rauw15}) where the hot shocked gas (with temperatures of a
few million Kelvin and a volume filling factor on the order of
$10^{-3}$ to a few $10^{-2}$) is collisionally ionized/excited and
emits X-ray/EUV photons due to spontaneous decay, radiative
recombinations and bremsstrahlung. The ambient, {\textit{cool}} wind then
re-absorbs part of the emission, mostly via K-shell processes.
The strength of this wind-absorption has a strong
frequency dependence. For energies beyond 0.5~keV (e.g., the
CHANDRA-bandpass), the absorption is quite modest (e.g.,
\citealt{Cohen11}), whilst for softer X-rays and the EUV regime the
absorption is significant, even for winds with low mass-loss rate
(e.g., \citealt{Cohen96}). In the latter case, only a small fraction
of the produced radiation actually leaves the wind. 

This simple model, sometimes extended to account for the post-shock
cooling zones of radiative and adiabatic shocks (see
\citealt{Feldmeier97b}, but also \citealt{OwockiSundqvist13}), is used
in the previously mentioned NLTE codes, particularly to account for
the influence of X-ray/EUV emission on the photo-ionization rates.

Since the detection of high ionization stages in stellar wind UV
spectra, such as {\OVI}, S\,{\sc vi}, and {\NV} \citep{Snow76,
LamersMorton76, Lamers78}, that cannot be produced in a cool wind
(thus, denoted by `superionization'), the responsible mechanism was
(and partly still is) subject to debate. Because the X-ray and
associated EUV luminosity emitted by the shocks is quite strong, it
can severely affect the degree of ionization of highly ionized
species, by Auger ionization \citep{Macfarlane93}, and even more by
direct ionization in the EUV \citep{pauldrach94c, pauldrach01}. A first
{\textit{systematic}} investigation of these effects on the complete FUV
spectrum, as a function of stellar parameters, mass loss, and X-ray
luminosity has been performed by \citet{Garcia05}.

In this paper, we present our approach for implementing wind-embedded
shocks into {\sc FASTWIND}, to allow for further progress as outlined
above, and report on corresponding tests and first results. In
Sect.~\ref{method}, our model for the X-ray emission and cool-wind
absorption is described, together with the coupling to the equations
of statistical equilibrium. In Sect.~\ref{model_grid} we present our
model grid which constitutes the basis of our further discussion.
Sect.~\ref{tests} provides some basic tests, and Sect.~\ref{results}
presents first results. In particular, we discuss how the ionization
fractions of specific, important ions are affected by X-ray emission,
and how these fractions change when the description of the emission
(filling factors, shock temperatures) is varied
(Sect.~\ref{results_fractions}). We compare with results from
other studies (Sect.~\ref{comparing}), and investigate the impact of
Auger compared to direct ionization (Sect.~\ref{Auger_impact}). We
discuss the impact of dielectronic recombination in \OV\ in
Sect.~\ref{DR_OV}, and comment on the radial behavior of the mass
absorption coefficient (as a function of wavelength), an important
issue for X-ray line diagnostics (Sect.~\ref{op_section}).  Finally,
we present our summary and conclusions in Sect.~\ref{conclusions}.

\section{Implementation of X-ray emission and absorption in FASTWIND}
\label{method}

Our implementation of the X-ray emission and absorption from
wind-embedded shocks follows closely the implementation by
\citet{pauldrach01} (for WM-{\sc basic}, see also
\citealt{pauldrach94c}), which in turn is based on the model for shock
cooling zones developed by \citet{Feldmeier97b} (see
Sect.\ref{Introduction}). Except for the description of the cooling
zones, this implementation is similar to the approaches by
\citet{hilliermiller98} ({\sc CMFGEN}) (who use a different definition
of the filling factor, see below), \citet{Oskinova06} (POWR), and
\citet[hereafter KK09]{krticka09}. In the following, we
summarize our approach.
\subsection{X-ray Emission}
\label{xrayemission}

Following \citet{Feldmeier97b}, the energy (per unit of volume, time and frequency), 
emitted by the {\textit{hot}} gas into the full solid angle 4$\pi$ can be written
as\footnote{The corresponding emissivity is lower by a factor
$1/4\pi$.}
\beq
\label{emission_xray}
\epsilon_{\nu} =
\fv(r)\protden(r)\elecden(r)\Lambda_{\nu}(\elecden(r),\Tshock(r))
\eeq
where \protden($r$) and \elecden($r$) are the proton and electron density 
of the (quasi-)stationary, `cool' (pre-shock) wind, \Tshock($r$) is the
shock temperature, and \fv($r$) the filling factor related to the
(volume) fraction of the X-ray emitting material.\footnote{The actual, local
pre-shock density may be different from its quasi-stationary
equivalent, but this difference gets absorbed in the \fv-factor.}
Indeed, this
definition differs from the formulation suggested by \citet[their
Eq.~2]{Hillier93}, since we include here their factor\footnote{accounting
for the density jump in a strong adiabatic shock} 16 into \fv. 
This definition is then identical with that
used in WM-{\sc basic}, POWR (presumably\footnote{We were not able to
find a definite statement, but \citet{Oskinova06} also refer to 
\citet{Feldmeier97b}.}) and by KK09, whilst the
relation to the filling factor used in {\sc CMFGEN}, $e_{\rm s}$, is given
by
\beq
\label{filfac}
\fv = 16\, e_{\rm s}^2.
\eeq
In principle, $\Lambda_{\nu}$ is the frequency dependent volume emission
coefficient (`cooling function') per proton and electron, calculated 
here using the Raymond-Smith code (\citealt{Raysmith77}, see also
\citealt{Smith01}), with abundances from the {\sc FASTWIND} input,
%\footnote{based on solar abundances from \citealt{asplund09}}, 
and neglecting the weak dependence on \elecden. We evaluate the
cooling function at a fixed electron density, $\elecden =
10^{10}$~cm$^{-3}$ (as also done, e.g., by \citealt{Hillier93} and
\citealt{Feldmeier97b}), and have convinced ourselves of the validity
of this approximation. 
We note here that the only spectral
features with a significant dependence on electron density are the
forbidden and intercombination lines of He-like emission complexes, and
even there (i) the density dependence is `swamped' by the dependence on UV
photo-excitation, and (ii) in any case the flux of the 
forbidden plus intercombination line complex (f+i lines are very closely
spaced) is conserved.

Contrasted to the assumption of a hot plasma with a fixed post-shock
temperature and density (as adopted in some of the above codes), in our
implementation we account for a temperature and density stratification
in the post-shock cooling zones, noting that the decreasing temperature
and increasing density should significantly contribute to the 
shape of the emitted X-ray spectrum \citep{krolray85}. To this end, we adopt
the structure provided by \citet{Feldmeier97b}, and integrate the
emitted energy (Eq.~\ref{emission_xray}) over the cooling zone, 
\beq
\label{emission_xray2} 
\overline{\epsilon}_{\nu} = \fv(r)\protden(r)\elecden(r)
\overline{\Lambda_{\nu}}(10^{10}~{\rm cm}^{-3},\Tshock(r)), 
\eeq 
with
\beq
\label{cooling_function}
\overline{\Lambda_{\nu} }(\Tshock(r)) = \pm \frac{1}{L_{c}} 
\int_{r}^{r\pm L_{c}}f^{2}(r')\, \Lambda_{\nu}(\Tshock(r) \cdot g(r'))\ 
\dd r',
\eeq 
where $r$ is the position of the shock front, and $L_{c}$ the spatial
extent of the cooling zone. In this formulation, the `+' sign
corresponds to a reverse shock, and the `$-$' sign to a forward one.
The functions $f$ and $g$ provide the
normalized density and temperature stratification inside the cooling zone,
and are calculated following \citet{Feldmeier97b}, accounting for
radiative and adiabatic cooling in the inner and outer wind,
respectively (see Sect.~\ref{radadcool}). We integrate over 1,000
subgrid points within $L_{c}$, finding identical results for both
$f(r)$ and $g(r)$ as well as for $\overline{\Lambda_{\nu}}$, compared to
the original work (Figs.~1/7 and 2/8 in \citealt{Feldmeier97b}).
Note that by setting $f = g = 1$, we are able to return to 
non-stratified, isothermal shocks.

In our implementation, the (integrated) cooling function and thus the
emissivity is evaluated in the interval between 1 eV and 2.5 keV, for a
bin-size of 2.5 eV. These emissivities are then resampled onto our
coarser frequency grid as used in {\sc FASTWIND}, in such a way as to
preserve $\int \varepsilon_{\nu}$ d$\nu$ in each of the coarser subintervals,
thus enabling correct photo-integrals for the rate equations.

The immediate post-shock temperature, \Tshock$(r)$, entering
Eq.~\ref{cooling_function}, follows from the Rankine-Hugoniot
equations:
\beq
\label{shock_temp}
\Tshock(r) = \frac{3}{16}\frac{\mu m_{\textnormal{H}}}{k_{\rm B}}\  \biggl(u^{2}+
\biggl[ \frac{14}{5}\adiabsound^{2}\ \biggl(1 - \frac{3}{14}\ 
\frac{\adiabsound^{2}}{u^{2}}\biggr)\biggr]\biggr)
\eeq 
where $u$ is the jump velocity, $\mu$ the mean atomic weight, and
\adiabsound\ the adiabatic upstream sound speed. For simplicity, we
calculate the shock temperature from a more approximate expression,
neglecting the term in the square bracket, i.e., assuming the strong
shock scenario ($u^{2} \gg \adiabsound^{2}$):
\beq
\label{shock_temp2}
\Tshock(r) = \frac{3}{16}\frac{\mu m_{\textnormal{H}}}{k_{\rm B}} u^{2}
\eeq 
To derive \Tshock, we thus need to specify the jump velocity $u$, adopted 
in accordance with \citet[their Eq.~3]{pauldrach94c} as
\beq
\label{jump_velo}
u(r) = u_{\infty} \biggl[\frac{v(r)}{\vinf}\biggr]^{\gamma_x}
\eeq 
where $u_{\infty}$ is the maximum jump speed which in our
implementation is an input parameter (on the order of 300 to 600~\kms,
corresponding to a maximum shock temperature, $\Tshockmax \approx
10^6$ to $5\cdot10^6$~K for O-stars), together with the exponent
$\gamma_x$ (in the typical range 0.5{\ldots}2) that couples the jump
velocity with the outflow velocity, controlling the shock strength. A
parameterization such as Eq.~\ref{jump_velo} is motivated primarily by
the observed so-called `black troughs' in UV P-Cygni profiles. Namely,
when modeled using a steady-state wind\footnote{but see
\citealt{Lucy82, POF93, Sundqvist12a} for the case of time-dependent,
non-monotonic velocity fields}, such black troughs can only be reproduced when
assuming a velocity dispersion that increases in parallel with the
outflow velocity, interpreted as a typical signature of wind-structure
(e.g., \citealt{GL89, Haser95}). Note, however, that
Eq.~\ref{jump_velo} only represents one possible implementation of the
radial distribution of wind-shock strengths, and that ultimately the 
user will be responsible for her/his choice of parameterization (see
also discussion in Sect.~\ref{conclusions}). 

The last required parameter is the onset radius of the X-ray
emission, \Rmin. This value is controlled by two input parameters,
$\Rmin^{\rm input}$ and a factor \mx\ (the latter in accordance 
with \citealt{pauldrach94c}). From these values, \Rmin\ is calculated
via
\beq
\Rmin = {\rm min}\,\bigl(\,\Rmin^{\rm input}, r(v_{\rm
min})\,\bigr) \qquad \mbox{with} \qquad v_{\rm min} = \mx \, \adiabsound
\eeq
For all radii $r > \Rmin$, the X-ray emission is switched on. \Rmin\
values from 1.1 to 1.5~\Rstar\ are, e.g., supported by
\citet{pauldrach94c}, from their analysis of the \OVI\ resonance
lines. \cite{Hillier93} analyzed the sensitivity to \Rmin, pointing to
indistinguishable X-ray-flux differences when the onset is varied
between 1.5 and 2~\Rstar. Recent analyses of X-ray {\textit{line}}
emission from hot star winds also point to values around 1.5~\Rstar
(e.g., \citealt{Leutenegger06, Oskinova06, Herve13, Cohen14}), though
\citet{rauw15} derived a value of 1.2~\Rstar\ for the wind of
$\lambda$~Cep.

%LP
%It is not our objective on this work to constrain values of X-ray 
%parameters for specic stars. However in Sect. \ref{comp_wmbasic} we show 
%a profiles comparision with WM-Basic, which has exhaustively presented along 
%the years, fits over observations. Therefore once we reach an excellent 
%agreement between the results produced by both 
%codes, we assume that the same methods can be used with FASTWIND in order to 
%obtain an appropriate X-ray description for a particular star. 
%
\subsection{X-ray absorption and Auger ionization}
\label{xrsandauger} Besides the X-ray emission, the absorption by
the `cold' background wind\footnote{The optical depths {\textit{inside}}
the shocked plasma are so low that absorption can be neglected there.}
and needs to be computed. 

In {\sc FASTWIND}, the `cool' wind opacity is computed in NLTE, and to
include X-ray absorption requires that we (i) extend the frequency grid and
coupled quantities (standard\footnote{= outer electron shell} opacities and
emissivities, radiative transfer) into the X-ray domain (until 2.5 keV
$\approx$ 5~\AA), and (ii) compute the additional absorption by
inner shell electrons, leading, e.g., to Auger ionization. So far, we
included only K-shell absorption for light elements using data from
\cite{Dalta72}. L- and M-shell processes for heavy elements -- which
are also present in the considered energy range -- have not been
incorporated until now, but would lead to only marginal effects, as
test calculations by means of WM-{\sc basic} have shown. 

We checked that the K-shell opacities by
\cite{Dalta72} are quite similar (with typical differences less than
5\%) to the alternative and more `modern' dataset from
\citet{Verner95}, at least in the considered energy range (actually,
even until 3.1~keV).\footnote{The major reason for
using data from \citet{Dalta72} was to ensure compatibility with results
from WM-{\sc basic}, to allow for meaningful comparisons. In the near
future, we will update our data following \citet{Verner95}.}

The reader may note that though the provided dataset includes
K-shell opacities from the elements C, N, O, Ne, Mg, Si, and S, the
last one (S) has threshold energies beyond our maximum energy, 2.5
keV, so that K-shell absorption and Auger ionization for this element
is not considered in our model.

After calculating the radiative transfer in the X-ray regime,
accounting for standard and K-shell opacities as well as standard and
X-ray emissivities, we are able to calculate the corresponding
photo-rates required to consider Auger-ionization in our NLTE
treatment. Here, we do not only include the transition between ions
separated by a charge-difference of two (such as, e.g., the ionization
from {\OIV} to {\OVI}), but we follow \cite{Kaastra93} who stressed
the importance of cascade ionization processes, enabling a sometimes
quite extended range of final ionization stages. E.g., the branching
ratio for \OIV\ to \OV\ vs. \OIV\ to \OVI\ is quoted as 96:9904 whilst
the branching ratios for \SiIII\ to \SiIV/\SiV/\SiVI\ are 3:775:9222,
i.e., here the major Auger-ionization occurs for the process {\sc III}
to {\sc VI}. In our implementation of Auger ionization, we have
accounted for all possible branching ratios following the data
provided by Kaastra \& Mewe.

Finally, we re-iterate that in addition to such inner shell
absorption/Auger ionization processes, direct ionization due to
X-rays/enhanced EUV radiation (e.g., of \OV\ and \OVI) is essential
and `automatically' included in our FASTWIND modeling. The impact of
direct vs.  Auger ionization will be compared in
Sect.~\ref{Auger_impact}.

\subsection{Radiative and Adiabatic cooling}
\label{radadcool}

As pointed out in Sect.~\ref{xrayemission}, the shock cooling zones
are considered to be dominated by either radiative or adiabatic
cooling, depending on the location of the shock front. More
specifically, the transition between the two cooling regimes is obtained
from the ratio between the radiative cooling time, $t_{c}$, i.e., the
time required by the shocked matter to return to the ambient wind
temperature, and the flow time, $t_{f}$, the time for the material to
cross $L_{c}$.\footnote{Expressions for these quantities
can be found in \citet{Feldmeier97b}, but see also \citet{Hillier93}.}
In the inner part of the wind, the cooling time is shorter than the
flow time, and the shocks are approximated as radiative. Further out in
the wind, at low densities, $t_{c} \gg t_{f}$, and the cooling is
dominated by adiabatic expansion (see also \citealt{simox66}). In our
approach, we switch from one treatment to the other when a unity ratio
is reached, where $t_c/t_f \propto \Tshock(r)^{1.5} r\,v^2(r)/\mdot$.
For typical O-supergiants and shock temperatures, the transition
occurs in the outermost wind beyond $r > 50~\Rstar$, whilst for
O-dwarfs the transition can occur at much lower radii, $r > 2.5~\Rstar$ or
even lower for weak-winded stars.

Basically, each cooling zone is bounded by a reverse shock at the
starward side and a forward shock at the outer side. Time-dependent wind
simulations (e.g., \citealt{Feldmeier95}) show that in the radiative
case the forward shock is much weaker than the reverse one, and is
thus neglected in our model. In the adiabatic case, we keep both the
reverse and the forward shock, and, because of lack of better knowledge,
assume equal \Tshock\ for both components ($\Theta = 1$ in the
nomenclature by \citealt{Feldmeier97b}), and an equal contribution of
50\% to the total emission.

\section{Model grid} 
\label{model_grid}

In this section, we describe the model grid used in most of the
following work. In order to allow for a grid of theoretical models
that enables us to investigate different regimes of X-ray emission for
different stellar types, and to perform meaningful tests, we use the
same grid as presented by \cite{pauldrach01} (their Table 5) for
discussing the predictions of their (improved) WM-{\sc basic}
code.\footnote{This grid, in turn is based on observational results from
\citet{Puls96}, which at that time did not include the effects of wind
inhomogeneities, so that the adopted mass-loss rates might be too
large, by factors from $\sim$3{\ldots}6.} Moreover, this grid has already been
used by \citet{Puls05} to compare the results from an earlier version
of {\sc FASTWIND} with the WM-{\sc basic} code.

For convenience, we present the stellar and wind parameters of this
grid in Table~\ref{tab_grid}.  For all models, the velocity field
exponent has been set to $\beta =0.9$. Note that the {\sc FASTWIND}
and WM-{\sc basic} models display a certain difference in the velocity
field\footnote{WM-{\sc basic} calculates the velocity field from a
consistent {\textit{hydrodynamic}} approach.}.

All entries displayed in Table~\ref{tab_grid} refer to
homogeneous winds, though for specific tests (detailed when required)
we have calculated micro-clumped models as well (i.e., assuming
optically thin clumps). We remind the reader that though clumping is
not considered in our standard model grid, a (micro-)clumped wind
could be roughly compared to our unclumped models as long as the
mass-loss rate of the clumped model corresponds to the mass-loss rate
of the unclumped one, divided by the square root of the clumping
factor.\footnote{Note, however, that K-shell
opacities scale linearly with density, i.e., $\propto \mdot$, and as
such are {\textit{not}} affected by micro-clumping.}

\begin{table}
\tabcolsep1.7mm
\begin{center}
\caption{Stellar and wind parameters of our grid models with homogeneous winds,
following \citet{pauldrach01}. For X-ray emission parameters, see text. 
%Right
%part: \lxray/\lbol\ (logarithmic) provided as input for WM-{\sc basic}
%(WMB), compared with the corresponding output value from {\sc
%FASTWIND} (FW), integrated in the frequency range between 0.1 to
%2.5~keV. See Sect.~\ref{comp_wmbasic}.
}
\label{tab_grid}
\begin{tabular}{lllcrcc}
\hline
\hline
\multicolumn{1}{l}{Model}
&\multicolumn{1}{l}{\Teff}
&\multicolumn{1}{l}{\logg}
%&\multicolumn{1}{c}{\Rstar/\rsun}
&\multicolumn{1}{c}{\Rstar}
&\multicolumn{1}{c}{\vinf}
&\multicolumn{1}{c}{\mdot}
&\multicolumn{1}{c}{\Rmin\ }
\\
\multicolumn{1}{l}{}
&\multicolumn{1}{l}{\tiny{(kK)}}
&\multicolumn{1}{l}{\tiny{(\cms)}}
&\multicolumn{1}{c}{\tiny{(\rsun)}}
&\multicolumn{1}{c}{\tiny{(\kms)}}
%&\multicolumn{1}{c}{(10$^{-6}$\mdot/yr)}
&\multicolumn{1}{c}{\tiny{(\mdu)}}
&\multicolumn{1}{c}{\tiny{(\Rstar)}}
\\
\hline
\multicolumn{7}{c}{Dwarfs} \\
\hline
\vspace{0.07mm}
D30 &  30  & 3.85   &  12  &   1800   &  0.008  & 1.24 \\
D35 &  35  & 3.80   &  11  &   2100   &  0.05   & 1.29 \\
D40 &  40  & 3.75   &  10  &   2400   &  0.24   & 1.20 \\
D45 &  45  & 3.90   &  12  &   3000   &  1.3    & 1.20 \\
D50 &  50  & 4.00   &  12  &   3200   &  5.6    & 1.23 \\
D55 &  55  & 4.10   &  15  &   3300   &  20     & 1.21 \\
\hline
\multicolumn{7}{c}{Supergiants} \\
\hline
\vspace{0.07mm}
S30 &  30  & 3.00   &  27  &   1500   & 5.0  & 1.51 \\
S35 &  35  & 3.30   &  21  &   1900   & 8.0  & 1.43 \\
S40 &  40  & 3.60   &  19  &   2200   & 10   & 1.33 \\
S45 &  45  & 3.80   &  20  &   2500   & 15   & 1.25 \\
S50 &  50  & 3.90   &  20  &   3200   & 24   & 1.25 \\
\hline
\end{tabular}
\end{center}
\end{table}

All models in the present work were calculated by means of the most
recent version (as described in \citealt{rivero12}) of the NLTE
atmosphere/spectrum synthesis code {\sc FASTWIND}, including the X-ray
emission from wind-embedded shocks as outlined in Sect.~\ref{method}.
Let us further point out that FASTWIND
calculates the temperature structure (of the photosphere and `cold'
wind) from the electron thermal balance \citep{Kubat99}, and its major
influence {\textit{in the wind}} is via recombination rates. In most
cases, this temperature structure is only slightly or moderately
affected by X-ray/EUV emission, since the overall
ionization balance {\textit{with respect to main ionization
stages}}\footnote{which dominate the heating/cooling of
the cold wind plasma via corresponding free-free, bound-free and
collisional (de-)excitation processes} remains rather unaffected 
(see Sect.~\ref{results}), except for extreme X-ray emission
parameters. In any case, the change of the net ionization rates for ions
with edges in the soft X-ray/EUV regime is dominated by modified
photo-rates (direct and Auger ionization), whilst the changes of
recombination rates (due to a modified temperature) are of second
order.

In FASTWIND, we used detailed model atoms for H, He, and N (described
by \citealt{Puls05} and \citealt{rivero12}) together with C, O, P 
(from the WM-{\sc basic} data base, see \citealt{pauldrach01}) and Si
(see \citealt{Trundle04}) as `explicit' elements. Most of the other
elements up to Zn are treated as background elements. For a
description of {\sc FASTWIND} and the philosophy of explicit and
background elements, see \citet{Puls05} and \citet{rivero12}. 

In brief, explicit elements are those used as
diagnostic tools and treated with high precision, by detailed atomic
models and by means of comoving frame transport for all line
transitions. The background elements (i.e., the rest) are needed `only'
for the line-blocking/blanketing calculations, and are treated in a
more approximate way, using parameterized ionization cross-sections
following \citet{seaton58} and a comoving frame transfer only for the
most important lines, whilst the weaker ones are calculated by
means of the Sobolev approximation.

We employed solar abundances from \cite{asplund09}, together with a
helium abundance, by number, $N_{\rm He}$/$N_{\rm H}$ = 0.1. 

Besides the atmospheric and wind parameters displayed in
Table~\ref{tab_grid}, our model of X-ray emission requires
the following additional input parameters: \fv, $u_{\infty}$,
$\gamma_x$, \mx, and $\Rmin^{\rm input}$, as described in the previous
section.

For most of the models discussed in Sect.~\ref{results}, we
calculated, per entry in Table~\ref{tab_grid}, 9 different sets of X-ray
emission: \fv\ (adopted as spatially constant) was set to 0.01, 0.03, and 0.05, whilst 
the maximum shock velocity, $u_{\infty}$,
was independently set to 265, 460, and 590~\kms, corresponding
to maximum shock temperatures of 1, 3, and 5$\cdot$10$^{6}$~K. 
 
For all models, we used $\gamma_x = 1.0$, $\Rmin^{\rm input} =
1.5~\Rstar$, and \mx = 20. This corresponds to an effective onset of
X-rays, \Rmin, between 1.2 and 1.5~\Rstar, or 0.1 and 0.2~\vinf,
respectively (see Table \ref{tab_grid}, last column).
Thus, our current grid comprises 9 times 11 = 99 models, and has
enough resolution for comparisons with previous results from other
codes and for understanding the impact of the X-ray radiation onto the
ionization fractions of various elements.

\section{Tests}
\label{tests}

In this section, we describe some important tests of our
implementation, including a brief parameter study. A comparison to
similar studies with respect to ionization fractions (also regarding
the impact of Auger ionization) will be provided in
Sect.~\ref{results}. Of course, we have tested much more than
described in the following sections, e.g., 

(i) the impact of $\gamma_x$ (see also \citealt{pauldrach01}),
particularly when setting $\gamma_x$ to zero (and consequently forcing
all shocks, independent of their position, to emit at the maximum
shock temperature, \Tshockmax). In this case and compared to our
standard grid with $\gamma_x = 1$, the dwarf models cooler than 50\,kK
display a flux increase of 2~dex shortward of 100\,\AA\ (already for
D50 this increase is barely noticeable), whilst the supergiant models
display a similar increase, but for wavelengths around 10\,\AA\ and
below.  In terms of ionization fractions, setting $\gamma_x$ to zero
results in an increase of highly ionized species (e.g., \OVI\ and
\NVI) by roughly one dex, from the onset of X-ray emission throughout
the wind. For all other dwarf models, this increase appears only out
to $\sim$4.0\,\Rstar. The same effect is present in the supergiant
models, but for a smaller radial extent.

(ii) We compared the ionization fractions of important atoms when
either treated as explicit (i.e., `exact') or as background (i.e,
approximate) elements (cf. Sect.~\ref{model_grid}), and we mostly
found an excellent agreement\footnote{In all cases, the agreement was
at least satisfactory.} for the complete model grid. 

(iii) During our study on the variations of the mass absorption
coefficient with \Teff\ and $r$ in the X-ray regime (see
Sect.~\ref{op_section}), we also compared our opacities with those
predicted by KK09 (their Fig.~15, displaying mass absorption
coefficient vs. wavelength), and we were able to closely reproduce
their results, at least shortward of 21~\AA\, (including the
dominating \OIV/\OV\ K-shell edge), but our model produces lower
opacities on the longward side, thus indicating a different He
ionization balance (see Sect.~\ref{op_section}). When comparing the
{\textit{averaged}} (between 1.5 and 5~\Rstar) absorption coefficients in
the wavelength regime shortward of 30~\AA, KK09 found a slight
decrease of 8\% after including X-rays in their models, because
of the induced ionization shift. This is consistent with our findings,
which indicate, for the same range of $r$ and $\lambda$, a
decrease by 9\%. 

\subsection{Impact of various parameters}
\label{test_parameters}

First, we study the impact of various parameters on the emergent
(soft) X-ray fluxes, in particular \Rmin, \fv, and \Tshockmax. For these
tests, we used the model S30 (see Table~\ref{tab_grid}, similar to the
parameters of $\alpha$~Cam (HD\,30614, O9.5Ia)) since the latter
object has been carefully investigated by \citet[their Table
9]{pauldrach01} as well.

Before going into further details, let us clarify that
the soft X-ray and EUV shock emission are composed almost entirely of
narrow lines, and that the binning and blending make the spectral
features look more like a pseudo-continuum, which is clearly visible
in the following figures (though most of them display the emergent
fluxes, and not the emissivities
themselves).\footnote{As shown by
\citet{pauldrach94c}, the total shock emissivity is roughly a factor
of 50 larger than the corresponding hot plasma free-free emission from
hydrogen and helium.}

\begin{figure}[t]
\resizebox{\hsize}{!}
{\includegraphics{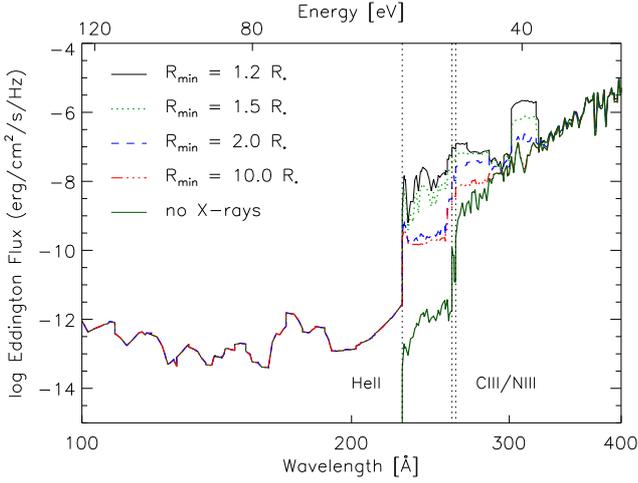}}
%file: paper_models/diff_rmin.pro
%
\caption{Emergent Eddington fluxes for model S30, with \Tshockmax\ =
	 3$\cdot$10$^{6}$~K and \fv\ = 0.03, for different onset radii
	 of X-ray emission, \Rmin, and for a model with an unshocked
	 wind. The vertical dotted lines refer to the
	 \HeII, \CII, and \NIII\ ionization edges, respectively.}
\label{diff_rmin}
\end{figure}

\paragraph{Impact of \Rmin.} The sensitivity of the X-ray fluxes on
\Rmin\ is displayed in Fig.~\ref{diff_rmin}, where the other
parameters were fixed at their center values within our small X-ray
grid (i.e., \fv\ = 0.03 and \Tshockmax\ =
3$\cdot$10$^{6}$~K\footnote{Note that particularly the shock
temperature is quite high for such a stellar model, but chosen deliberately to
allow for somewhat extreme effects.}). 

Indeed, the only visible differences are present in the range between
the \HeII\ edge and roughly 330~\AA. Shortward of the \HeII\ edge, all
fluxes are identical (though only shown down to 100~\AA, to allow for
a better resolution), since the (cool) wind becomes optically thick
already far out in the wind at these wavelengths (\HeII, \OIV, etc.
continua, and K-shell processes). For $\lambda$ > 350\,\AA, on the
other hand, the shock emissivity becomes too low to be of significant
impact. 

\begin{figure}[t]
\resizebox{\hsize}{!}
{\includegraphics{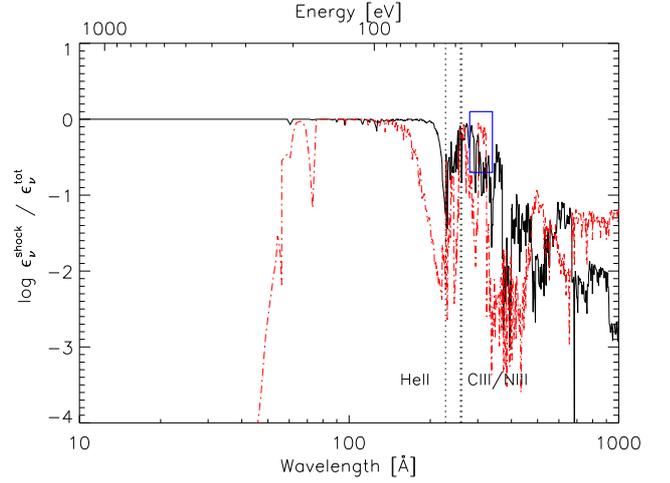}}
\caption{Ratio of shock emissivity to total emissivity for model S30
	 from Fig.~\ref{diff_rmin}, with \Rmin = 1.2~\Rstar. Solid
	 line: emissivity ratio at the outer boundary, $r \approx$
	 130~\Rstar; dash-dotted: emissivity ratio at the lower
	 boundary of X-ray emission, $r \approx$ 1.2~\Rstar. The box
	 located between 300 and 320~\AA\ highlights the strong shock
	 emissivity leading to the corresponding `emission feature'
	 present in Fig.~\ref{diff_rmin}.} 
\label{emiss}
\end{figure}

In this context, it is interesting to note that in
$\epsilon$~CMa (B2II, the only massive hot star with EUVE data) the
observed EUV emission lines in the range between 228 to 350 \AA\ {\textit{
each}} have a luminosity comparable to the total X-ray luminosity in
the ROSAT bandpass \citep{Cassinelli95}, which stresses the
importance of this wavelength region also from the observational
side.

In Fig.~\ref{emiss}, we show the ratio of the shock emissivity to the
total emissivity (including averaged line processes and Thomson
scattering), evaluated at the outer boundary of the wind (solid) and at
1.2~\Rstar\ (dash-dotted), corresponding to the onset of X-ray
emission in this model. There are a number of interesting features
visible: 

\noindent (i) The total emissivity in the outer wind is dominated by
shock emission from just shortward of the \HeII\ edge until 2.5~keV
(the highest energy we consider in our models). The emissivity in the
lower wind, however, is dominated by shock emission only until 200~eV,
whilst for larger energies the (local) shock contribution decreases
drastically, because the assumed shock temperatures ($\propto
(v(r)/\vinf)^2$) are rather low here ($\la$ 100 kK). The question is
then: Which processes dominate the {\textit{total}} emissivity at high
energies in the lower wind? Indeed, this is the re-emission from
electron-scattering, being proportional to the mean intensity, and
being quite high due to the large number of incoming photons from {\textit{
above}}, i.e., from regions where the shock temperatures are high! This
effect becomes also visible in the local radiative fluxes at these
frequencies, which are negative, i.e., directed inwards.

\noindent (ii) Both in the outer and inner wind, the shock emission is
also significant longward from the \HeII\ edge, until $\lambda
\approx$~350~\AA, thus influencing the ionization balance of important
ions. Whilst the fluxes of models without shock emission and those
with $\Rmin \ga 2 \Rstar$ display a significant absorption edge for
\CIII\ and \NIII\ (see Fig.~\ref{diff_rmin}), these edges have almost
vanished in the models with \Rmin\ = 1.2 {\ldots} 1.5 \Rstar, because
of the dominant shock emissivity increasing the degree of ionization.
Even more, all models display fluxes in this region which lie well
above those from models without shock emission, because of the higher
radiation temperatures compared to the cool wind alone.

\noindent
(iii) Beyond 350~\AA, the shock emissivity becomes almost irrelevant
(below 10\%), so that the corresponding fluxes are barely affected.

\noindent
(iv) For the two models with \Rmin\ = 1.2 and 1.5 \Rstar, a prominent
emission feature between roughly 300 and 320 \AA\ is visible in
Fig.~\ref{diff_rmin}. A comparison with Fig.~\ref{emiss} (note the
box) shows that this emission is due to dominating shock emission of
the lower wind, increasing the temperatures of the radiation field
beyond those of the unshocked wind.

Coming back to Fig.~\ref{diff_rmin}, significant flux differences
between the shocked and the unshocked models are visible for all 
values of \Rmin\ (even for \Rmin\ = 2 or 10~\Rstar) below $\lambda \la
350$~\AA, particularly below the \NIII\ and \CIII\ edges, because of
the higher ionization.

On the other hand, the models with \Rmin\ = 1.2 and 1.5~\Rstar\ are
almost indistinguishable, at least regarding the pseudo-continuum
fluxes. This turns out to be true also for \HeII~1640 and \HeII~4686, 
though these lines become sensitive to the choice of \Rmin\ if we
change \Rmin\ from 1.5 to 2~\Rstar, due to the different intensities
around the \HeII\ edge and around \HeII~303 (Lyman-alpha) in the
line-forming region. We will come back to this point in
Sect.~\ref{impact_helium}.

\begin{figure}[t]
\resizebox{\hsize}{!}
{\includegraphics{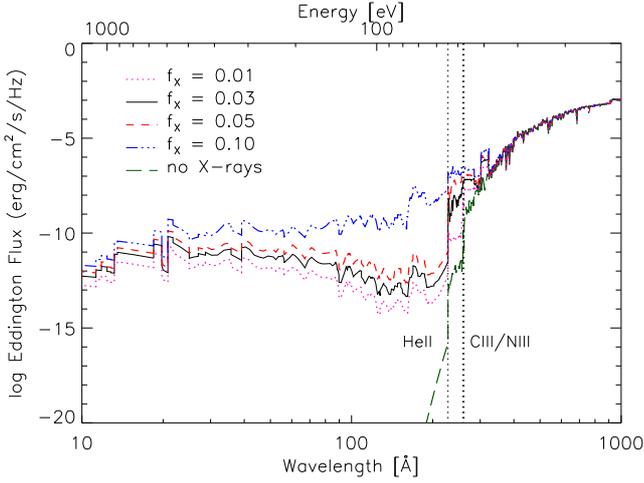}}
%LP: diff_f.pro
%
%
\caption{Emergent Eddington fluxes for model S30, with
	 \Tshockmax\ = 3$\cdot$10$^{6}$~K and $\Rmin^{\rm input}$ =
	 1.5~\Rstar, for different values of \fv, and for a model with
	 an unshocked wind. The histogram-like
	 flux-distribution at highest energies results from our
	 resampling of X-ray emissivities (see
	 Sect.~\ref{xrayemission}).}
\label{diff_f}
\end{figure}

\paragraph{Impact of \fv.} In Fig.~\ref{diff_f}, we investigate the 
impact of \fv, which has a most direct influence on the strength of
the X-ray emission (cf. Eqs.~\ref{emission_xray} and
\ref{emission_xray2}). Having more X-ray photons leads to higher X-ray
fluxes/luminosities and to less XUV/EUV-absorption from the cool wind,
because of higher ionization stages. The latter effect becomes
particularly visible for the model with \fv\ = 0.1, which was used to
check at which level of X-ray emission we start to change the overall
ionization stratification. Most importantly, helium
(with \HeII\ as main ion beyond 1.2~\Rstar\ for S30 models with
typical values $0.03 \la \fv \la 0.05$) becomes more ionized,
reaching similar fractions of \HeII\ and \HeIII\ between 2.2~\Rstar\
($\sim$0.5~\vinf) and 8.7~\Rstar\ ($\sim$0.8~\vinf). And also the main ionization
stage of oxygen (which is \OIV\ in S30 models with typical X-ray
emission parameters) switches to \OV\ between 1.8~\Rstar\ ($\sim$0.4~\vinf)
and 4.0~\Rstar\ ($\sim$0.7~\vinf) when \fv\ is set to 0.1. The change in the
ionization of helium (and oxygen) becomes clearly visible in the much
weaker \HeII\ edge and much higher fluxes in the wavelength range
below 228~\AA, compared to models with lower \fv.

\begin{figure}[t]
\resizebox{\hsize}{!}
{\includegraphics{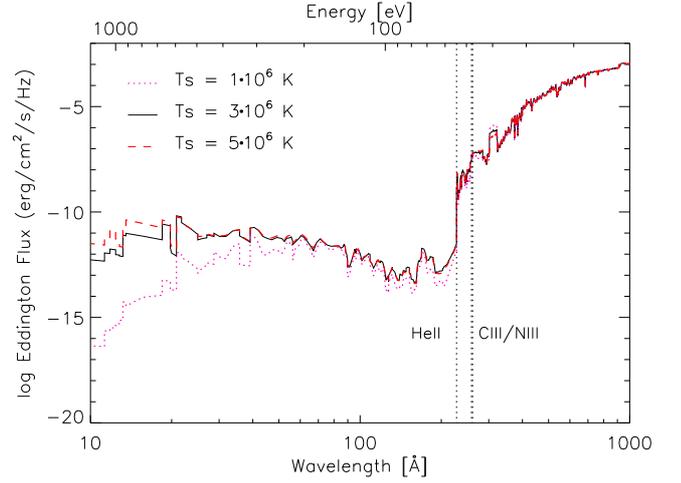}}
%LP: diff_tshock.pro
%
%
\caption{Emergent Eddington fluxes for model S30, with \fv\ = 0.03
	 and $\Rmin^{\rm input}$ = 1.5~\Rstar, for different values of
	 maximum shock temperature, \Tshockmax.} 
\label{diff_tshock}
\end{figure}

\paragraph{Impact of \Tshockmax.} As displayed in Fig.~\ref{diff_tshock}
(see also \citealt{pauldrach01}), the change in the maximum shock
temperature, \Tshockmax, becomes mostly visible for the fluxes shortward
of $\approx$ 60~\AA\ (of course, the hard X-ray band is even more
affected, but not considered in our models).  Though for the highest
maximum shock temperature considered here, 
\Tshockmax\ = $5 \cdot 10^{6}$~K (corresponding to
$u_{\infty} \approx 590$~\kms), we significantly increase the
population of the higher ionized atomic species, this is still not
sufficient to change the main ionization stages present in the wind.

\subsection{Scaling relations for \lxray} 
\label{lxscal}

From analytical considerations, \cite{OwockiCohen99} showed that for a
constant volume filling factor (and neglecting effects of radiative
cooling, see also below) the optically thin\footnote{with respect to
the cool wind absorption} wind X-ray luminosity depends on the square
of the mass-loss rate, \lxray\ $\propto$ (\mdot/\vinf)$^{2}$, whilst
the X-ray luminosity of optically thick winds scales linearly with the
mass-loss rate, \lxray\ $\propto$ \mdot/\vinf, provided that one
compares models with the same shock temperatures and assumes a
spatially constant X-ray filling factor. These relations become
somewhat modified if there is a dependence of \Tshock\ on the wind
terminal velocity, as adopted in our `standard' X-ray description (see
also KK09). 

\begin{figure}[t]
\psfrag{\lxray}[][][1.8]{$L_{\rm x}$ [erg/s]}
\psfrag{Mdot/vinf}[][][1.8]{\mdot/\vinf [in M$_\odot$/yr and 1000 kms$^{-1}$]}
\psfrag{eq1}[l][l][1.3]{$L_{\rm x} \propto 
   \Bigl(\frac{\displaystyle{\mdot}}{\displaystyle{\vinf}}\Bigr)^2$}
\psfrag{eq2}[l][l][1.3]{$L_{\rm x} \propto 
   \Bigl(\frac{\displaystyle{\mdot}}{\displaystyle{\vinf}}\Bigr)^2$}
\psfrag{eq3}[l][l][1.3]{$L_{\rm x}$ $\propto 
   \Bigl(\frac{\displaystyle{\mdot}}{\displaystyle{\vinf}}\Bigr)$}
\resizebox{\hsize}{!}
  {\includegraphics{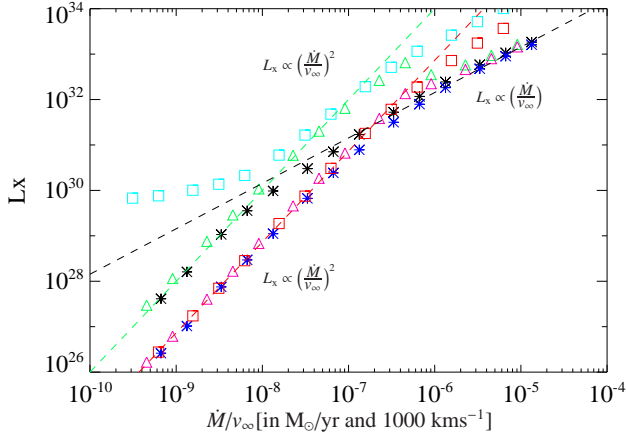}}
\caption{Emergent X-ray luminosities (in erg\,s$^{-1}$) as a function of
	 \mdot/\vinf. Supergiant models S30 (asterisks), S40
	 (triangles) and S50 (squares) with \Teff = 30, 40 and 50 kK,
	 respectively, and mass-loss rates between $10^{-9}$ and $2
	 \cdot 10^{-5} M_{\odot}/{\rm yr}$. All models have the same
	 X-ray properties, $f_x$ = 0.025, $\gamma_x$ = 0.5, $m_x$ =
	 20, and a maximum jump-velocity, $u_{\infty}$ = 400~\kms,
	 corresponding to maximum shock temperatures of $2.3 \cdot
	 10^6$~K. The X-ray luminosities have been calculated in the
	 range 0.1 $-$ 2.5 keV (black, green, turquoise), and in the
	 range 0.35 to 2.5 keV (blue, red and magenta). The dashed
	 lines (no fits) serve as guidelines to check the predicted
	 behavior for optically thin (red and green) and optically
	 thick (black) conditions. Note the strong deviation of models
	 S50 (turquoise squares) from the predicted optically thin
	 scaling, when integrating until 100 eV, due to `normal'
	 stellar/wind radiation just in this energy range. (See text.)}
\label{scalrel}
\end{figure}

Note, however, that in a more recent study, \citet{OwockiSundqvist13}
derived, again from analytic considerations, scaling relations for
\lxray\ for radiative and adiabatic shocks embedded in a cool wind. At
first glance, their assumptions seem quite similar to those adopted by
\citet{Feldmeier97b} (which is the basis of our treatment), but in the
end they predict different scaling relations for {\textit{radiative}}
shocks than resulting from the modeling here. This discrepancy might
lead to somewhat different scaling relations for $L_{\rm x}$, and
needs to be investigated in forthcoming work; for now, we simply
compare our models to the earlier results by \cite{OwockiCohen99} (a
similar test was done by KK09).  

To this end, we calculated S30, S40 and S50 wind models with a fixed
X-ray description: \fv\ = 0.025, \mx\ = 20, and $\gamma_x$ = 0.5. For our
tests we used, for all models, a constant maximum jump velocity,
$u_{\infty}$ = 400~\kms\ (corresponding to maximum shock temperatures
of $2.3 \cdot 10^6$~K), in order to be consistent with the above
assumptions.

For these models (with parameters, except for \mdot, provided in
Table~\ref{tab_grid}), we varied the mass-loss rates in an interval
between $10^{-9}$ and $2 \cdot 10^{-5} M_{\odot}/{\rm yr}$. and
integrated the resulting (soft) X-ray luminosities in two different
ranges: 0.1 to 2.5 keV and 0.35 to 2.5 keV. 

From \mdot\ $\ga$ 10$^{-7}$\msunyr\ on, the wind becomes successively
optically thick at higher and higher energies (though, e.g., for \mdot =
10$^{-6}$\msunyr\ it is still optically thin below $\sim 10$~\AA, i.e.,
above 1.24 keV). Indeed, the X-ray luminosities of our corresponding
models are linearly dependent on (\mdot/\vinf), as can be seen
in Fig.~\ref{scalrel} by comparing with the black dashed line. For
lower \mdot, the wind is optically thin at most high energy
frequencies, and also here our results follow closely the predictions
($\lxray \propto (\mdot/\vinf)^2$), when comparing with the red or
green dashed lines.

\begin{figure}[t]
\resizebox{\hsize}{!}
 {\includegraphics{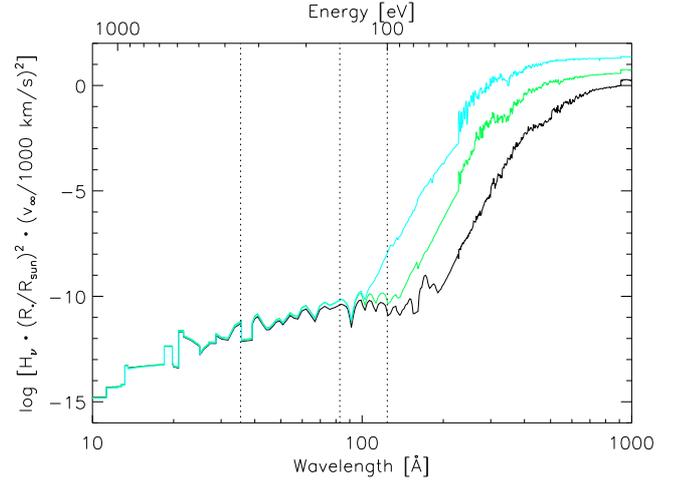}}
\caption{Logarithmic, scaled Eddington flux (in units of
	 erg\,cm$^{-2}$\,s$^{-1}$\,Hz$^{-1}$) as a function of
	 wavelength/energy, for the supergiant models S30 (black), S40
	 (green) and S50 (turquoise) with identical mass-loss rates,
	 $10^{-8} M_{\odot}/{\rm yr}$.  All models have the same X-ray
	 properties, as denoted in Fig.~\ref{scalrel}. The
	 Eddington fluxes have been scaled by $(\Rstar/\rsun)^2$ and
	 $(\vinf/1000~\kms)^2$, to ensure {\textit{theoretically}} similar
	 values of optically thin X-ray emission. 
%Obviously, our calculations provide identically results, in
%accordance with such predictions. Note, however, that the X-ray
%flux/luminosity of model S50 becomes contaminated by `normal'
%stellar/wind radiation longwards of 100 \AA, such potentially
%contaminating the {\textit{integrated}} X-ray emission (cf.
%Fig.~\ref{scalrel}). 
         The dotted lines denote energies of 350, 150, and 100 eV,
	 corresponding to 35, 83, and 124~\AA. (See text.)}
\label{compflux1}
\end{figure}

A second finding of Fig.~\ref{scalrel} relates to the optically thin
scaling for model S50, when either starting the integration at 100~eV
(turquoise squares) or at 350~eV (red squares). Whilst for S30
(asterisks) and S40 (triangles) the X-ray luminosities just increase
by roughly one dex when including the range from 100 to 350~eV but
still follow the predicted scaling relation, the
S50 models show an increase of four orders of magnitude for the lowest
\mdot/\vinf\ values in this situation (and do {\textit{not}} follow the
predictions).

To clarify this effect, Fig.~\ref{compflux1} shows the scaled (scaling
proportional to $\Rstar^2$ and $\vinf^2$) Eddington flux as a function
of wavelength and energy, for the supergiant models S30 (black), S40 
(green) and S50 (turquoise) with identical, low mass-loss rates,
$10^{-8} M_{\odot}/{\rm yr}$.  Additionally, energies of 100, 150 and
350~eV have been marked by dotted vertical lines. Beyond 150 eV, all
models, independent of their specific parameters, display the same
scaled fluxes, thus verifying the optically thin scaling of X-ray
luminosities (in this case, only with respect to \vinf). For the S50
model, however, the energy range below 150~eV (and, for other
parameter-sets, also below even higher energies) is contaminated by
`normal' stellar/wind radiation (which increases as function of \Teff;
see also \citealt{Macfarlane94}, their Fig.~5), leading to the strong
deviation from the optically thin X-ray scaling law as visible in
Fig.~\ref{scalrel}. In so far, the {\textit{total}} X-ray luminosity
(regarding the wind emission) of hotter objects might be overestimated
when integrating until 100~eV.

In summary, we conclude that our implementation follows the predicted
scaling relations, but we also suggest to choose a lower (in energy)
integration limit of 0.15~keV (or even 0.3~keV, to be on the safe side)
when comparing the X-ray luminosities of different stars (both with
respect to models {\textit{and}} observations). 

In this context, we note that there is a clear distinction
between the {\textit{observable}} soft X-ray and the longer-wavelength,
soft X-ray and XUV/EUV emission that is almost never directly
observed, but -- as already outlined  -- is
very important for photoionizing relevant ions. `Modern' X-ray
observatories such as XMM-{\sc Newton}/RGS and CHANDRA/HETG do not have a
response below 0.35 keV and 0.4 keV, respectively\footnote{though ROSAT observed down
to 0.1~keV, and also EUVE made a few important measurements relevant
for massive stars, in particular for $\epsilon$~CMa (B2II), e.g., 
\citet{Cassinelli95}}, and even a modest ISM column makes it functionally
impossible to see X-ray emission below 0.5 keV. 

\begin{table}
\begin{center}
\caption{Left part: X-ray emission parameters used to compare {\sc
FASTWIND} and WM-{\sc basic} models ($u_{\infty}/\vinf = 0.3$ and
$\gamma_x$ = 1.0). For stellar and wind parameters see
Table~\ref{tab_grid}. Right part: \lxray/\lbol\
(logarithmic) provided as input for WM-{\sc basic} (WMB), compared
with the corresponding output value from {\sc FASTWIND} (FW),
integrated in the frequency range between 0.1 to 2.5~keV. See
Sect.~\ref{comp_wmbasic}.}
%
%\vspace{0.3cm}
\label{tab_wm}
\begin{tabular}{lcccc|cc}
\hline 
\hline
\multicolumn{1}{l}{Model}
&\multicolumn{1}{c}{\fv}
&\multicolumn{1}{c}{\Rmin}
&\multicolumn{1}{c}{$u_{\infty}$}
&\multicolumn{1}{c|}{\Tshockmax}
&\multicolumn{1}{c}{\lxray/\lbol}
&\multicolumn{1}{c}{\lxray/\lbol}
\\
\multicolumn{1}{l}{}
&\multicolumn{1}{c}{\tiny{(\%)}}
&\multicolumn{1}{c}{\tiny{(\Rstar)}}
&\multicolumn{1}{c}{\tiny{(\kms)}}
&\multicolumn{1}{c|}{\tiny{(10$^{6}$ K)}}
&\multicolumn{1}{c}{\tiny{({\sc WMB})}}
&\multicolumn{1}{c}{\tiny{({\sc FW})}}\\
\hline
\multicolumn{7}{c}{Dwarfs} \\
\hline
D30 &  2.00  &  1.24  &   532   & 3.90  &$-$9.4&$-$9.4\\
D35 &  0.96  &  1.29  &   622   & 5.27  &$-$8.3&$-$8.5\\
D40 &  1.44  &  1.21  &   715   & 6.98  &$-$7.0&$-$7.0\\  
D45 &  1.38  &  1.20  &   894   & 10.9  &$-$6.4&$-$6.5\\
D50 &  2.11  &  1.22  &   950   & 12.4  &$-$5.6&$-$5.8\\
\hline                                           
\multicolumn{7}{c}{Supergiants} \\
\hline
S30 &  1.99  &  1.50  &   453   & 2.93  &$-$6.3&$-$6.4\\
S35 &  1.24  &  1.43  &   577   & 4.54  &$-$6.2&$-$6.3\\
S40 &  0.80  &  1.33  &   663   & 6.00  &$-$6.3&$-$6.5\\   
S45 &  0.93  &  1.25  &   754   & 7.76  &$-$6.2&$-$6.3\\
S50 &  3.13  &  1.26  &   941   & 12.1  &$-$5.2&$-$5.4\\
\hline
\end{tabular}
\end{center}
\end{table}

\subsection{Comparison with WM-{\sc basic} models}
\label{comp_wmbasic}

Finally, we checked also the quantitative aspect of our results, by
comparing with analogous\footnote{remember the difference in the
velocity fields} WM-{\sc basic} models. As already pointed out, the X-ray
description in both codes is quite similar, and there is only one
major difference. In WM-{\sc basic}, the user has to specify a certain value
for \lxray/$L_{\rm Bol}$ (e.g., $10^{-7}$ as a prototypical value),
and the code determines iteratively the corresponding \fv, whilst
the latter parameter is a direct input parameter in the updated version 
of {\sc FASTWIND}. In both cases, we used a frequency range between 0.1 to
2.5~keV.

Thus, we first calculated WM-{\sc basic} models with stellar/wind
parameters from Table~\ref{tab_grid}, and with X-ray emission
parameters from Table~\ref{tab_wm}. For the maximum jump velocity we
assumed, as an extreme value, $u_{\infty}/\vinf = 0.3$,
together with X-ray luminosities as displayed in the
sixth column of Table~\ref{tab_wm}. These values then correspond to
the \fv\ values provided in the second column of the same table,
acquired from the WM-{\sc basic} output. We note here that the input
values of \lxray/$L_{\rm Bol}$ (to WM-{\sc basic}) were not chosen on
physical grounds, but were estimated in such a way as to result in
similar values for \fv, in the range between 0.01 to 0.03. 

%The other way round, while a dwarf model at
%40~kK (D40) with a filling factor of 1.4\% results in \lxray/$L_{\rm
%Bol} \approx 10^{-7}$ (given the other X-ray emission parameters), a
%similar model at 30~kK produces less than \lxray/$L_{\rm Bol}$ less than
%$10^{-9})$, and whilst required the `canonical value' would result in
%\fv\ values where our description would certainly break down.

To check the overall consistency, we calculated a similar set of {\sc
FASTWIND} models, now using the \fv\ values from Table~\ref{tab_wm} as
{\textit{input}}. In case of consistent models, the resulting \lxray\ values
(from the output) should be the same as the corresponding input values
used for WM-{\sc basic}. Both these values are compared in the last two
columns of Table~\ref{tab_wm}. Obviously, the agreement is quite
good (not only for the supergiants, but also for the dwarfs), 
with differences ranging from 0.0
to 0.2 dex, and an average deviation of 0.13 dex.

\begin{figure}[h]
\resizebox{\hsize}{!}
{\includegraphics{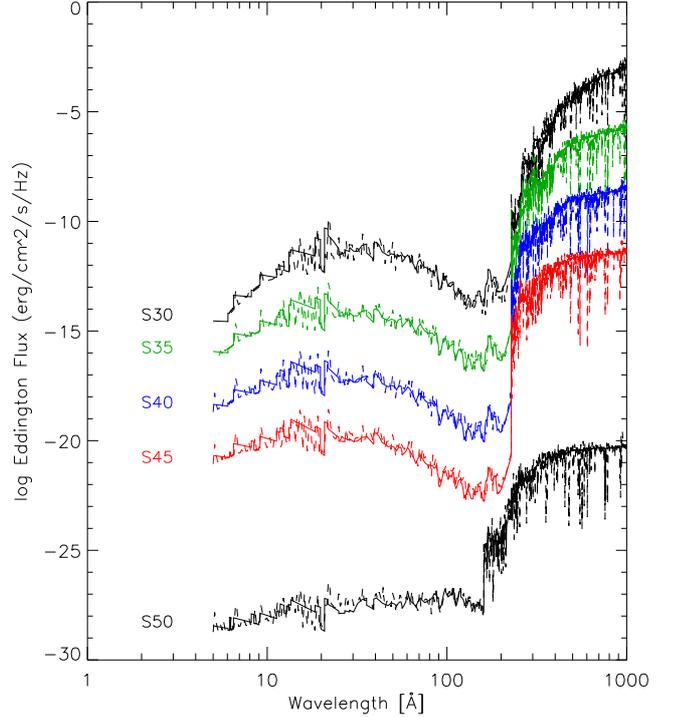}}
\caption{Logarithmic Eddington fluxes as a function of wavelength for
	 supergiant models (see Table~\ref{tab_grid} and
	 Table~\ref{tab_wm}). The solid lines refer to results from
	 our updated version of {\sc FASTWIND}, and the dashed ones to
	 WM-{\sc basic} results \citep{pauldrach94c, pauldrach01}. For
	 clarity, the S35, S40, S45, and S50 model fluxes have been
	 shifted by $-3$, $-6$, $-9$, and $-18$ dex, respectively.}
\label{comp_eddington_flux}
\end{figure}

In a second step, we compared the fluxes resulting from 
this procedure in Fig.~\ref{comp_eddington_flux}. For clarity, the
fluxes were shifted by $-3$, $-6$, $-9$, and $-18$ dex (S35, S40, S45, S50),
where the solid lines correspond to the {\sc FASTWIND} and the dashed
lines to the WM-{\sc basic} results.

The comparison shows a remarkably good agreement, with no striking
differences. Smaller differences in the lower wavelength range
($\lambda <$ 100~\AA) are related to a different frequency sampling
(without an effect on the total X-ray luminosity). At longer
wavelengths, these differences are related to the fact that WM-{\sc basic}
provides high-resolution fluxes, whilst {\sc FASTWIND} calculates
fluxes using averaged line-opacities\footnote{for details, see
\citealt{Puls05}}. Most important, however, is our finding that the
fluxes are not only similar at high frequencies (indicating similar
emissivities and cool-wind opacities), but also longward from the
\HeII\ edge, indicating a similar ionization equilibrium (modified in
the same way by the emission from shocked material).

At this stage, we conclude that our implementation provides
results that are in excellent agreement with the alternative code
WM-{\sc basic}, both with respect to integrated fluxes 
as well as frequency edges, which moreover follow the predicted scaling 
relations. Having thus verified our implementation, we will now examine 
important effects of the X-ray radiation within the stellar wind.

\begin{figure*}[t]
\resizebox{\hsize}{!}
{\includegraphics[angle=90]{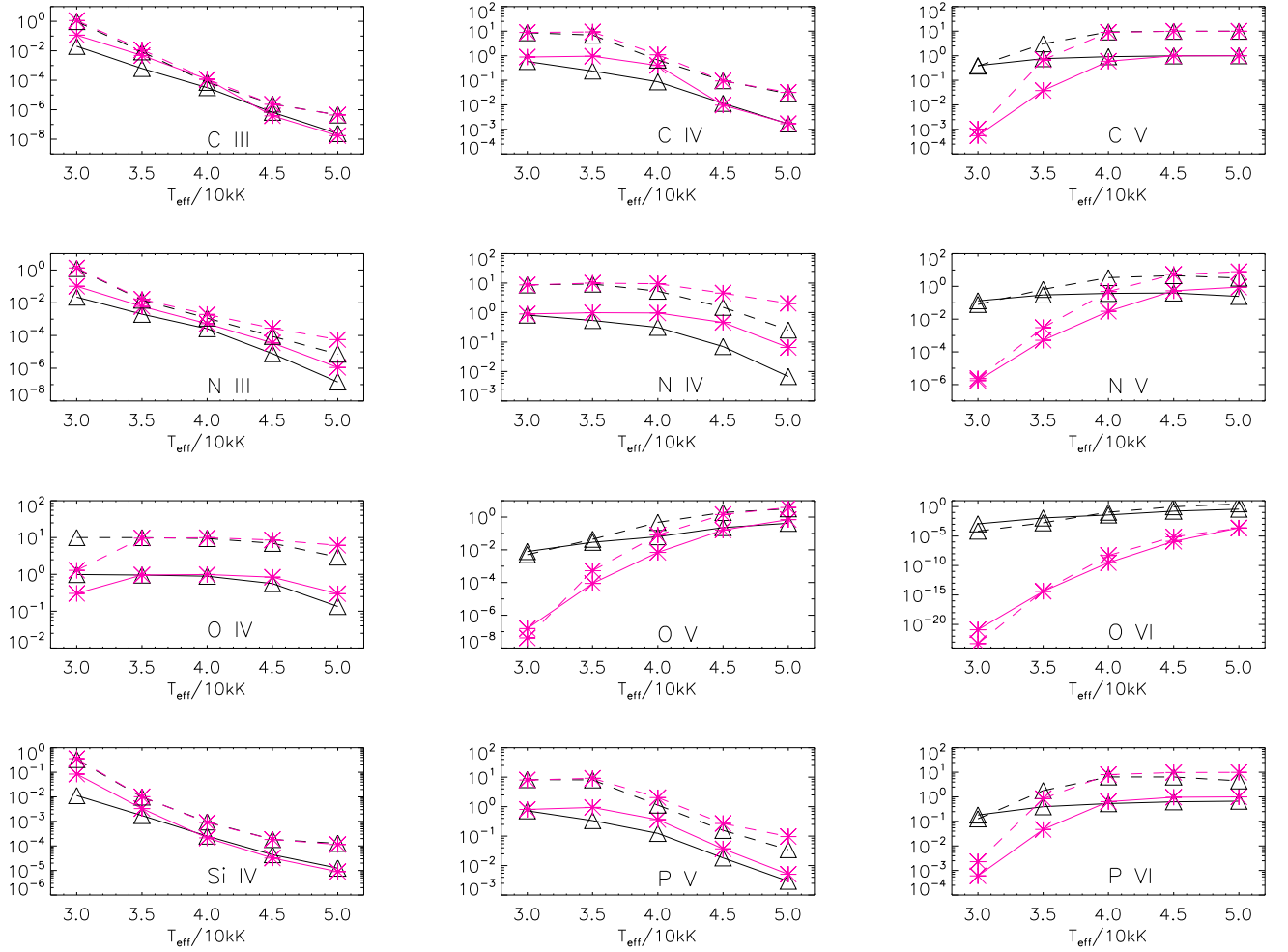}}
%file: paper43.pro
%
\caption{Ionization fractions of important ions at $v(r)$ = 0.5~\vinf,
	 as a function of \Teff, for models with typical X-ray
	 emission (triangles, \fv\ = 0.03, \Tshockmax\ = $3 \cdot
	 10^{6}$~K, corresponding to $u_{\infty}$ = 460~\kms), and
	%LP without X-rays (crosses). The solid lines refer to supergiant
	 without X-rays (asterisks). The solid lines refer to supergiant
	 models, and the dashed ones to dwarf models. {\textit{For
	 clarity, the ionization fractions of dwarf models have been
	 shifted by one dex.}} For stellar parameters and onset radius,
	 \Rmin, see Table~\ref{tab_grid}.}
\label{xrays43}
\end{figure*}

\section{Results}
\label{results}

In this section, we discuss the major results of our model
calculations. In particular, we study the impact of X-ray emission
on the ionization balance of important elements, both with respect
to direct (i.e., affecting the valence electrons) and Auger
ionization. 
We also discuss the impact of dielectronic recombination and
investigate the radial behavior of the high-energy mass absorption
coefficient, an essential issue with respect to the analysis of X-ray
line emission.

Note that all following results refer to our specific choice of the
run of the shock temperature (see Eqs.~\ref{shock_temp2} and
\ref{jump_velo}), which, in combination with our grid-parameter
$\gamma_x$ = 1, leads to shock temperatures of $T_{\rm s} (\vinf/2) =
0.25~\Tshockmax$ in the intermediate wind at $v(r) = 0.5~\vinf$. 

\subsection{Ionization fractions}
\label{results_fractions}

\subsubsection{General effects}
\label{generaleffects}

Though only indirectly observable (particularly via UV resonance
lines), ionization fractions provide useful insight into the various
radiative processes in the atmosphere. In the following, we compare,
for important ions (i.e., for ions with meaningful wind lines), the
changes due to the {\textit{combined}} effects of direct and Auger
ionization, whilst the specific effects of Auger ionization will be
discussed in Sect.~\ref{Auger_impact}. These comparisons will be
performed for our supergiant (solid) and dwarf models (dashed) from
Table~\ref{tab_grid}, and for the center values of our X-ray emission parameter
grid (Sect.~\ref{model_grid}), \fv\ = 0.03, \Tshockmax\ = $3 \cdot
10^{6}$~K, that are prototypical in many cases.\footnote{Note that
such maximum shock temperatures might be too high for models around
\Teff = 30~kK, and that certain effects (as discussed in the
following) might thus be overestimated in this temperature range.}
Comments on the reaction due to different parameters will be given in
the next section. All ionization fractions have been evaluated at a
representative velocity, $v(r)$ = 0.5~\vinf, and are displayed in
Fig.~\ref{xrays43}. To check the influence of X-ray emission, one
%LP simply needs to compare the triangles (with) and the crosses (without
simply needs to compare the triangles (with) and the asterisks (without
X-ray emission).  

\smallskip 
\noindent 
{\textit{Carbon.}} Though our model atom for carbon will be improved soon,
already the present one (from the WM-{\sc basic} data base) is
certainly sufficient to study the impact of shock radiation. The upper
panels of Fig.~\ref{xrays43} display the results, which indicate an
effect only for `cooler' supergiant models, with \Teff\ $<$ 40~kK. For
these objects, \CIII\ and \CIV\ become somewhat depleted (less than a
factor of ten), whilst \CV\ (which is, without X-ray emission, a trace
ion at 30~kK) becomes significantly enhanced. For dwarfs in this
temperature range, only \CV\ becomes increased, since the emission
(scaling with $\rho^2$) is still too weak to affect the major
ions.\footnote{But note that the actual filling factor in dwarfs might
be {\textit{much larger}} than 0.03, e.g., \citet{Cassinelli94, Cohen97,
Cohen08a, Huenemoerder12}.} For models with \Teff\ $>$ 40~kK, on the
other hand, the temperature is already hot enough that the ionization
balance is dominated by the `normal' stellar radiation field, and no
effect due to X-ray emission is visible.

Nitrogen (2nd row) and oxygen (third row of Fig.~\ref{xrays43}) suffer
most from the inclusion of shock radiation. In the following, we
concentrate on the differences produced by X-ray ionization in general,
whilst in subsequent sections we will consider specific effects.

\smallskip
\noindent
{\textit{Nitrogen.}} In the `cool' range, the behavior of \NIII,
\NIV\ and \NV\ is very similar to the corresponding carbon ions (i.e.,
a moderate depletion of \NIII\ and \NIV, and a significant increase of
\NV, particularly at \Teff\ between 30 and 35~K), whereas in the hot
range it is different. Here, \NIII\ and \NIV\ continue to become
depleted, but \NV\ increases only as long as \Teff\ $<$ 45~kK, and
decreases again at 45 and 50~kK. In other words, when \NV\ is already
the main ion for non-X-ray models, it becomes (slightly) depleted when
the X-rays are switched on, in contrast to \CV\ which remains
unmodified beyond 40~kK. This difference, of course, relates to the
fact that \CV\ has a stable noble-gas (He-) configuration, with a
high-lying ionization edge (31.6~\AA), compared to the \NV\ edge at
roughly 126~\AA\ that allows for a more efficient, direct ionization by
emission from the shock-heated plasma. 

\smallskip 
\noindent 
{\textit{Oxygen.}} For almost every temperature considered in our grid, the
inclusion of X-rays has a dramatic effect on the ionization of oxygen.
At 30~kK, \OIV\ becomes the dominant ion\footnote{this is also true
for models with different X-ray emission parameters}, when for non-X-ray models
the main ionization stage is still \OIII, whereas at the hot end \OIV\
becomes somewhat depleted. The behavior of \OV\ is similar to \NV\
(though the final depletion is marginal), and \OVI\ displays, at all
temperatures, the largest effect. At cool temperatures, the ionization
fraction changes by 15 orders of magnitude, but even at the hottest
\Teff\ there is still an increase by three to four dex. As is well
known, this has a dramatic impact on the corresponding resonance
doublet.

\smallskip 
\noindent 
{\textit{Silicon.}} In almost all hot stars, the dominant ion of silicon is \SiV\
(again a noble-gas configuration), and \SiIV\ forms by recombination, giving
rise to the well-known \SiIV\ luminosity/mass-loss effect \citep{Walborn84,
  Pauldrach90}. The bottom left panel of Fig.~\ref{xrays43} displays an
analogous dependence.  Whilst for dwarfs (low $\rho^2$) no X-ray effects are
visible for \SiIV, this ion becomes depleted for cool supergiants (\Teff\
$\la$ 35~kK), at most by a factor of ten.

\smallskip 
\noindent 
{\textit{Phosphorus.}}
During recent years, it turned out that the observed \PV\ doublet at
$\lambda$\,1118,1128 is key\footnote{because it is the only UV
resonance line(-complex) that basically never saturates, due to the low
phosphorus abundance} for deriving mass-loss rates from hot star
winds, in parallel with constraining their inhomogeneous structure
\citep{fullerton06, oskinova07, Sundqvist11, Surlan13, Sundqvist14}.
Thus, it is of prime importance to investigate its dependence on
X-rays, since a strong dependence would contaminate any quantitative
result by an additional ambiguity.  

As already found in previous studies (e.g., KK09, \citealt{bouret12}),
also our results indicate that \PV\ is not strongly modified by X-ray
emission (middle and right lower panels of Fig.~\ref{xrays43}), though
more extreme X-ray emission parameters, e.g., \fv\ = 0.05 and/or \Tshockmax\ =
$5 \cdot 10^6$~K, can change the situation (see
section~\ref{dependencefvtshock}). Even more, the apparently small
change in the ionization fraction of \PV\ at typical X-ray emission 
parameters (decrease by a factor of two to three) can still be of
significance, given the present discussion on the precision of derived
mass-loss rates (with similar uncertainties).

Regarding the ionization of \PVI, cold models (30 and 35~kK) change
drastically when X-ray emission has been included, both for
supergiants and dwarfs. Since we find less \PVI\ in hot models with
shocks (compared to models without), this indicates that the
ionization balance is shifted towards even higher stages (\PVII).

In this context, we note that \citet{krticka12} investigated the
reaction of \PV\ when incorporating additional, strong XUV emissivity
(between 100 and 228~\AA) and micro-clumping into their models. The
former test was driven by a previous study by \citet{WC10} who argued
that specific, strong emission lines in this wavelength range 
could be of significant impact. Indeed, \citet{krticka12} were able to
confirm that under such conditions\footnote{enhanced
emissivity in the XUV range; note, however, that the lines refered to
by \citet{WC10} are included in `standard' plasma emission
codes} \PV\ becomes strongly depleted, in parallel with changes in
the ionization fractions of, e.g., \CIV, \NIV, and \OIV\ (see also
Sect.~\ref{dependencefvtshock}). Further work is certainly required to
identify the source of such additional emissivity (if present), and,
in case, incorporate this mechanism into our models. 

\begin{figure}
\resizebox{\hsize}{!}
{\includegraphics{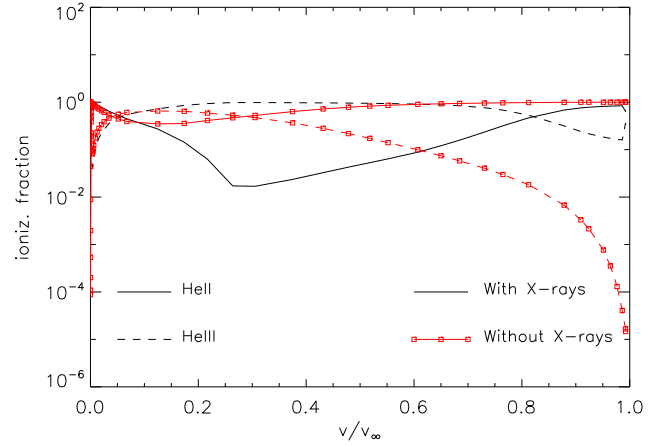}}
%file: he_s30_ionfract.pro
%
\caption{Helium ionization fractions as a function of local velocity, 
	 for an S30 model with (\fv\ = 0.03 and \Tshockmax\ = $3 \cdot
	 10^6$~K) and without X-rays. See text.}
\label{he_s30}
\end{figure}

\subsubsection{Impact on helium}
\label{impact_helium}

During our analysis, we noted that also helium can be affected by
shock emission (see also Sect.~\ref{test_parameters}), a finding that
has been rarely discussed in related literature. In particular, \HeII\
(and \HeI) can become depleted in the intermediate wind, though only
for our `cooler' supergiant models with 30~kK $\la$ \Teff\ $\la$
40~kK. The effect is strongest for S30 models, but barely noticeable
already at S40, independent of the specific X-ray emission parameters. For all
our dwarf models, no changes are visible at all.

Figure~\ref{he_s30} displays the helium ionization fractions for an S30
model with typical X-ray emission parameters, as a function of local velocity.
The depletion of \HeII\ (and, in parallel, of \HeI\ that is not
displayed) is significant in the region between
0.2\vinf\ $\la v(r) \la$ 0.8\vinf, and results from the increased
ionization due to the increased radiation field (in the \HeII\ Lyman
continuum) in models with shocks (note also the corresponding increase
of \HeIII). 

\begin{figure}
\resizebox{\hsize}{!}
{\includegraphics{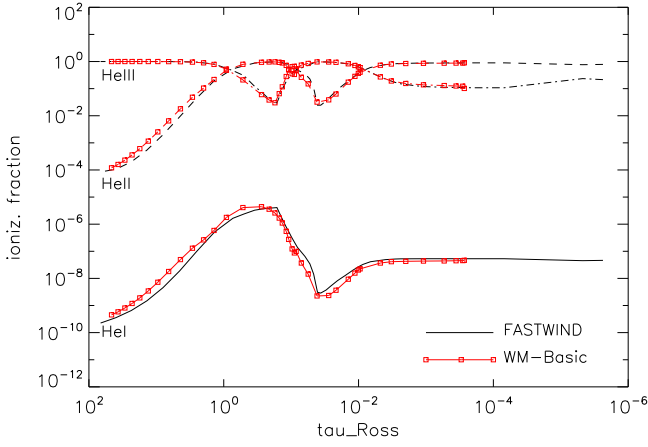}}
%file: he_s30_tadziu.pro
%
\caption{Helium ionization fractions as a function of \taur, for S30
	 models calculated by {\sc FASTWIND} and WM-{\sc basic}, both
	 with X-ray emission parameters from Table~\ref{tab_wm}. The agreement
	 is excellent.}
\label{he_s30_tadziu}
\end{figure}

In Fig.~\ref{he_s30_tadziu}, we compare the helium ionization
fractions from our solution and a corresponding WM-{\sc basic} S30
model, but now with X-ray emission parameters as tabulated in
Table~\ref{tab_wm} (the major difference is a filling factor of 0.02
instead of 0.03). Here, we display the fractions as a function of
\taur, to enable a comparison of the photospheric regions as well.
Again, the depletion of \HeII\ (now located between \taur $\approx$
0.1{\ldots}0.01) is visible, and our results coincide perfectly with
those predicted by WM-{\sc basic}.

\begin{figure}[t]
\resizebox{\hsize}{!}
{\includegraphics[angle=90]{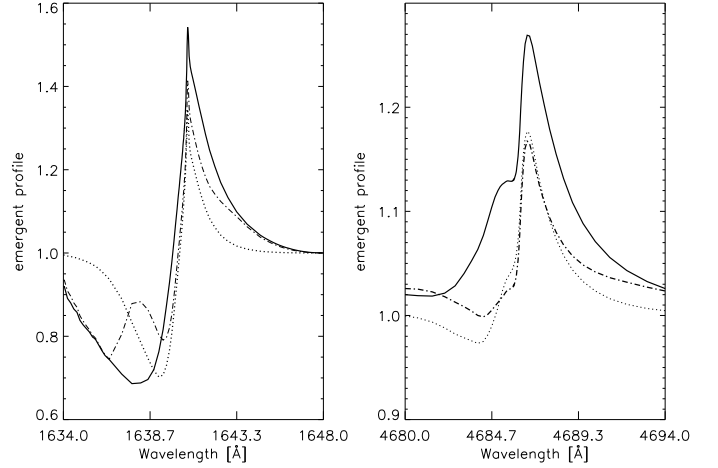}}
%file: plot_he.pro
%
\caption{Synthetic \HeII~1640 and \HeII~4686 profiles for our S30
	 model. Each profile corresponds to a different X-ray
	 description. Solid: \fv\ = 0.03, \Tshockmax\ = $3 \cdot
	 10^{6}$~K, \Rmin\ = 1.5 ~\Rstar ; dash-dotted: as solid,
	 but with \Rmin\ = 2~\Rstar; dotted: no shock emission.}
\label{lines_he_s30}
\end{figure}

Since the ionization balance changes already at quite low velocities,
this might affect at least two important strategic lines, \HeII~1640
and \HeII~4686.\footnote{Most other \HeII\ and \HeI\ lines are formed
in the photosphere, and remain undisturbed.} From
Fig.~\ref{lines_he_s30}, we see that this is actually the case:
\HeII~4686 displays stronger emission, whilst \HeII~1640 displays a
stronger emission in parallel with absorption at higher velocities,
compared to the non-X-ray model (dotted). This is readily understood
since \HeII~4686 is predominantly a recombination line, such that the
increase in \HeIII\ leads to more emission; to a lesser extent, this
is also true for \HeII~1640. The lower level of this line, $n=2$
(responsible for the absorption), is primarily fed by pumping from the
ground-state via \HeII~303. We have convinced ourselves that the
increased pumping because of the strong EUV radiation field leads to a
stronger population of the $n=2$ state (even if \HeII\ itself is
depleted), so that also the increased absorption is explained. 

As already pointed out in Sect.~\ref{test_parameters}, changing \Rmin\
from 1.5 to 1.2~\Rstar\ does not make a big difference. Increasing
\Rmin\ to 2~\Rstar, however, changes a lot, as visible from the
dash-dotted profiles in Fig.~\ref{lines_he_s30}.  Except for
slightly more emission (again because of increased \HeIII\ in regions
with $r > 2 \Rstar$), the difference to profiles from models without
shock emission becomes insignificant, simply because both lines
predominantly form below the onset radius.

\subsubsection{Dependence on filling factor and shock temperature}
\label{dependencefvtshock}

As we have seen already above, each ion reacts somewhat differently to
the imposed shock radiation. In this section we describe how a change
of important X-ray characteristics affects important ions. The figures
related to this section are enclosed in Appendix
\ref{diffxraysdescription}. The upper figure on each page shows
specific ionization fractions with and without X-rays, as a function
of \Teff, for our supergiant and dwarf models (S30 to S50 and D30 to
D50, respectively). The ionization fractions have been evaluated at
the location where the impact of shock radiation is most evident
for the considered ion. Each of these figures contains nine panels,
where both the filling factor and the maximum shock temperature are
varied according to our grid, i.e., \fv\ = 0.1, 0.3, 0.5 and \Tshockmax\ =
1,3,5$\cdot 10^6$~K. Note that the onset radius, \Rmin, was
set to its default value for all models. The lower two figures on each
page display the ionization fractions for our dwarf (left) and
supergiant models (right), evaluated at the same location as above,
but now overplotted for all values of \fv\ (different colors) and
\Tshockmax\ (different symbols), and without a comparison to the non-X-ray 
case. Thus, the upper figure allows to evaluate the X-ray effects
in comparison to models without shock emission, whilst the lower
two figures provide an impression on the differential effect, i.e.,
the range of variation. 

\smallskip 
\noindent 
{\textit{Carbon.}} \CIII\ and \CIV\ are significantly affected in
supergiant models with 30~kK $\la$ \Teff\ $\la$ 40~kK, for
intermediate to large values of \fv\ and \Tshockmax. The depletion of
\CIII\ and \CIV\ reaches a factor of $10$ (or even more) in cooler
supergiant models when the highest values of X-ray emission parameters are
adopted, which is reflected in a corresponding increase of \CV. On the
other hand, \CIII\ and \CIV\ are barely modified in supergiant models
with the lowest values of \fv\ or \Tshockmax, which is also true for
dwarf models with any value of our parameter grid (see
Figs.~\ref{app_civ}/\ref{civ_diff}). The ionization fraction of \CV\
increases also for the lowest values of X-ray emission parameters, once more
for the cooler models (here, also dwarfs are affected). \CV\ remains
unmodified beyond 40~kK due to its stable noble-gas configuration, as
previously noted.

\smallskip 
\noindent 
{\textit{Nitrogen.}} The behavior of \NIII, \NIV\ and \NV\ in the colder
models is similar to the corresponding carbon ions, for all different
X-ray descriptions. For higher \Teff, increasing \fv\ enhances the
depletion of \NIII\ and \NIV\ in both supergiants and dwarfs, whilst
the impact of \Tshockmax\ is rather weak. At the largest values of
X-ray emission parameters, both stages become highly depleted (one to
two orders of magnitude) for all models but D30 and D35. 

Shock radiation is essential for the description of \NV\ at almost any
temperature, particularly for models with \Teff\ $<$ 45~kK
(Figs.~\ref{app_nv}/\ref{nv_diff}). Here, the increase of \NV\
(compared to non-X-ray models) can reach 4 to 5 dex at the lowest
temperatures. At 45~kK, only a weak impact of shock radiation can be
noted, whilst for 50~kK a high depletion of \NV\ for extreme
parameters values becomes obvious. Once more, the impact of \fv\ is
more prominent than of \Tshockmax, mainly for the coldest models where
\NV\ becomes enhanced by one order of magnitude when increasing \fv\
from 0.01 to 0.05 and keeping \Tshockmax\ constant. The hottest models
with moderate to high parameters (\fv\ $\ga$ 0.02 and \Tshockmax\ $\ga
2\cdot 10^6$~K) indicate that also \NVI\ becomes strongly affected by
changes in the X-ray ionization. 

\smallskip 
\noindent 
{\textit{Oxygen.}} 
Independent of the X-rays description, the depletion of \OIV\ for hot
models happens only in a specific range of the wind, between 0.4 to
0.8~\vinf\ (similar to the case of \HeII\ discussed in the previous
section). Also for X-ray emission parameters different from the
central value of the grid, the behavior of \OV\ is still very similar
to \NV, where mainly the cold models are quite sensitive to variations
of \fv\ (Figs.~\ref{app_ov}/\ref{ov_diff}). The shock radiation
increases the ionization fraction of \OV\ by 5 to 6 dex (when \fv\
varies between 0.01 and 0.05, independent of \Tshockmax) for the
coolest models, whilst these factors decrease as \Teff\ approaches 40
to 45~kK. Models with \Teff\ = 45~kK are barely affected, independent
of the specific X-ray emission parameters. Similar to the case for \NV\ at
highest values of \fv, \Tshockmax, and \Teff, the corresponding
depletion of \OV\ points to the presence of a significant fraction of
higher ionization stages.

As pointed out already in Sect.~\ref{generaleffects} (see also
Sect.~\ref{Auger_impact}), the X-ray radiation is essential for the
description of \OVI, which shows, particularly in the cold models, a
high sensitivity to both \fv\ and \Tshockmax\
(Figs.~\ref{app_ovi}/\ref{ovi_diff}).

\smallskip 
\noindent 
{\textit{Silicon.}} Also when varying the X-rays description, \SiIV\ still
remains unaffected from shock emission in dwarf models. On the other
hand, for cool supergiants (\Teff\ $\la$ 35~kK), \SiIV\ becomes even
more depleted when \fv\ increases (though \Tshockmax\ has a negligible
influence). No variation is seen in \SiV, as expected due to its
noble-gas configuration. 

\smallskip 
\noindent 
{\textit{Phosphorus.}} 
%Due to the relevance of \PV\ as an indicator for
%mass-loss and wind-inhomogeneities, it is particularly important to
%check how it changes when different X-ray conditions are adopted. 
\PV\ shows a sensitivity to both \fv\ and \Tshockmax, but in this case
\Tshockmax\ is more relevant. Though no difference between models with
and without shocks is seen for the lowest values of \Tshockmax,
particularly the supergiant models develop a depletion with increasing
shock temperature, even at lowest \fv. As noted already in
Sect.~\ref{generaleffects}, for extreme X-ray emission parameters the
depletion of \PV\ is significant for all models (both supergiants and
dwarfs), except for D30 (Figs.~\ref{app_pv}/\ref{pv_diff}). Finally,
even \PVI\ becomes highly depleted for hot models (\Teff\ $\ga$ 40~kK)
at intermediate and high values of \Tshockmax, which indicates the
presence of even higher ionization stages. 

To summarize our findings: When increasing the values for \fv\ and
\Tshockmax, the effects already seen in Fig.~\ref{xrays43} become even
more pronounced, as to be expected. For most ions, the impact of \fv\
appears to be stronger than the choice of a specific \Tshockmax\
(provided the latter is still in the range considered here), though
\PV\ and \OVI\ (for the cooler models) show quite a strong reaction to
variations of the latter parameter. Overall, the maximum variation of
the ionization fractions within our grid reaches a factor of 10 to 100
(dependent on the specific ion), where lower stages (e.g., \CIV, \NIV,
\OIV\ and \PV) become decreased when \fv\ and \Tshockmax\ are increased,
whilst the higher stages (e.g., \NV, \OV, \OVI) increase in parallel
with the X-ray emission parameters. Only for \SiIV, the impact of X-rays
remains negligible in all models except for S30 and S35.

\subsubsection{Comparison with other studies}
\label{comparing}

Since the most important indirect effect of shock emission is the
change in the occupation numbers of the {\textit{cool}} wind, it is
worthwhile and necessary to compare the ionization fractions resulting
from our implementation with those presented in similar studies. 

To this end, (i) we recalculated the models described in KK09, (ii)
compared with two models (for HD\,16691 and HD\,163758) presented in
\cite{bouret12} who used {\sc CMFGEN} and SEI\footnote{Sobolev with
exact integration \citep{Lamersetal87}}-fitting to calculate/derive
the ionization fractions of phosphorus, and (iii) compared with the
ionization fractions predicted by WM-{\sc basic}.

\begin{figure*}[t]
\resizebox{\hsize}{!}
{\includegraphics[angle=90]{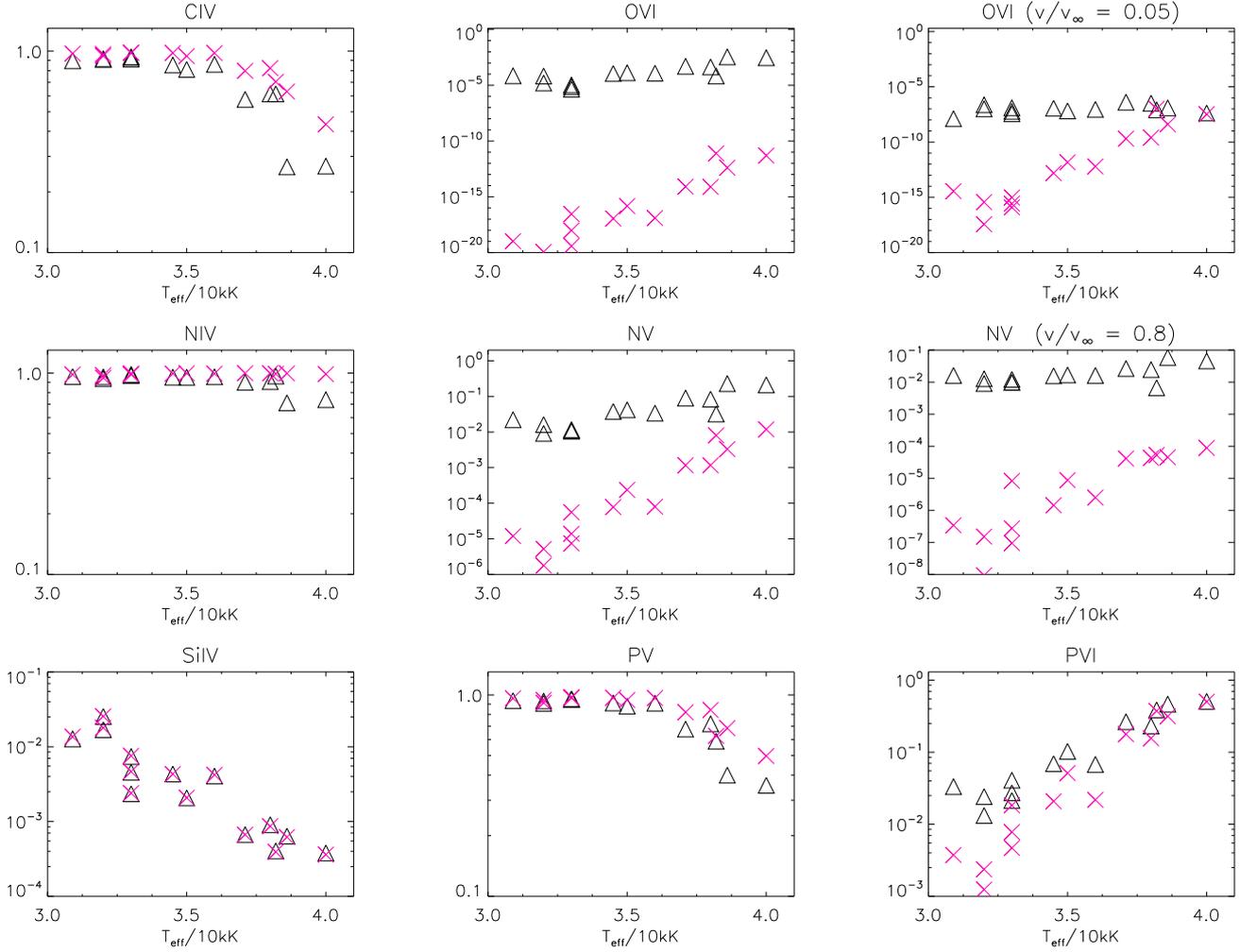}}
% file:Krticka/models/final_kr.pro
%
\caption{Ionization fractions of selected ions as a function of \Teff,
	 for 14 O-star models as detailed in \citet[KK09]{krticka09}
	 and recalculated by us using {\sc FASTWIND}. If not indicated
	 otherwise, fractions are displayed at $v(r) = 0.5 \vinf$. As in
	 previous figures, triangles represent models with shocks, and
	 crosses those without. This figure reproduces, in most parts,
	 the layout of Figure~8 from KK09, such that differences and
	 similarities between our and their results can be easily
	 recognized. For details, see text.}
\label{final_kr}
\end{figure*}

\begin{figure}[t]
\resizebox{\hsize}{!}
{\includegraphics{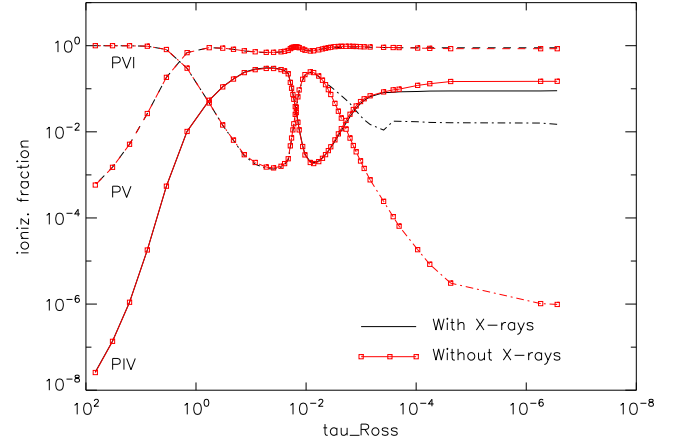}}
%file: compare_pv_krticka.pro
%
\caption{Radial stratification of phosphorus ionization fractions, as
	 a function of \taur, for our model of HD\,203064 at
	 \Teff\ = 34.5~kK (see KK09 for stellar, wind and X-ray
	 emission parameters). In our implementation, \PV\ is barely
	 modified by the X-ray radiation field, whilst a considerable
	 impact is seen for \PVI.}
\label{compare_pv_krticka}
\end{figure}

\begin{figure}
\resizebox{\hsize}{!}
{\includegraphics{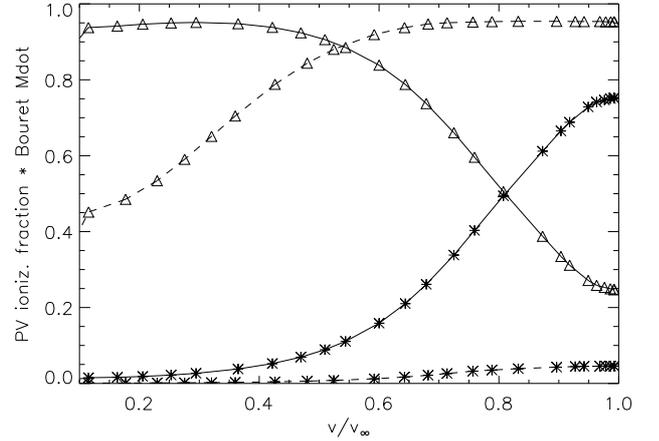}}
%file: plot_pv_bouret.pro
%
\caption{Ionization fractions of \PIV\ (asterisks) and \PV\ (triangles)
	 as a function of normalized velocity, for an S35 (solid)
	 and S40 (dashed) model. Both models have been calculated with
	 a clumping factor $\fcl = 20$, and a mass-loss rate
	 reduced by a factor of $\sim$4 compared 
	 to the values provided in Table~\ref{tab_grid}. Compare with Fig.~10
	 in \citet{bouret12}.}
\label{plot_pv_bouret}
\end{figure}

Regarding the first point, we recalculated the 14 O-star models (in
the temperature range between 30 and 40~kK) presented by KK09, using
parameters from their Tables~2 and 3, both without and with shock
emission (\fv\ = 0.02 and $u_{\infty}$/\vinf=0.3), by means of {\sc
FASTWIND}, using H, He, C, N, O, Si and P as explicit ions. Figure
\ref{final_kr} shows our results for the ionization fractions of
selected ions, as a function of \Teff, and evaluated at $v(r) = 0.5
\vinf$. The layout of this figure is similar to Figure~8 in KK09, and
has been augmented by \OVI\ evaluated at $v(r) = 0.05 \vinf$ and \NV\
evaluated at $v(r) = 0.8 \vinf$, corresponding to their Figures~9 and
10.  

Indeed, there are only few ions which display similar fractions over
the {\textit{complete}} temperature range of the O-star models considered
by KK09 (which still misses the hotter O-stars beyond 40~kK). For
\CIV, an agreement is present only for the coolest regime (\Teff\
$\leq$ 32~kK) where both studies predict \CIV\ as the main ion,
independent whether X-rays are present or not.  Whilst the fractions
for non-X-ray models are comparable also for hotter temperatures, the
X-ray models by KK09 show a much larger depletion of \CIV\ (fractions
of $10^{-2}$ to $10^{-3}$ for \Teff\ $>$ 34~kK) than our models (still
above $10^{-1}$).

For \OVI, agreement between both results is present only at the hottest
temperatures, whilst between 30~kK $<$ \Teff\ $\la$ 37~kK our models
display a factor of $\sim$100 lower fractions, for both the non-X-ray 
models and the models with shock emission. The same factor is
visible in the lower wind ($v(r) = 0.05 \vinf$) for the X-ray models,
but the non-X-ray models are similar here. 

For nitrogen (\NIV\ and \NV), on the other hand, the results are quite
similar in most cases. The exception is \NV\ for models without shocks, 
where our results are lower (by $\sim$1~dex) in the intermediate and 
outer wind ($v(r) = 0.8 \vinf$). 

For \SiIV, both results fairly agree for the X-rays models, though we do not
see a significant effect from including the shock emission in our
calculations (in other words, X-ray and non-X-ray models yield more or
less identical results). In contrast, the models by KK09 indicate a
small depletion of \SiIV\ when including the shock emission, by
factors of roughly 2 to 3. Thus, our non-X-ray models have less
\SiIV\ than those by KK09.

Again, phosphorus (in particular, \PV) has to be analyzed in more
detail. Comparing the last two panels of Fig.~\ref{final_kr} with
Fig.~8 from KK09, we see that our ionization fractions for \PV\ agree
with KK09 in the coolest models, and in the hottest models regarding
\PVI. In the other temperature ranges, however, differences by a
typical factor of 2 (regarding \PV) and 2 to 5 (regarding \PVI) are
present. In their Fig.~12, KK09 display the radial stratification of
the phosphorus ionization fractions for their model of HD\,203064, 
whilst the corresponding results from our implementation are shown in
Fig.~\ref{compare_pv_krticka}. Both codes yield quite similar
fractions for \PIV\ and \PV\ (with and without X-rays) in the external
wind. The same is true for \PVI\ in the model with X-rays, but we have
considerably less \PVI\ for the non-X-ray model. Prominent differences
are visible in the lower wind and close to the lower boundary. We
attribute this difference to a boundary condition (in the models by
KK09) at quite low optical depths, where the electron temperature is
still close to the effective one.\footnote{Indeed, we were not able to
find statements or figures related to the {\textit{photospheric}} structure
of their models in any of the papers by Krti\v{c}ka and co-workers, so our
argument is somewhat speculative.} 
Thus far, it is conceivable that quite a low ionization stage (\PIV)
dominates their internal atmosphere (followed by \PV\ and negligible \PVI),
whilst in our case it is just the other way round, and \PVI\ 
dominates, because of the much higher temperatures. 

To check these discrepancies further, we compared our results also
with calculations performed with {\sc CMFGEN}. Particularly, we
concentrated on two supergiant models at roughly 35~kK and 40~kK
(HD\,163758 and HD\,16691, respectively), as described by
\cite{bouret12}. In these models, an X-ray emitting plasma with {\textit{
constant}} shock-temperature, \Tshock$(r)$ = 3$\cdot 10^6$~K, a filling
factor corresponding to $L_x/L_{\rm bol} = 10^{-7}$, and an onset
radius corresponding to 200 to 300~\kms\ was used (J.-C. Bouret, priv.
comm.). In Fig.~\ref{plot_pv_bouret}, we present our results for \PIV\
and \PV\ which can be compared with their Fig.~10, displaying \PV\
alone. Though our models\footnote{S35 and S40, but using a clumped
wind with reduced mass-loss rates to ensure comparable wind
structures} do not have identical parameters (in particular, our shock
temperatures increase with velocity), the ionization fractions behave
quite similar:
%comparison of clumped vs. unclumped Lx/Lbol
%epsilon scales with <rho>^2 = (rho_cl/fcl)^2 (since fx defined w.r.t. smooth wind.)
%since <rho> depends on Mdot, and Mdot_cl=Mdot/sqrt(fcl) here, 
%eps_cl = eps_uncl/fcl = eps/20.
%tau propto kappa*<rho>  scales with tau_uncl/sqrt(fcl), thus less
% absorption
%for optically thin wind, maximum reduction of Lx by factor of 20 (if
%clumped and unclumed wind thin)
%for tau > 1, less absorption, thus lower reduction (clumped wind is
% thinner than unclumped wind).
%
%D30  0.27d-10 --   -10.57 -- normal grid -- 0.40d-9 -- -9.39 -- factor of 15
%D35  0.63d-9  --   -9.19  -- normal grid -- 0.48d-8 -- -8.32 -- factor of 7.4
%D40  0.65d-8  --   -8.18  -- normal grid -- 0.12d-6 -- -6.93 -- factor of 17.8
%D45  0.30d-7  --   -7.52  -- normal grid -- 0.38d-6 -- -6.42 -- factor of 12.5
%D50  0.14d-6  --   -6.85  -- normal grid -- 0.13d-5 -- -5.89 -- factor of 9.1
%D55  0.26d-6  --   -6.57  -- normal grid -- 0.18d-5 -- -5.73 -- factor of 6.91
%S30  0.92d-7  --   -7.03  -- normal grid -- 0.59d-6 -- -6.23 -- factor of 6.30
%S35  0.10d-6  --   -6.98  -- normal grid -- 0.62d-6 -- -6.21 -- factor of 6.36
%S40  0.88d-7  --   -7.05  -- normal grid -- 0.57d-6 -- -6.25 -- factor of 6.31
%S45  0.82d-7  --   -7.08  -- normal grid -- 0.59d-6 -- -6.23 -- factor of 7.08
%S50  0.11d-6  --   -6.92  -- normal grid -- 0.19d-5 -- -5.71 -- factor of 16.22
%For the last model, there is not only the reduction of tau, but also
% and increase because of recombining He. Thus, strong reduction.
In the cooler model (solid), the ionization of \PV\ decreases with
velocity, and in the hotter one (dashed), it increases outwards. This
is because in the cooler model, \PV\ is the dominant ion at low
velocities, recombining to \PIV, whilst in the hotter model
\PVI\ dominates at low velocities, recombining to \PV\ in the run
of the wind. Of course, there are some quantitative
differences, particularly in the intermediate wind\footnote{J.-C.
Bouret provided us with an output of the ionization fractions for
\PIV\ and \PV.}, but we attribute these to a different stratification
of the clumping factor, $\fcl$, and to the different description of the
X-ray emitting plasma (concerning the reaction of \PV\ on various
X-ray emission parameters, see Fig.~\ref{pv_diff}).

\begin{figure*}[t]
\resizebox{\hsize}{!}
{\includegraphics[angle=90]{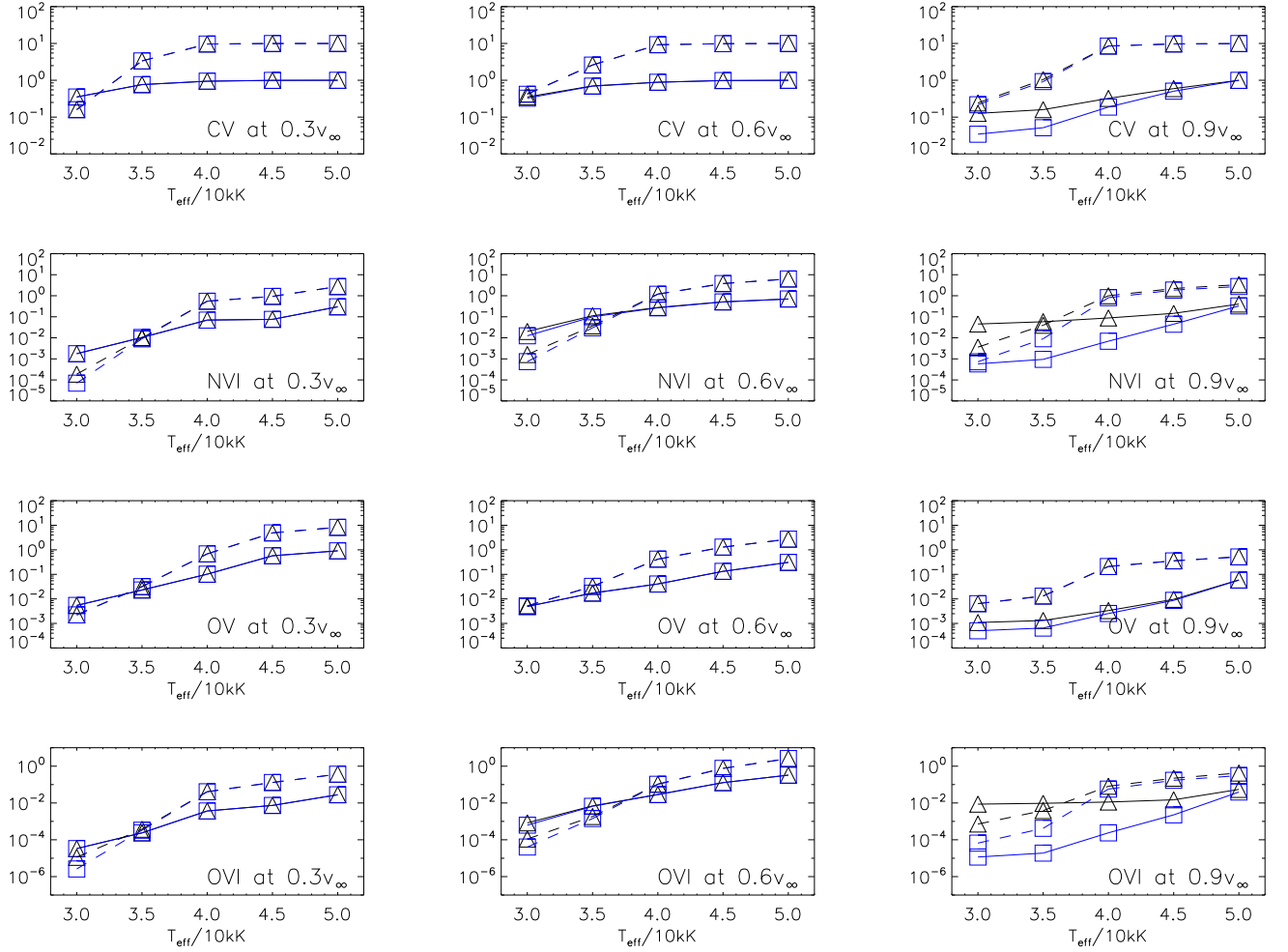}}
%LP: auger43.pro
%
\caption{Ionization fractions of ions most affected by Auger
	 ionization, at different depth points. All models have
	 typical X-ray emission parameters (\fv\ = 0.03 and
	 \Tshockmax\ = 3$\cdot 10^6$~K). The triangles represent
	 models including Auger ionization (standard approach, similar
	 to Fig.~\ref{xrays43}), and squares models without (i.e.,
	 `only' direct ionization has been considered).  Solid lines
	 refer to supergiant models, and dashed ones to dwarf models.
	 {\textit{For clarity, the ionization fractions of dwarf models
	 have been shifted by one dex.}} }
\label{auger43}
\end{figure*}

As a final test, we compared our solutions to the predictions by
WM-{\sc basic}, using our dwarf and supergiant models
(Table~\ref{tab_grid} and X-ray emission parameters from Table~\ref{tab_wm}).
The results are displayed in Figs.~\ref{tadziu_33_dw} and
\ref{tadziu_33_sg} (Appendix~\ref{comparisonwm}). Note that the range
of comparison extends now from 30 to 50~kK, i.e., to much hotter
temperatures than in the comparison with KK09. 

Overall, the agreement between {\sc FASTWIND} and WM-{\sc basic} is
satisfactory, and all trends are reproduced. However, also here we
find discrepancies amounting to a factor of 10 in specific cases,
particularly for \SiIV. Typical differences, however, are on the order
of a factor of two or less. We attribute these discrepancies to
differences in the atomic models, radiative transfer and the
hydrodynamical structure, but conclude that both codes yield rather
similar results, maybe except for \SiIV\ which needs to be
reinvestigated in future studies. 

In Fig.~\ref{all_profiles} we see how some of the encountered
differences (compared at only {\textit{one}} depth point, $v(r) = 0.5
\vinf$, except for \NV) translate to differences in the emergent
profiles. As prototypical and important examples, we have calculated
line profiles for \NIV~1720, \NV~1238,1242, \OV~1371, \OVI~1031,1037
and \PV~1117,1128, and compare them with corresponding WM-{\sc basic}
solutions for models S30, D40, S40, D50 and S50 (for model D30, all
these line are purely photospheric, and thus not compared). Both the
WM-{\sc basic} and the {\sc FASTWIND} profiles have been calculated 
with a radially increasing microturbulence, with maximum value
\vturb(max) = 0.1\vinf, which allows for reproducing the blue absorption
edge and the `black trough' (see Sect.~\ref{xrayemission}) in case of
saturated P Cygni profiles.

This comparison clearly shows that in almost all considered cases the
agreement is satisfactory (note that WM-{\sc basic} includes the
photospheric `background', whilst {\sc FASTWIND} only accounts for the
considered line(s)), and that larger differences are present only (i) for
\NIV\ and \OV\ in the outer wind, where {\sc FASTWIND} produces more
(\NIV) and less (\OV) absorption, respectively, and (ii) for strong
\PV\ lines, where {\sc FASTWIND} predicts higher emission.

\subsection{Impact of Auger Ionization}
\label{Auger_impact}

All X-ray models discussed so far include the effects from direct {\textit{
and}} Auger ionization, which was shown to play an important role for the
ionization balance in stellar winds (e.g., \citealt{Cassolson79,
Olcast81, Macfarlane94, pauldrach94c}).  In the following, we
investigate the contribution of the latter effect to the total
ionization in more detail, particularly since there is still a certain
debate on this question. 

Figure~\ref{auger43} displays how specific ions are affected throughout the 
wind, for dwarf and supergiant models with different \Teff\ and typical 
X-ray emission parameters  (\fv\ = 0.03 and \Tshockmax\ = 3$\cdot 10^6$~K). Each 
ion is shown at three different locations: 
$v(r)$ = 0.3~\vinf\ (close to the onset of the shock emission), 
$v(r)$ = 0.6~\vinf\ (intermediate wind) and $v(r)$ = 0.9~\vinf\ (outer
wind). 

Two general comments: (i) Significant effects are to be expected only
for quite high ionization stages, since in the majority of cases Auger
ionization couples ions with a charge difference of two (but see
Sect.~\ref{xrsandauger}). E.g., \CIV\ should remain (almost)
unmodified, since \CII\ is absent in O- and at least early B-stars,
and the K-shell absorption of \CIV\ (with threshold at 35.7~\AA),
resulting in the formation of \CV\ (charge difference of one!), is in
most cases (but see below) negligible compared to the direct
ionization of \CIV\ (with threshold at $\sim$192~\AA\ for the
ground-state ionization; remember that the radiation field is stronger
at longer wavelengths, which favors direct vs. Auger ionization). In
contrast, \OVI\ should become significantly affected, since \OIV\ is
strongly populated in O-stars, and the transition threshold for the
direct ionization from \OV\ (at $\sim$109~\AA) is now closer to the
K-shell edge. Consequently, the transition rates (depending on the
corresponding radiation field) are more similar than in the case of
\CIV.  

(ii) In the same spirit, Auger ionization should become negligible, at
least in most cases, for the hotter O-stars (see also
Sect.~\ref{tests}).  Once \Teff\ is high, more direct ionization is
present (because of the stronger radiation field at the corresponding,
lower-frequency edges), and consequently the impact of Auger
ionization should decrease. Though this argumentation is basically
correct, the actual results depend, of course, also on the
wind-strength, since higher densities lead to more X-ray emission (for
identical \fv), which increases the impact of Auger-ionization. E.g.,
if we check for the behavior of \NVI\ at 0.9~\vinf\ in
Fig.~\ref{auger43}, we see
that for D40, D45 and D50 there is indeed no effect, whilst for S40
and S45 Auger ionization still has a certain influence.

Now to more details. At first note that all ions from C, N, O, Si
and P that are {\textit{not}} displayed in Fig.~\ref{auger43} are barely
changed by Auger ionization, with a maximum difference of 
$\pm 0.08$~dex (corresponding to factors of 0.8 to 1.2) 
in the fractions calculated with and without Auger.
 
For carbon, \CV\ is the only ion which under specific conditions
becomes affected by Auger ionization. As visible in the first line of
Fig.~\ref{auger43}, cold supergiant models display an increase of \CV\
in the outer wind when Auger is included, since in this case the
radiation field at the corresponding K-shell edge becomes quite strong,
compared to the radiation field around 192~\AA\ (see
Fig.~\ref{comp_eddington_flux}). This increase is compensated by a
similar decrease of \CIV, which, in absolute numbers, is quite small
though.

\NVI\ (second line in Fig.~\ref{auger43}) is the only nitrogen ion
where larger changes can be noted. In cool dwarfs, it becomes
influenced already at 0.3~\vinf, and also in the intermediate wind,
which is also true for model S30. In the outer wind, differences
appear clearly for all models, except for dwarfs with \Teff\ $\ga$
40~kK.  The corresponding change in \NIV, on the other hand, is
marginal, again because \NVI\ itself has a low population, even when
Auger is included.

\begin{figure}[t]
\resizebox{\hsize}{!}
{\includegraphics{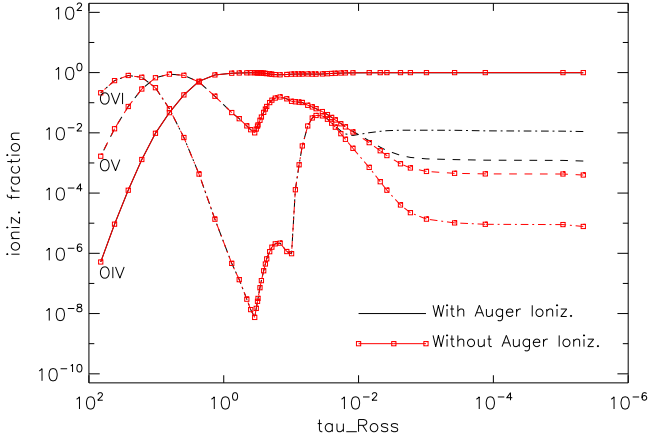}}
%file: compare_ww_auger_paper.pro
%
\caption{Radial stratification of oxygen ionization fractions, as
         a function of \taur, for an S40 model with
         \fv\ = 0.03 and \Tshockmax\ = 3$\cdot 10^6$~K. Auger 
         ionization notably affects the presence of \OVI\ in the outer
         wind ($\taur \leqslant 10^{-2}$ corresponding to 
         $r \geqslant$ 4~\Rstar\ or $v(r) \geqslant$ 0.7~\vinf). 
         The model without Auger ionization has more \OV\ than \OVI, 
         and vice versa when the effect is included.} 
\label{comp_auger_s40}
\end{figure}

\OV\ behaves similar to \NV\ (mostly no changes), but now a weak effect
appears in the outer wind of cool supergiants (third line of
Fig.~\ref{auger43}), and even for \OVI\ (compare to the reasoning
above), changes in the lower and intermediate wind are barely visible
(if at all, then only for the S30 model, see last line of
Fig.~\ref{auger43}). In the outer wind, however, considerable
differences in \OVI\ (up to three orders of magnitude) can be clearly
spotted for all supergiants and cooler dwarf models, similar to the case of
\NVI. Only for the hottest models, the effect becomes weak.
Fig.~\ref{comp_auger_s40} shows an example for an S40 model where the
second-most populated oxygen ion (\OV) changes to \OVI\ after the
inclusion of Auger ionization. 

Finally, the K-shell edges for phosphorus (not implemented so far) and
silicon (with quite low cross-sections) are located at such high
energies ($>$ 2~keV or $>$ 6~\AA) that the corresponding Auger rates
become too low to be of importance, at least for the considered
parameter range. 

To conclude, in most cases the effects of Auger ionization are only
significant in the outer wind (for a different run of shock
temperatures, they might become decisive already in the lower or
intermediate wind), and for highly ionized species. The
effect is essential for the description of \NVI\ and \OVI,
particularly in the outer wind. Thus, and with respect to strategic UV
resonance lines, it plays a decisive role only in the formation of
\OVI~1031,1037 (but see also \citealt{Zsargo08b}).

\begin{figure}[t]
\resizebox{\hsize}{!}
{\includegraphics{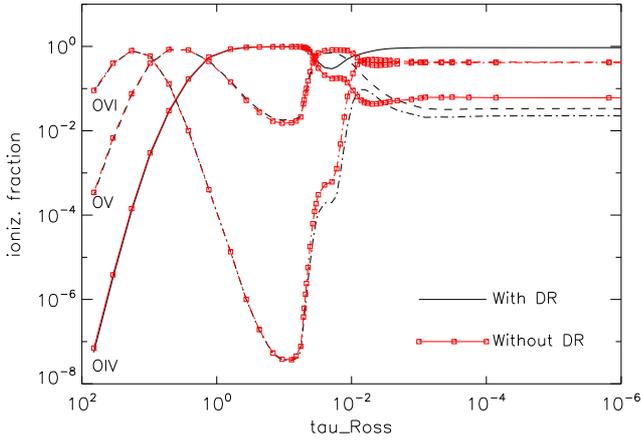}}
%file: compare_drpaper.pro
%
\caption{Ionization fractions of oxygen, as a function of \taur, for a
	 D45 model with \fv\ = 0.03 and \Tshockmax\ = 3$\cdot 10^6$~K,
	 with and without dielectronic recombination (DR). Note the
	 large differences for all the stages when $\taur \leqslant
	 10^{-2}$ ($v(r) \geqslant 0.05\vinf$),
	 particularly the change in the main ionization
         stage (from \OV/\OVI\ to \OIV) when DR is included.} 
\label{ww_DR}
\end{figure}

\subsection{Dielectronic Recombination of \OV}
\label{DR_OV}

After comparing the results from our first models accounting for
shock-emission with corresponding WM-{\sc basic} results, it turned
out that in a specific parameter range (for dwarfs around 45~kK) both
codes delivered largely different fluxes around the \OIV\ edge at
$\sim$160~\AA, which could be tracked down to completely different ionization
fractions of oxygen. In particular, our models displayed more \OV\
and less \OIV\ than calculated by WM-{\sc basic}.

After investigating the origin of this discrepancy, it turned out that
we, inadvertently, had not included the data for dielectronic
recombination\footnote{This process can be summarized as `the capture
of an electron by the target leading to an intermediate doubly excited
state that stabilizes by emitting a photon rather than an electron'
\citep{rivero12}.} (hereafter DR) in our oxygen atomic
model.\footnote{For Si, P and \CV\ $\rightarrow$ \CIV\ corresponding
data are still missing in our database.} Thus, DR processes had not
been considered for oxygen. 

A series of studies had recently reconsidered the effects of DR with
respect to {\textit{nitrogen}} (\citealt{rivero11, rivero12,rivero122}),
however no significant effects were found, particularly
concerning the formation of the prominent \NIII\ $\lambda \lambda\
$4634-4640-4642 emission lines that was previously attributed to DR
processes \citep{brucato71, mihalas73}.

Nevertheless, we subsequently included DR also into our oxygen atomic
model, and were quite surprised by the consequences. Though in a
large region of our model grid the changes turned out to be negligible
for the fluxes, in all supergiant models and in the dwarf models
around 45~kK the ionization fractions were strongly affected, leading
to a decrease of \OV, typically by a factor of 10 to 50. 

For our most problematic D45 model, DR proved to be essential to even
predict the correct main ion throughout the wind, and to produce a
reliable SED around the \OIV\ edge. Fig.~\ref{ww_DR} displays the 
impact of DR for this model. Indeed, the population of every
ionization stage becomes modified in the wind, but for \OIV\ this
difference is large enough to change it to the main stage of the
model. The reason for such drastic impact in the region around D45 is
based on the fact that only here the X-ray ionization is potentially
able to allow for the dominance of \OV\ (see Fig.~\ref{xrays43}),
which then can be compensated by quite strong dielectronic
recombination rates.\footnote{As an independent check of our findings,
we also calculated WM-{\sc basic} models without DR, and they turned
out to be consistent with our non-DR models.}

Nevertheless, since in the majority of models \OV\ becomes severely
depleted (see above), independent of whether it is a main ion or not, and
also \OVI\ is affected, this leads to considerable changes in the
corresponding UV lines. Thus, we conclude that DR is inevitable for a
correct treatment of oxygen. Moreover, because of this strong impact,
the precision of corresponding data needs to be re-checked.  

As a final remark, let us note that the inclusion of DR has also an
impact on non-X-ray models, but to a much lower extent.

\subsection{Mass Absorption Coefficient}
\label{op_section}

As already mentioned in Sect.~\ref{Introduction}, also the X-ray {\textit{
line}} emission (observed by means of CHANDRA and XMM-{\sc Newton}) has
been modeled and analyzed during recent years, by various groups. Such
analysis particularly allows us to obtain constraints on the presence,
structure, and degree of wind inhomogeneities at X-ray wavelengths
(e.g., \citealt{Oskinova06, Sundqvist12b, leutenegger13}), to
independently `measure' the mass-loss rates of O-star winds (e.g.,
\citealt{Herve13, Cohen14, rauw15}), and even to derive nitrogen and
oxygen abundances\footnote{Primarily, these abundance determinations
involve measuring the strengths of corresponding emission lines in the
soft X-ray regime, maybe correcting them for absorption. Note,
however, that this diagnostics is {\textit{not}} a wind absorption
diagnostics, but that absorption is only a correction needed to derive
line luminosities.} \citep{Oskinova06, ZhekovPalla07, Naze12, 
leutenegger13a}. One of the assumptions made by various authors is to
consider the mass absorption coefficient of the cool wind material,
$\kappa_{\nu}(r)$, as spatially constant, which simplifies the
analysis \citep{OwockiCohen06, leutenegger13, Cohen14}. Other groups
include detailed predictions for the spatial and frequency dependence
of $\kappa_{\nu}(r)$, calculated by means of POWR (e.g.,
\citealt{Oskinova06}) or CMFGEN (e.g., \citealt{Herve13, rauw15}), and
there is an ongoing discussion whether the assumption of a spatially
constant $\kappa_{\nu}$ is justified and in how far it affects the
precision of the deduced mass-loss rates. Though \citet{Cohen10,
Cohen14} have investigated the variation of $\kappa_{\nu}(r)$ and its
influence on the derived parameters based on selected CMFGEN-models
(also accounting for variations in the CNO-abundances), a systematic
study has not been performed so far, and in this section we will do
so.

At first, let us consider why and under which conditions
$\kappa_{\nu}$ should become more or less spatially constant. The
prime reason for this expectation is the fact that the K-shell cross
sections (at threshold and with respect to wavelength dependence) of
the various ions of a specific atom are quite similar, and that the
corresponding edges (for these ions) lie close together. Provided now
that (i) {\textit{all}} ions which are present in the wind are actually
able to absorb via K-shell processes, and (ii) that there are no
`background' opacities from other elements, $\kappa_{\nu}(r)$ 
indeed becomes (almost) spatially constant, since the total opacity is then
the simple sum over the K-shell opacities from all contributing atoms,
\beqa
\kappa_{\nu}(r) &\approx& \sum_k \Bigl(\sum_j
\frac{n_{k,j}(r)}{\rho(r)} \sigma_{k,j}(\nu) \Bigr) \approx \nonumber \\
&\approx& \sum_k \Bigl(\sum_j \frac{n_{k,j}(r)}{\rho(r)} \Bigr)\, 
\sigma_{k}(\nu) \approx  \nonumber \\
&\approx& \sum_k \frac{n_k(r)}{\rho(r)} \sigma_k(\nu)
\approx \frac{\sum_k \alpha_k \sigma_k(\nu)}{m_{\rm H}(1+4Y_{\rm He})} 
:= \kappa_{\rm \nu}^{\rm appr},
\label{kappa_approx}
\eeqa
with $\alpha_k$ the elemental abundance, $Y_{\rm He}$ the helium
abundance (both quantities normalized to hydrogen), and $m_{\rm H}$
the hydrogen mass. $k$ denotes the atomic species, and $j$ the ion,
$n_{k,j}$ is the occupation number of ion ($k$,$j$), and 
$\sigma_{k,j} \approx \sigma_k$ the K-shell cross section, being almost
independent of $j$. In the last step of the above derivation, we have
assumed that the atmosphere consists mostly of hydrogen and helium.

Thus, we have to check under which conditions restrictions (i) and
(ii) might no longer be valid. For the light and abundant elements
CNO, K-shell absorption is no longer possible for \CV, \NVI\ and
\OVII. For these ions, only `ordinary', outer-shell ionization is
present, but also here the cross-sections are not too different from
the K-shell processes (both with respect to strength and location of
edge).
%E.g., we already mentioned that the K-shell
%edge for \CIV\ (at $\approx$ 36~\AA) is far away from the outer-shell
%ionization edge of this ion (at $\approx$ 192~\AA), whilst the edge of
%\CV\ is close to the \CIV\ edge, at $\approx$~30~\AA. 
Thus, even for
highly ionized winds (hot or with strong X-ray emission) where \CV,
\NVI, and \OVII\ are actually present somewhere, the above approximation
is still justified. In so far, restriction (i) should play no
role, since even higher ionization stages are not too be expected to
be significantly populated.

\begin{figure}
\begin{minipage}{9cm}
\resizebox{\hsize}{!}
  {\includegraphics{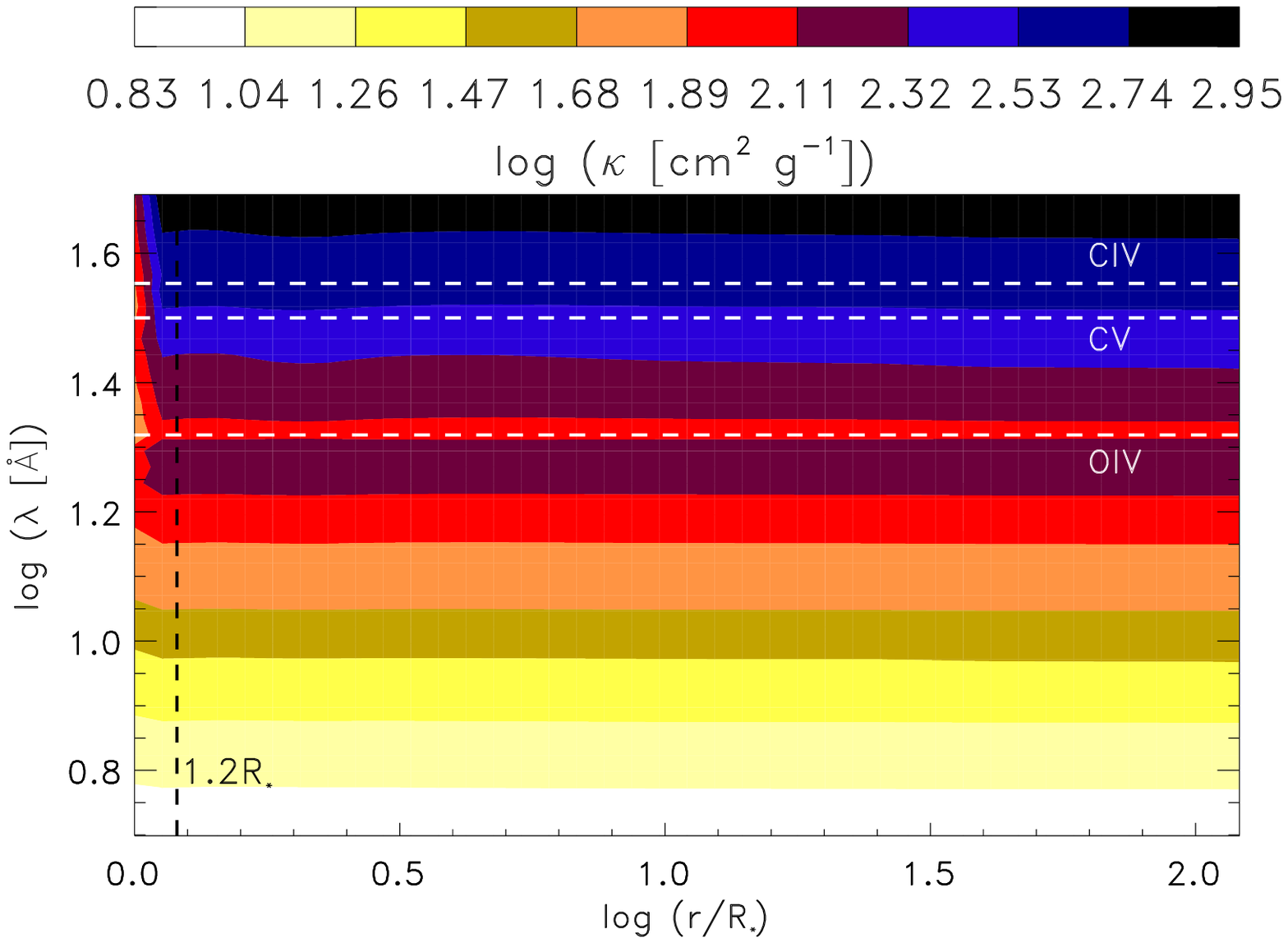}} 
\end{minipage}

\begin{minipage}{9cm}
\resizebox{\hsize}{!}
  {\includegraphics{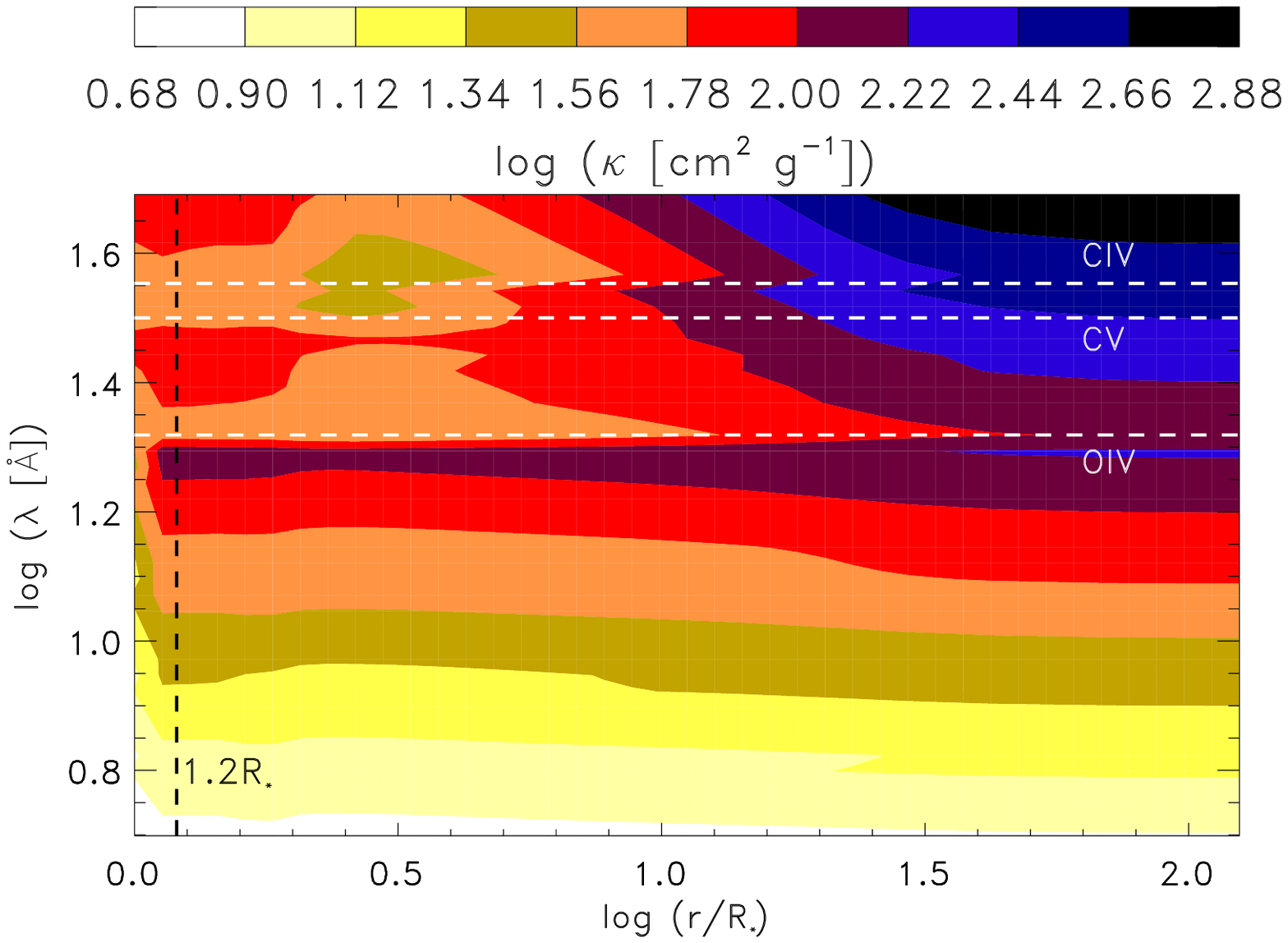}} 
\end{minipage}
%file: plot_opac_xray_reverse.pro
%
\caption{Contour plots illustrating the radial dependence of the mass
absorption coefficient, $\kappa_{\nu}(r)$, as a function of
wavelength. The upper panel refers to model D30, and the lower to
model S40, both with typical X-ray emission parameters (\Tshockmax\ = $ 3 \cdot
10^{6}$~K and \fv\ = 0.03). The positions of the \CV\ edge (outer-shell
ionization) and the \CIV\ and \OIV\ K-shell edges are indicated.}
\label{cont_plot}
\end{figure}

Regarding restriction (ii), the situation is different. The prime
background is given by the \HeII\ bound-free opacity, which becomes
strong in `cool' and/or helium-recombined winds\footnote{Additionally,
the outer-shell ionization of \OIV\ with edge at $\approx$~160~\AA\
and the bound-free opacities from other, strongly abundant ions can
play a minor role, particularly if \HeII\ is weak or absent.}, where
in the following we always refer to the recombination of \HeIII\ to
\HeII. Note that already \citet{Hillier93} showed the importance of
outer-wind helium recombination on wind opacity and emergent soft
X-ray emission.

Let's check on the maximum influence of the \HeII\ bound-free opacity
at important K-shell edges. For a crude estimate, we approximate its
frequency dependence by $(\nu_0/\nu)^3 = (\lambda/\lambda_0)^3$, and
assume the worst case that \HeII\ is the only He ion present in the
wind. Then, a lower limit for the opacity ratio at specific K-shell
edges can be approximated by 
\beqa
\frac{\kappa_{k}}{\kappa_{\rm HeII}}\bigl(\lambda_0({k})\bigr)
&\approx& \frac{n_{k}}{n_{\rm HeII}} \frac{\sigma_0({k})}
{\sigma_0({\rm HeII})} \Bigl(\frac{\lambda_0({\rm HeII})}
{\lambda_0({k})}\Bigr)^3 \ga \nonumber \\ 
&\ga& \frac{\alpha_{k}}{\alpha_{\rm He}} \frac{\sigma_0({k})}
{\sigma_0({\rm HeII})} \Bigl(\frac{228\ \mbox{\AA}}{\lambda_0({k})}\Bigr)^3,
\eeqa
where $\sigma_0$ is the cross section at the corresponding edge. Using
solar abundances from \citet{asplund09}, $\lambda_0({\rm C}) \approx$
35~\AA\ and $\lambda_0({\rm O}) \approx$ 20~\AA, $\sigma_0 \approx$
1.6, 0.9, and 0.5$\cdot 10^{-18}$~cm$^2$ for the threshold cross
sections of \HeII, carbon (K-shell), and oxygen (K-shell),
respectively, we find $\kappa_{\rm C}/\kappa_{\rm HeII}(35~\AA) \ga$ 0.42 and
$\kappa_{\rm O}/\kappa_{\rm HeII}(20~\AA) \ga$ 2.3. Thus, for cool and/or
He-recombined winds, the \HeII\ opacity dominates at the carbon K-shell
edge, whilst at the oxygen edge the K-shell opacities are quite a bit larger
than the background. Thus, we would predict that somewhat below $\approx$
20~\AA\ (beyond 620 eV) restriction (ii) becomes valid, and that
$\kappa_{\nu}$ should become depth-independent. Vice versa, the mass
absorption coefficient should vary with radius longward from the oxygen or
carbon K-shell edge, whenever the background mass absorption
coefficient varies, mostly due to changes in the \HeII\ ionization
throughout the wind.

In the following, we discuss these issues by means of our grid models
-- all of them with shock emission described by our typical parameters
(\Tshockmax\ = $3 \cdot 10^6$~K and \fv\ = 0.03). In particular, we
will provide estimates for suitable means of $\kappa_{\nu}$, as a
function of \Teff.

Figure \ref{cont_plot} displays contour-plots of the radial dependence
of the mass absorption coefficient in a D30 (upper panel) and in an
S40 model (lower panel), as a function of wavelength. In accordance
with our expectation from above, in both panels we note that
$\kappa_{\nu}$ becomes constant when $r \ga$ 1.2~\Rstar \footnote{to
be on the safe side. In most cases, this limit -- arising from
fluctuations in the opacity background -- is even lower.} and the
wavelength is lower than 20~\AA\ (log $\lambda \la$ 1.3).  

Longward of the \OIV\ K-shell edge ($\lambda >$ 21~\AA), the radial
variation of $\kappa_{\nu}$ depends on effective temperature and wind
density.  For the D30 model, $\kappa_{\nu}$ increases significantly
with wavelength, but nevertheless does not vary
with radius, because in this case the dominating ionization fraction
of \HeII\ remains constant throughout the wind. In contrast, somewhat
hotter models (e.g., D35), but particularly models with denser winds
such as S40 display a different behavior. Here, the lower wind is
dominated by \HeIII, so that the background is weak, and one can
already discriminate the \CIV\ and \CV\ edges around 10~\Rstar\
(indicated as dashed lines). Compared to the dwarf models, the total
$\kappa_{\nu}$ in the inner wind is much lower, shows much more
structure, and is influenced by the carbon and nitrogen opacities.
Once helium begins to recombine in the outer wind, the background
begins to dominate again, and the K-shell features vanish.

\begin{figure}[t]
\resizebox{\hsize}{!}
{\includegraphics{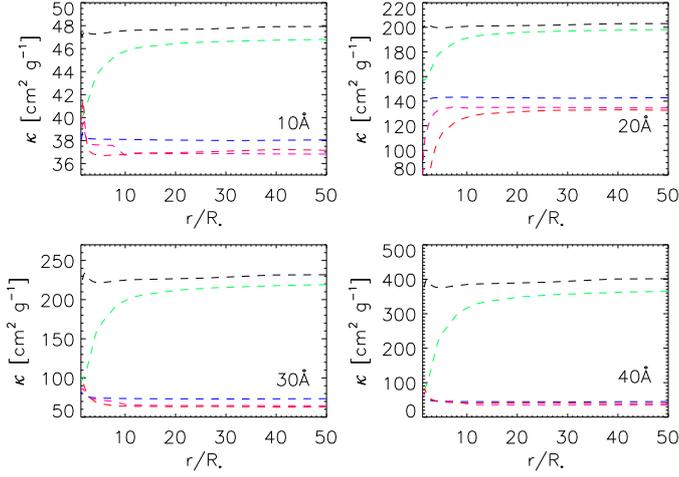}}
\caption{Radial variation of the mass absorption coefficient in dwarf
	 models, for specific values of wavelength. Black: \Teff\ =
	 30kK, green: 35\,kK; blue: 40\,kK; magenta: 45\,kK; red: 50\,kK. All
	 models calculated with \Tshockmax\ = $ 3\cdot 10^{6}$~K and
	 \fv\ = 0.03. Note the different scales for $\kappa_{\nu}$.}
\label{opacity_dwarfs}
\end{figure}

Fig.~\ref{opacity_dwarfs} illustrates the radial variation of the mass
absorption coefficient for different wavelengths, and for our dwarf
models with \Teff\ from 30 to 50~kK. Independent of \Teff, the radial
variation of $\kappa_{\nu}$ is marginal at (and below) 10~\AA. Around
20~\AA, the variations in the inner/intermediate wind (until
10~\Rstar) are somewhat larger, due to changes in the oxygen
ionization, where the specific positions of the corresponding edges
play a role (see also Fig.~\ref{average_op_dw}). At 30~\AA, we see a
separation between D30 (black) with high values of $\kappa_{\nu}$
(\HeII\ dominating), hot models with low values of $\kappa_{\nu}$
(\CV\ + low background, since helium completely ionized), and D35
(green) with a significantly varying $\kappa_{\nu}$, due to the
recombination of \HeIII\ in the external wind. At 40~\AA, finally, the
behavior is similar, and only the $\kappa_{\nu}$ values for the
cooler models are larger, because of the increasing \HeII\ background.

\begin{figure}
\resizebox{\hsize}{!}
{\includegraphics{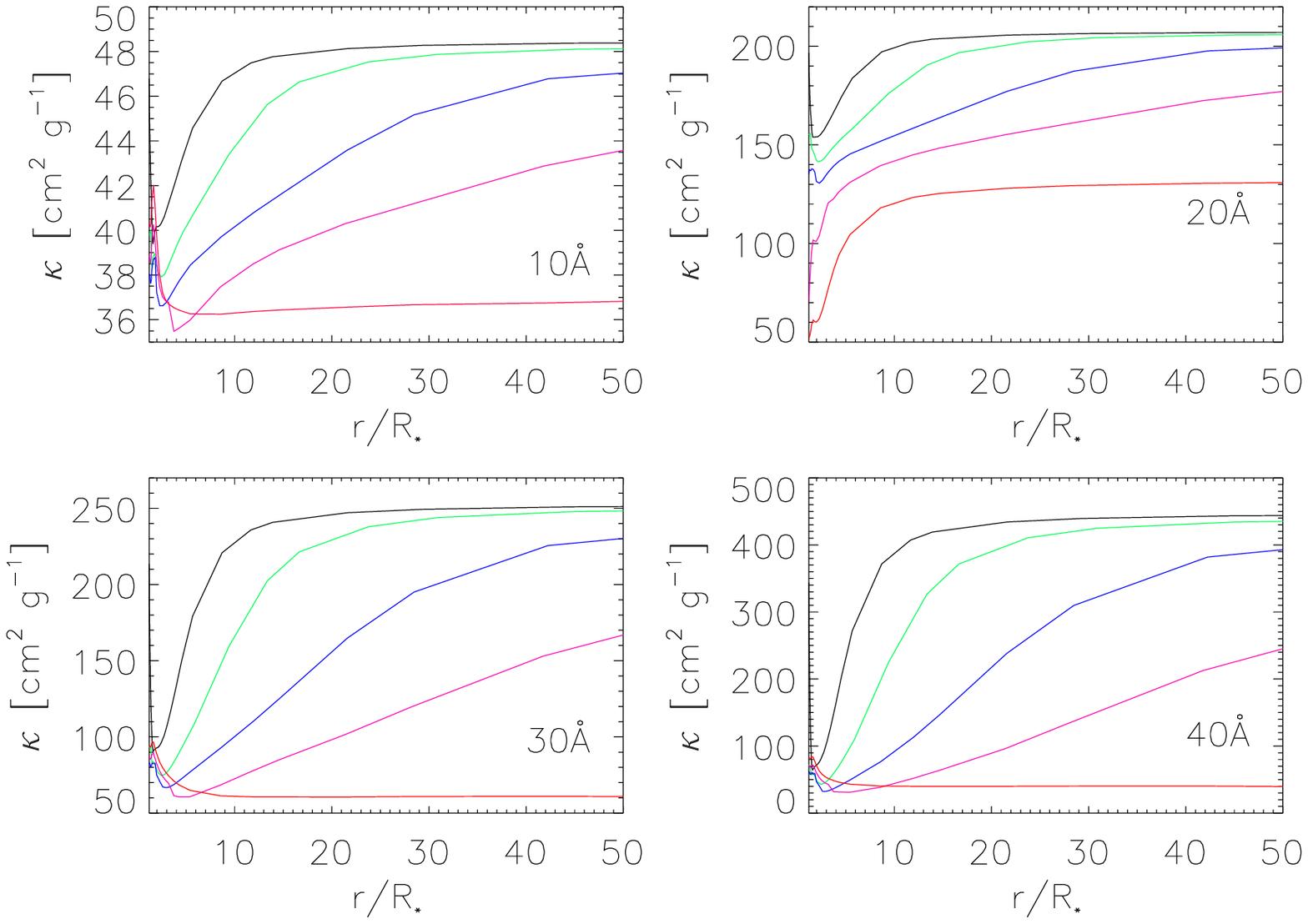}}
\caption{As Fig.~\ref{opacity_dwarfs}, but for supergiant models.}
\label{opacity_supergiants}
\end{figure}

\begin{figure}[t]
\resizebox{\hsize}{!}
{\includegraphics{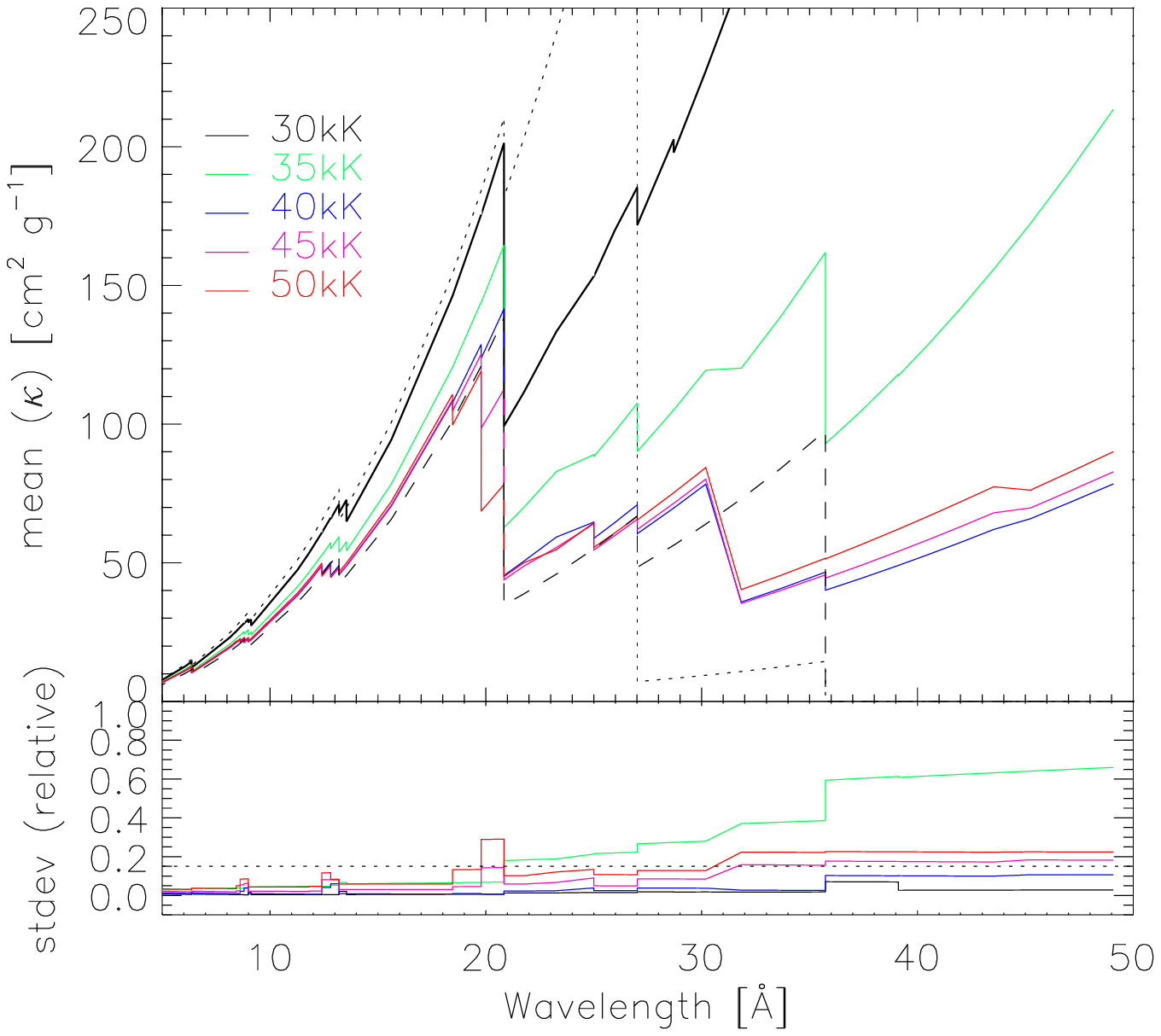}}
\caption{Upper panel: Density-weighted mean (Eq.~\ref{meankappa}) of
	 the mass absorption coefficient, $\bar \kappa_\nu$, for the
	 interval between 1.2 and 110~\Rstar, as a function of
	 wavelengths, and for dwarf models with \Tshockmax\ =
	 3$\cdot$10$^{6}$~K and \fv\ = 0.03. Solar abundances
	 following \citet{asplund09} have been adopted.  Dashed:
	 approximate, radius-independent $\kappa_\nu^{\rm appr}$
	 (Eq.~\ref{kappa_approx}), using solar abundances and K-shell
	 opacities only, with cross-sections from \CIV\ (with
	 threshold at 35.7~\AA), \NIV\ (27.0~\AA), \OIV\ (20.8~\AA),
	 Ne\,{\sc iv} (13.2~\AA), Mg\,{\sc iv} (9.0~\AA), and \SiIV\
	 (6.4~\AA). The \CV\ edge (at 31.6~\AA) appears as
	 unresolved in our frequency grid. Dotted: as dashed, but with
	 nuclear processed CNO abundances as derived for $\zeta$~Pup
	 by \citet{bouret12}. Note that the nitrogen abundance is more
	 than a factor of 10 larger than the solar one. Dashed and
	 dotted lines serve also as a guideline for comparison with
	 similar figures.\newline
         Lower panel: Relative
	 standard deviation, $\sqrt{{\rm Var}(\kappa_\nu)}/\bar
	 \kappa_\nu$ (see Eq.~\ref{varkappa}), for the same models.
	 The dotted line denotes a relative scatter of 15\%.}
\label{average_op_dw}
\end{figure}

The analogous situation for supergiants is shown in
Fig.~\ref{opacity_supergiants}. Whilst for dwarfs the variation of
$\kappa_{\nu}$ (when present) vanishes at around 10~\Rstar, here it is
visible throughout the wind until large radii, for all but the coolest
(black) and the hottest (red) model. Note that the limiting values (at
the outermost radius) are similar to those of the corresponding dwarf
models at \Teff\ = 30 and 35~kK (recombined) and at \Teff\ = 50~kK
(\HeIII). In contrast, for models with \Teff\ = 40 and 45~kK the
opacity continues to increase outwards, since the recombination is
still incomplete.

\citet{Herve13} provided a similar figure to investigate the radial
variation of $\kappa_{\nu}$, in this case for a model of $\zeta$ Pup
calculated by CMFGEN. Though the stellar parameters roughly agree with
our S40 model, they considered a clumpy wind (with volume filling
factor $f_V = 0.05$), and nuclear processed CNO abundances. Because
this model shows an earlier recombination of helium, a larger nitrogen
and a weaker oxygen K-shell edge, the actual values of $\kappa_{\nu}$
are somewhat different from our results (except at shortest
wavelengths), but the basic trends are quite similar. In particular,
our results support \cite{Herve13}'s idea of parameterizing the run of
$\kappa_\nu$: In any of the $\kappa_\nu(r)$-curves displayed in
Figs.~\ref{opacity_dwarfs} and \ref{opacity_supergiants}, these curves
either increase or slightly decrease, but eventually reach a plateau from
a certain radius on (which differs for each model). This radius
then separates two different regimes of $\kappa_\nu$ that might be
parameterized in an appropriate way (see \citealt{Herve13} for details).

Instead of a parameterization, it is also possible to calculate
meaningful averages of $\kappa_\nu$ and the corresponding scatter. The
size of this scatter then allows us to conclude when (w.r.t.
wavelength and $\Teff$) a spatially constant mass absorption
coefficient might be used to analyze X-ray line profiles. Instead of a
straight average, we use here a density-weighted average (and a
corresponding variance), to account for the fact that the optical
depth, $\tau_\nu$, is the quantity which needs to be calculated with
high precision:
\beqa
\tau_\nu = \int^{R_{\rm max}}_{R_{\rm min}} \kappa_\nu(r) \rho(r) \dd r &=:&
\bar \kappa_\nu \int^{R_{\rm max}}_{R_{\rm min}} \rho(r) \dd r \qquad \Rightarrow
\nonumber \\
\label{meankappa}
\bar \kappa_\nu &=& \int^{R_{\rm max}}_{R_{\rm min}} \kappa_\nu(r)
f(r) \dd r, \\
\label{varkappa}
{\rm Var}(\kappa_\nu) &=& \int^{R_{\rm max}}_{R_{\rm min}} (\kappa_\nu(r)-\bar \kappa_\nu)^2 
f(r) \dd r\\
{\rm with \quad p.d.f.} \quad f(r)\dd r &=& \rho(r)\dd r/\biggl[\int^{R_{\rm max}}_{R_{\rm min}}
\rho(r) \dd r\biggr]. \nonumber
\eeqa
In this approach\footnote{Note that the quantity $R_{\rm min}$
indicates the lower boundary for the averaging process, and must not
be confused with the onset radius of the X-ray emission.}, the
density weights correspond to a probability distribution function
(p.d.f.).

Fig.~\ref{average_op_dw} displays such mean mass absorption
coefficients, $\bar \kappa_\nu$, as a function of wavelength, averaged
over the interval between 1.2 and 110.0~\Rstar, for our dwarf
models.\footnote{The impact of the chosen interval will be
discussed below.} The lower panel denotes the {\textit{relative}} standard
deviation, $\sqrt{{\rm Var}(\kappa_\nu)}/\bar \kappa_\nu$. Also here,
cold and hot models are clearly separated, with D35 in between (cf.
with Fig.~\ref{opacity_dwarfs}): for $\lambda \ga 21$~\AA, the cold
models are affected by a strong \HeII-background, whilst this
background is weak for the hotter ones. In this `long wavelength'
region, the radial variation of $\kappa_\nu$ is large for model D35,
due to recombining helium. There is also a considerable scatter
between 18 and 21~\AA, because of radial changes in the oxygen
ionization. Overall, however, the assumption of a constant mass
absorption coefficient (suitably averaged) is not too bad for the {\textit{
complete}} wavelength range (scatter below 20\%), if we exclude
model D35.  Below 18~\AA, the scatter becomes negligible, except at
the Ne, Mg, and Si edges. 

Even if $\kappa_\nu(r)$ can be approximated by a single number, $\bar
\kappa_\nu$, the question is then about its value. For a comparison,
the dashed line in Fig.~\ref{average_op_dw} displays the (analytic)
estimate, $\kappa_\nu^{\rm appr}$ as provided by
Eq.~\ref{kappa_approx}, using solar abundances and K-shell opacities
only, with cross-sections from \CIV, \NIV, \OIV, Ne\,{\sc iv},
Mg\,{\sc iv}, and \SiIV. At least for the hotter models, this estimate
is quite appropriate when comparing to the actual case, except for a
somewhat erroneous description of the carbon edge(s): Since \CV\
dominates in the hotter models and there is a $\sim4$~\AA\ difference
between the \CIV\ K-shell and the \CV\ edge, this region is badly
described by our approximation. For the cooler models, on the other
hand, the difference
between the dashed and the solid curves is (mostly) due to the helium
background, {\textit{which varies as a function of \Teff, \logg, and
wind-density}}, thus affecting the actual value of $\bar \kappa_\nu$.
Even below 18~\AA, this background is still non-negligible for model
D30, with a maximum deviation of roughly 30\% close to the oxygen
edge. Nevertheless, we conclude that for all dwarf models with $\Teff
\ge 35$~kK the assumption of a constant mass absorption coefficient
approximated by $\kappa_\nu^{\rm appr}$ is justified when $\lambda \le
18$~\AA\ (at least within our present assumptions, i.e., solar
abundances and unclumped winds with not too large optical depths, such
that the averaging down to 1.2 \Rstar\ is reasonable; for different
models, see below). In all other cases, results from NLTE-atmosphere
modeling should be preferred. 

\begin{figure}
\resizebox{\hsize}{!}
{\includegraphics{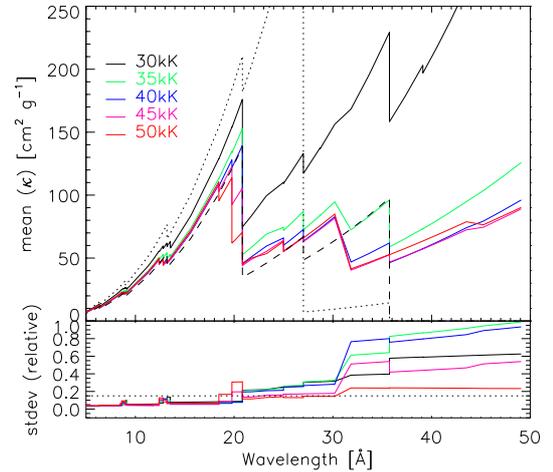}}
\caption{As Fig.~\ref{average_op_dw}, but for supergiant models.  
%         All the supergiant models have a large average opacity
%	 standard deviation, except S50 which is dominated by K-shell
%	 opacity along the wind.  } 
         }
\label{average_op_super}
\end{figure}

The situation for our supergiant models is displayed in
Fig.~\ref{average_op_super}. Below 20~\AA, the situation is similar to
the dwarf case, though here the background is lower, even for the
coolest model, and the approximation of $\bar \kappa_\nu$ by
$\kappa_\nu^{\rm appr}$ might now be applied at all temperatures. For
$\lambda > 30$~\AA, however, almost all models (except for S50) can no
longer be described by a radially constant $\kappa$, since all
models with $\Teff \le 45$~kK show recombining helium of different
extent, leading to strong variations throughout the wind.

Thus far, we considered models with solar abundances and unclumped
winds. To illustrate the variation of the total and K-shell opacities
with abundance (already investigated for particular models by, e.g.,
\citealt{Cohen10, Cohen14}), the dotted lines in
Figs.~\ref{average_op_dw} and \ref{average_op_super} denote the
approximate K-shell opacities, $\kappa_\nu^{\rm appr}$, for the case
of highly processed CNO material, based on the abundances derived for
$\zeta$ Pup by \citet{bouret12}. Here, the carbon and oxygen
abundances are depleted by 0.8 and 0.6~dex, respectively, whilst the
nitrogen abundance is extremely enhanced (by $\sim1.3$~dex), compared
to the solar values. Such a composition leads to weak C and O K-shell
edges, but to an enormous nitrogen edge (dotted vs. dashed line). 

Now, if the individual abundances are known during an analysis, there
will be no problem, and $\kappa_\nu$ might be approximated by either
$\kappa_\nu^{\rm appr}$ below 18~\AA\ or calculated by means of
NLTE-model atmospheres, simply accounting for these abundances.
However, considerable uncertainties even in the low wavelength regime
might result when the abundances are {\textit{not}} known. From comparing
the dashed and the dotted line, we estimate this uncertainty as
roughly 50\% for $\bar \kappa_\nu$, and thus for $\tau_\nu$ and \mdot\
(when the mass-loss rate shall be derived). A similar value has
already been estimated by \citet{Cohen14}. In the range between the
oxygen and the carbon edge (20 to 35~\AA), the situation is even
worse, and we conclude that the corresponding absorption coefficients
are prone to extreme uncertainties when the abundances have to be
adopted without further verification. In particular, getting
$\kappa_\nu$ right around 25~\AA\ is important for measuring the N
emission lines at and near that wavelength (e.g., N\,{\sc vi} 24.9,
N\,{\sc vii} 24.78), and thus measuring the N abundance directly. But 
at longer wavelengths, where $\kappa_\nu$ will vary even more strongly
with radius, and even though nitrogen emission lines are not directly 
affected, the (direct) ionization of CNO etc. {\textit{is}} affected, and so 
optical and UV line strengths are affected too, as discussed in the
previous sections.

The impact of clumping is less severe. Comparing
Fig.~\ref{average_op_f20} (Appendix) with Figs.~\ref{average_op_dw}
and \ref{average_op_super}, we see that models accounting for
optically thin clumping (`micro-clumping') with typical clumping
factors\footnote{$f_{\rm cl} = 20$ corresponding to a volume filling
factor, $f_{\rm V}$, = 0.05} and adequately reduced mass-loss rates
give rather similar results compared with unclumped models. Again, the
scatter of $\kappa_\nu$ is negligible below 18~\AA. `Only' the region
longward of 20~\AA\ is stronger contaminated by the \HeII\
background, since the clumped models recombine earlier than the
unclumped ones. Note that the K-shell mass absorption coefficients
themselves are {\textit{not}} affected by optically thin clumping, since
the opacities scale linearly with density.

Finally, Fig.~\ref{average_op_r10} (Appendix) investigates the
consequences of averaging $\kappa_\nu$ in the outer wind alone (in the
interval between 10 and 110 \Rstar), which would be adequate if the
wind would become optically thick at such radii (which for short
wavelengths and O-star winds is quite unlikely, because of the low
value of $\kappa_\nu$). Anyway, below 18~\AA\ the differences to the
original values are small. Note that the hot dwarf models now behave
almost exactly as estimated by $\kappa_\nu^{\rm appr}$, because \HeII\
vanishes in the outer regions of these objects. 

Further conclusions on this topic are provided in the next section.

\section{Summary and conclusions} 
\label{conclusions}

In this paper, we described the implementation of X-ray emission from
wind-embedded shocks into the unified, NLTE atmosphere/spectrum
synthesis code {\sc FASTWIND}, discussed various tests, and presented
some first result.

Our implementation follows closely corresponding work by
\citet{pauldrach01} (for WM-{\sc basic}), which in turn is based on the
shock cooling zone model developed by \citet{Feldmeier97b}, with the
additional possibility to consider isothermal shocks. The (present) 
description of the shock-distribution and strength is provided by four
input, `X-ray emission parameters', controlling the filling factor,
the run of the shock temperature, and the radial onset of the
emitting plasma. We account for K-shell absorption and Auger
ionization, allowing for more than one final ionization stage due to
cascade ionization processes.

Most of our test calculations are based on a grid of 11 models
(supergiants and dwarfs within \Teff = 30 to 55 kK), each of them with
9 different X-ray emission parameter sets, but many more models have
been calculated for various comparisons, including 
models with optically thin clumping. 

A first test investigated the reaction when varying important
X-ray emission parameters. For radially increasing shock-strengths,
the emergent flux remains almost unaffected if the onset radius is
lowered compared to its default value (roughly 1.5~\Rstar), whilst
increasing the onset has a considerable effect in the range between
$\sim$350~\AA\ and at least the \HeII\ edge. Filling factor and
maximum shock temperature affect the ionization fractions,
particularly of the highly ionized species. We confirm some earlier
predictions for scaling relations for X-ray luminosities (as a
function of $\mdot/\vinf$) in the case of optically thin and thick
continua\footnote{though a discrepancy with recent work by
\citet{OwockiSundqvist13} was identified, which needs to be
investigated further}, but we noted that for our hottest models these
luminosities can become contaminated by `normal' stellar radiation,
for energies below $\sim$150~eV. Thus, we suggested to choose a lower
integration limit of 0.15 keV (or even 0.3 keV, to be on the safe
side) when comparing the X-ray luminosities of different stars or
theoretical models. Finally, we found an excellent agreement between
{\sc FASTWIND} and WM-{\sc basic} fluxes, demonstrating a similar
ionization balance, and a satisfactory agreement between corresponding
X-ray luminosities. Overall, the impact of typical shock emission
affects the radiation field in the wind for all wavelengths $\lambda <
350$~\AA, thus modifying all photo-rates for ions with ionization
edges in this regime.

Investigating the ionization fractions within our model grid allowed
us to study the impact of shock radiation for the proper
description of important ions, i.e., those with meaningful wind lines
(e.g., \CIV, \NIV, \NV, \OV, \OVI, \SiIV, and \PV). If we denote
models with \Teff = 30 to 35~kK as `cool', models with \Teff = 35 to
45 kK as `intermediate' and models with \Teff = 45 to 55~kK as `hot'
(note the overlap), we can summarize our findings as follows.
{\textit{Not}} (or only marginally) affected by shock emission (with typical
parameters and our parameterization of the shock-strengths) are
\begin{itemize}
\item[$\bullet$] in dwarfs: \CIII, \CIV, \NIII\ (cool), \NIV\ (cool),
\OIV\ (intermediate), \SiIV, \PV\ (cool+intermediate)
\item[$\bullet$] in  supergiants: \CIII\ (hot), \CIV\ (hot), \NIV\
(cool), \OIV\ (intermediate), \SiIV\ (hot).
\end{itemize}
In almost all other cases, lower stages (\CIII, \CIV, \NIII, \NIV,
\OIV\ (hot), \SiIV, and \PV) are depleted, i.e., corresponding wind
lines become weaker, and higher stages (\NV, \OIV\ (cool), \OV, \OVI)
become enhanced, i.e., corresponding wind lines become stronger when
accounting for shock emission.

We studied in some detail how the ionization fractions change when
the two most important parameters, filling factor and maximum shock
temperature, are varied. For most ions, the filling factor has a
larger influence than \Tshockmax, but particularly \OVI\ and \PV\ (the
latter only for higher filling factors and shock temperatures)
show a strong reaction to both parameters. 

Due to the importance of \PV\ with respect to mass-loss and
wind-structure diagnostics, we re-investigated its behavior, and
confirm previous results that for typical X-ray emission parameters
this ion is only weakly or moderately affected (by factors of two for
intermediate and hot supergiants at $v(r)/\vinf=0.5$ and by factors of
10 at $v(r)/\vinf=0.8$), though for a strong X-ray radiation field the
depletion can reach much higher factors. A comparison of \PV\
ionization fractions with results from {\sc CMFGEN} \citep{bouret12}
provided a reasonable agreement.

Not only metals, but also He can be affected by shock emission, due to the
location of the \HeII\ edge and \HeII\,303 in the EUV. Significant
effects, however, have only been found in the winds of cool
supergiants, where particularly \HeII\,1640 (emission and
high-velocity absorption) and \HeII\,4686 (emission) become stronger,
due to increased recombination cascades and increased pumping of the
$n=2$ level in case of \HeII\,1640.

When comparing our ionization fractions with those calculated by
WM-{\sc basic}, we found a good, though not perfect, agreement, which
turned out to be true also for various UV-lines. When comparing with
\citet{krticka09}, on the other hand, a similar agreement over the
complete covered temperature range was found only for few ions; for the
majority, such agreement is present only at specific temperatures. 

It is well known that Auger ionization can play an important role for
the ionization balance of specific ions. To further investigate this
issue, we compared the ionization fractions of all ions considered in
this study when including (default) or excluding this process in our
NLTE treatment. Overall, it turned out that only \NVI\ and \OVI\ (as
previously known) are significantly affected by Auger ionization, but,
at least in our models (with radially increasing shock temperatures),
only in the outer wind.  For the inner and intermediate wind, direct
EUV/XUV ionization due to shock emission dominates, which is generally
true for all other considered ions.\footnote{In clumped winds, the
presence of a low-density interclump medium is essential for the
formation of \OVI\ \citep{Zsargo08b}.} 

As an interesting by-product of our investigation, it turned out that
dielectronic recombination of \OV\ can have a considerable influence 
on the ionization balance of oxygen (\OIV\ vs. \OV), particularly for
dwarfs around 45~kK.

In the last part of this paper, we provided an extensive discussion of
the (high-energy) mass absorption coefficient, $\kappa_\nu$, regarding
its spatial variation and dependence on \Teff. This topic is
particularly relevant for various approaches to analyze X-ray emission
lines. To summarize and conclude, we found that (i) the approximation
of a radially constant $\kappa_\nu$ can be justified for $r \ga 1.2
\Rstar$ and $\lambda \la 18$~\AA, and also for many models at longer
wavelengths. (ii) In order to estimate the actual value of this
quantity, however, the \HeII\ background\footnote{and, to a lesser
extent, also the bound-free background from highly abundant metals}
needs to be considered from detailed modeling, at least for
wavelengths longer than 18 to 20~\AA. Moreover, highly processed CNO
material can change the actual value of $\kappa_\nu$ considerably,
particularly for $\lambda \ga 20$~\AA, and estimates for the optical
depth, $\tau_\nu$, become highly uncertain in this regime if the
individual abundances are unknown.

In this context, it is reassuring to note that, e.g., the mass-loss
determinations by \citet{Cohen14} using X-ray line spectroscopy (via
determining the optical depths of the cool wind material, under the
assumption of spatially constant $\kappa_\nu$) rely on 16 lines
observed by CHANDRA, where 14 out of these 16 lines are shortward of
19~\AA. The issues summarized above will be a much bigger problem for
\OVII\ and nitrogen X-ray emission line measurements (O\,{\sc vii} at
21.6-22.1~\AA, N\,{\sc vii} at 24.78~\AA\ and N\,{\sc vi} at 24.9~\AA)
that are planned to independently constrain, with high precision, the
nitrogen/oxygen content in (a few) massive O-stars
\citep{leutenegger13a}. To this end, a detailed modeling of
$\kappa_\nu$ (particularly regarding the helium ionization) will
certainly be advisable for such an analysis.

Having finalized and carefully tested our implementation of emission
from wind-embedded shocks, we are now in a position to continue our work on
the quantitative spectroscopy of massive stars. As outlined in the
introduction, we will concentrate on determining the carbon and oxygen
abundances in O- and early B-stars observed during the two VLT-{\sc
flames} surveys conducted within our collaboration, by means of
optical and, when available, UV spectroscopy. During such an analysis,
the X-ray emission parameters need to be derived in parallel with the
other, main diagnostics, at least in principle. We then have
to check in how far the derived abundances depend on corresponding
uncertainties.

Note further that any such UV analysis also needs to consider the
effects of optically thick clumping \citep[e.g.,][]{oskinova07, Sundqvist11,
Surlan13, Sundqvist14}. In parallel with the implementation of
wind-embedded shocks presented here, we have updated {\sc fastwind} to
also account properly for such optically thick clumping (porosity in
physical and velocity space), following \citet{Sundqvist14}; these
models will be presented in an upcoming (fourth) paper of this series.

Regarding quantitative spectroscopic studies accounting for X-ray
ionization effects, the parameterization represented by
Eq.~\ref{jump_velo} is certainly not the final truth, and actually
also not the best encapsulation of the results from present-day
numerical simulations. Though this probably does not matter too much
for most applications, it might be worth thinking about a better
representation, and how our results would change if the stronger and
weaker shocks were allowed to be more spatially mixed. 

LDI simulations \citep[e.g.,][]{Feldmeier97a, DO03, Sundqvist13}
indicate that the velocity dispersion peaks quite close to \Rmin\
($\sim$ 1.5-2.0~\Rstar) and then falls off. And the same simulations
show also some strong shocks near \Rmin. From the observational side,
f/i ratios of ions that form at higher temperatures (e.g., Si\,{\sc
xiii}) indicate a substantial amount of high-temperature plasma 
($\sim$10$^7$~K) near \Rmin\ (e.g., \citealt{WC01, WC07}), 
and
\citet{Leutenegger06} found an onset radius of
$1.1^{+0.4/-0.1}$~\Rstar\ for the S\,{\sc xv} line. On the other
hand, \citet{CohenLi14} showed that the shock temperature distribution
is very strongly skewed toward weak shocks, and our parameterization
Eq.~\ref{jump_velo} allows us to include that feature already now.

\begin{acknowledgements}
The authors like to thank the referee, David Cohen, for helpful
comments and suggestions. LPC gratefully acknowledges support from the
Brazilian Coordination for the Improvement of Higher Education
Personnel (CAPES), under grant 0964-13-1. JOS acknowledges funding
from the European Union's Horizon 2020 research and innovation
programme under the Marie Sklodowska-Curie grant agreement No 656725.
Many thanks also to J.-C. Bouret for providing us with the ionization
fractions of \PIV\ and \PV\ from `his' models of HD\,163758 and
HD\,16691.
\end{acknowledgements}

\bibliographystyle{aa}
\bibliography{carneiro}

\begin{thebibliography}{108}
\expandafter\ifx\csname natexlab\endcsname\relax\def\natexlab#1{#1}\fi

\bibitem[{{Asplund} {et~al.}(2009){Asplund}, {Grevesse}, {Sauval}, \&
  {Scott}}]{asplund09}
{Asplund}, M., {Grevesse}, N., {Sauval}, A.~J., \& {Scott}, P. 2009, \araa, 47,
  481

\bibitem[{{Bouret} {et~al.}(2012){Bouret}, {Hillier}, {Lanz}, \&
  {Fullerton}}]{bouret12}
{Bouret}, J.-C., {Hillier}, D.~J., {Lanz}, T., \& {Fullerton}, A.~W. 2012, \aa,
  544, A67

\bibitem[{{Bruccato} \& {Mihalas}(1971)}]{brucato71}
{Bruccato}, R.~J. \& {Mihalas}, D. 1971, \mnras, 154, 491

\bibitem[{{Butler} \& {Giddings}(1985)}]{ButlerGiddings85}
{Butler}, K. \& {Giddings}, J.~R. 1985, Newsl. Anal. Astron. Spectra, 9

\bibitem[{{Cassinelli} \& {Olson}(1979)}]{Cassolson79}
{Cassinelli}, J. \& {Olson}, G. 1979, \apj, 229, 304

\bibitem[{{Cassinelli} {et~al.}(1995){Cassinelli}, {Cohen}, {Macfarlane},
  {Drew}, {Lynas-Gray}, {Hoare}, {Vallerga}, {Welsh}, {Vedder}, {Hubeny}, \&
  {Lanz}}]{Cassinelli95}
{Cassinelli}, J.~P., {Cohen}, D.~H., {Macfarlane}, J.~J., {et~al.} 1995, \apj,
  438, 932

\bibitem[{{Cassinelli} {et~al.}(1994){Cassinelli}, {Cohen}, {Macfarlane},
  {Sanders}, \& {Welsh}}]{Cassinelli94}
{Cassinelli}, J.~P., {Cohen}, D.~H., {Macfarlane}, J.~J., {Sanders}, W.~T., \&
  {Welsh}, B.~Y. 1994, \apj, 421, 705

\bibitem[{{Cassinelli} \& {Swank}(1983)}]{cassinelli83}
{Cassinelli}, J.~P. \& {Swank}, J.~H. 1983, \apj, 271, 681

\bibitem[{{Chlebowski} {et~al.}(1989){Chlebowski}, {Harnden}, \&
  {Sciortino}}]{Chlebo89}
{Chlebowski}, T., {Harnden}, F., \& {Sciortino}, S. 1989, \apj, 341, 427

\bibitem[{{Cohen} {et~al.}(1997){Cohen}, {Cassinelli}, \&
  {MacFarlane}}]{Cohen97}
{Cohen}, D.~H., {Cassinelli}, J.~P., \& {MacFarlane}, J.~J. 1997, \apj, 487,
  867

\bibitem[{{Cohen} {et~al.}(1996){Cohen}, {Cooper}, {Macfarlane}, {Owocki},
  {Cassinelli}, \& {Wang}}]{Cohen96}
{Cohen}, D.~H., {Cooper}, R.~G., {Macfarlane}, J.~J., {et~al.} 1996, \apj, 460,
  506

\bibitem[{{Cohen} {et~al.}(2011){Cohen}, {Gagn{\'e}}, {Leutenegger},
  {MacArthur}, {Wollman}, {Sundqvist}, {Fullerton}, \& {Owocki}}]{Cohen11}
{Cohen}, D.~H., {Gagn{\'e}}, M., {Leutenegger}, M.~A., {et~al.} 2011, \mnras,
  415, 3354

\bibitem[{{Cohen} {et~al.}(2008){Cohen}, {Kuhn}, {Gagn{\'e}}, {Jensen}, \&
  {Miller}}]{Cohen08a}
{Cohen}, D.~H., {Kuhn}, M.~A., {Gagn{\'e}}, M., {Jensen}, E.~L.~N., \&
  {Miller}, N.~A. 2008, \mnras, 386, 1855

\bibitem[{{Cohen} {et~al.}(2010){Cohen}, {Leutenegger}, {Wollman},
  {Zsarg{\'o}}, {Hillier}, {Townsend}, \& {Owocki}}]{Cohen10}
{Cohen}, D.~H., {Leutenegger}, M.~A., {Wollman}, E.~E., {et~al.} 2010, \mnras,
  405, 2391

\bibitem[{{Cohen} {et~al.}(2014{\natexlab{a}}){Cohen}, {Li}, {Gayley},
  {Owocki}, {Sundqvist}, {Petit}, \& {Leutenegger}}]{CohenLi14}
{Cohen}, D.~H., {Li}, Z., {Gayley}, K.~G., {et~al.} 2014{\natexlab{a}}, \mnras,
  444, 3729

\bibitem[{{Cohen} {et~al.}(2014{\natexlab{b}}){Cohen}, {Wollman},
  {Leutenegger}, {Sundqvist}, {Fullerton}, {Zsarg{\'o}}, \& {Owocki}}]{Cohen14}
{Cohen}, D.~H., {Wollman}, E.~E., {Leutenegger}, M.~A., {et~al.}
  2014{\natexlab{b}}, \mnras, 439, 908

\bibitem[{{Crowther} {et~al.}(2002){Crowther}, {Hillier}, {Evans}, {Fullerton},
  {De Marco}, \& {Willis}}]{Crowther02}
{Crowther}, P.~A., {Hillier}, D.~J., {Evans}, C.~J., {et~al.} 2002, \apj, 579,
  774

\bibitem[{{Daltabuit} \& {Cox}(1972)}]{Dalta72}
{Daltabuit}, E. \& {Cox}, D. 1972, \apj, 177, 855

\bibitem[{{Dessart} \& {Owocki}(2003)}]{DO03}
{Dessart}, L. \& {Owocki}, S.~P. 2003, \aap, 406, L1

\bibitem[{{Evans} {et~al.}(2008){Evans}, {Hunter}, {Smartt}, {Lennon}, {de
  Koter}, {Mokiem}, {Trundle}, {Dufton}, {Ryans}, {Puls}, {Vink}, {Herrero},
  {Sim{\'o}n-D{\'{\i}}az}, {Langer}, \& {Brott}}]{evans08}
{Evans}, C., {Hunter}, I., {Smartt}, S., {et~al.} 2008, The Messenger, 131, 25

\bibitem[{{Evans} {et~al.}(2011){Evans}, {Taylor}, {H{\'e}nault-Brunet},
  {Sana}, {de Koter}, {Sim{\'o}n-D{\'{\i}}az}, {Carraro}, {Bagnoli}, {Bastian},
  {Bestenlehner}, {Bonanos}, {Bressert}, {Brott}, {Campbell}, {Cantiello},
  {Clark}, {Costa}, {Crowther}, {de Mink}, {Doran}, {Dufton}, {Dunstall},
  {Friedrich}, {Garcia}, {Gieles}, {Gr{\"a}fener}, {Herrero}, {Howarth},
  {Izzard}, {Langer}, {Lennon}, {Ma{\'{\i}}z Apell{\'a}niz}, {Markova},
  {Najarro}, {Puls}, {Ramirez}, {Sab{\'{\i}}n-Sanjuli{\'a}n}, {Smartt},
  {Stroud}, {van Loon}, {Vink}, \& {Walborn}}]{evans11}
{Evans}, C.~J., {Taylor}, W.~D., {H{\'e}nault-Brunet}, V., {et~al.} 2011, \aap,
  530, A108

\bibitem[{{Feldmeier}(1995)}]{Feldmeier95}
{Feldmeier}, A. 1995, \aap, 299, 523

\bibitem[{{Feldmeier} {et~al.}(1997{\natexlab{a}}){Feldmeier}, {Kudritzki},
  {Palsa}, {Pauldrach}, \& {Puls}}]{Feldmeier97b}
{Feldmeier}, A., {Kudritzki}, R.-P., {Palsa}, R., {Pauldrach}, A.~W.~A., \&
  {Puls}, J. 1997{\natexlab{a}}, \aap, 320, 899

\bibitem[{{Feldmeier} {et~al.}(1997{\natexlab{b}}){Feldmeier}, {Puls}, \&
  {Pauldrach}}]{Feldmeier97a}
{Feldmeier}, A., {Puls}, J., \& {Pauldrach}, A.~W.~A. 1997{\natexlab{b}}, \aap,
  322, 878

\bibitem[{{Fullerton} {et~al.}(2006){Fullerton}, {Massa}, \&
  {Prinja}}]{fullerton06}
{Fullerton}, A.~W., {Massa}, D.~L., \& {Prinja}, R.~K. 2006, \apj, 637, 1025

\bibitem[{{Gabler} {et~al.}(1989){Gabler}, {Gabler}, {Kudritzki}, {Puls}, \&
  {Pauldrach}}]{Gabler89}
{Gabler}, R., {Gabler}, A., {Kudritzki}, R.~P., {Puls}, J., \& {Pauldrach}, A.
  1989, \aap, 226, 162

\bibitem[{{Garcia}(2005)}]{Garcia05}
{Garcia}, M. 2005, PhD thesis, University of La Laguna (Teneriffe)

\bibitem[{{Giddings}(1981)}]{Giddings81}
{Giddings}, J.~R. 1981, PhD thesis, , University of London, (1981)

\bibitem[{{Gr{\"a}fener} {et~al.}(2002){Gr{\"a}fener}, {Koesterke}, \&
  {Hamann}}]{Graf02}
{Gr{\"a}fener}, G., {Koesterke}, L., \& {Hamann}, W.-R. 2002, \aap, 387, 244

\bibitem[{{Groenewegen} \& {Lamers}(1989)}]{GL89}
{Groenewegen}, M.~A.~T. \& {Lamers}, H.~J.~G.~L.~M. 1989, \aaps, 79, 359

\bibitem[{{Hamann} \& {Oskinova}(2012)}]{HamannOskinova12}
{Hamann}, W.-R. \& {Oskinova}, L. 2012, in COSPAR Meeting, Vol.~39, 39th COSPAR
  Scientific Assembly, 716

\bibitem[{{Harnden} {et~al.}(1979){Harnden}, {Branduardi}, {Gorenstein},
  {Grindlay}, {Rosner}, {Topka}, {Elvis}, {Pye}, \& {Vaiana}}]{harnden79}
{Harnden}, Jr., F.~R., {Branduardi}, G., {Gorenstein}, P., {et~al.} 1979,
  \apjl, 234, L51

\bibitem[{{Haser}(1995)}]{Haser95}
{Haser}, S.~M. 1995, PhD thesis, Ludwig-Maximilians-Universit{\"a}t M{\"u}nchen

\bibitem[{{Hauschildt}(1992)}]{haus92}
{Hauschildt}, P.~H. 1992, Journal of Quantitative Spectroscopy and Radiative
  Transfer, 47, 433

\bibitem[{{Herv{\'e}} {et~al.}(2013){Herv{\'e}}, {Rauw}, \&
  {Naz{\'e}}}]{Herve13}
{Herv{\'e}}, A., {Rauw}, G., \& {Naz{\'e}}, Y. 2013, \aap, 551, A83

\bibitem[{{Hillier} {et~al.}(1993){Hillier}, {Kudritzki}, {Pauldrach}, {Baade},
  {Cassinelli}, {Puls}, \& {Schmitt}}]{Hillier93}
{Hillier}, D.~J., {Kudritzki}, R.~P., {Pauldrach}, A.~W., {et~al.} 1993, \aap,
  276, 117

\bibitem[{{Hillier} \& {Miller}(1998)}]{hilliermiller98}
{Hillier}, D.~J. \& {Miller}, D.~L. 1998, \apj, 496, 407

\bibitem[{{Hubeny}(1998)}]{hubeny98}
{Hubeny}, I. 1998, in Astronomical Society of the Pacific Conference Series,
  Vol. 138, 1997 Pacific Rim Conference on Stellar Astrophysics, ed. K.~L.
  {Chan}, K.~S. {Cheng}, \& H.~P. {Singh}, 139

\bibitem[{{Huenemoerder} {et~al.}(2012){Huenemoerder}, {Oskinova}, {Ignace},
  {Waldron}, {Todt}, {Hamaguchi}, \& {Kitamoto}}]{Huenemoerder12}
{Huenemoerder}, D.~P., {Oskinova}, L.~M., {Ignace}, R., {et~al.} 2012, \apjl,
  756, L34

\bibitem[{{Hunter} {et~al.}(2008){Hunter}, {Brott}, {Lennon}, {Langer},
  {Dufton}, {Trundle}, {Smartt}, {de Koter}, {Evans}, \& {Ryans}}]{hunter08}
{Hunter}, I., {Brott}, I., {Lennon}, D.~J., {et~al.} 2008, \apjl, 676, L29

\bibitem[{{Hunter} {et~al.}(2007){Hunter}, {Dufton}, {Smartt}, {Ryans},
  {Evans}, {Lennon}, {Trundle}, {Hubeny}, \& {Lanz}}]{hunter07}
{Hunter}, I., {Dufton}, P.~L., {Smartt}, S.~J., {et~al.} 2007, \aap, 466, 277

\bibitem[{{Kaastra} \& {Mewe}(1993)}]{Kaastra93}
{Kaastra}, J. \& {Mewe}, R. 1993, \aa, 97, 443

\bibitem[{{Krolik} \& {Raymond}(1985)}]{krolray85}
{Krolik}, J. \& {Raymond}, J. 1985, \apj, 298, 660

\bibitem[{{Krti{\v c}ka} \& {Kub{\'a}t}(2001)}]{KK01}
{Krti{\v c}ka}, J. \& {Kub{\'a}t}, J. 2001, \aap, 369, 222

\bibitem[{{Krti{\v c}ka} \& {Kub{\'a}t}(2009)}]{krticka09}
{Krti{\v c}ka}, J. \& {Kub{\'a}t}, J. 2009, \mnras, 394, 2065

\bibitem[{{Krti{\v c}ka} \& {Kub{\'a}t}(2012)}]{krticka12}
{Krti{\v c}ka}, J. \& {Kub{\'a}t}, J. 2012, \mnras, 427, 84

\bibitem[{{Kub{\'a}t} {et~al.}(1999){Kub{\'a}t}, {Puls}, \&
  {Pauldrach}}]{Kubat99}
{Kub{\'a}t}, J., {Puls}, J., \& {Pauldrach}, A.~W.~A. 1999, \aap, 341, 587

\bibitem[{{Kudritzki} \& {Puls}(2000)}]{KP00}
{Kudritzki}, R.-P. \& {Puls}, J. 2000, \araa, 38, 613

\bibitem[{{Lamers} \& {Rogerson}(1978)}]{Lamers78}
{Lamers}, H. \& {Rogerson}, J. 1978, \aa, 66, 417

\bibitem[{{Lamers} {et~al.}(1987){Lamers}, {Cerruti-Sola}, \&
  {Perinotto}}]{Lamersetal87}
{Lamers}, H.~J.~G.~L.~M., {Cerruti-Sola}, M., \& {Perinotto}, M. 1987, \apj,
  314, 726

\bibitem[{{Lamers} \& {Morton}(1976)}]{LamersMorton76}
{Lamers}, H.~J.~G.~L.~M. \& {Morton}, D.~C. 1976, \apjs, 32, 715

\bibitem[{{Leutenegger} {et~al.}(2013{\natexlab{a}}){Leutenegger}, {Cohen},
  {Neely}, {Owocki}, \& {Sundqvist}}]{leutenegger13a}
{Leutenegger}, M.~A., {Cohen}, D., {Neely}, J., {Owocki}, S., \& {Sundqvist},
  J. 2013{\natexlab{a}}, in Massive Stars: From alpha to Omega, 42

\bibitem[{{Leutenegger} {et~al.}(2013{\natexlab{b}}){Leutenegger}, {Cohen},
  {Sundqvist}, \& {Owocki}}]{leutenegger13}
{Leutenegger}, M.~A., {Cohen}, D.~H., {Sundqvist}, J.~O., \& {Owocki}, S.~P.
  2013{\natexlab{b}}, \apj, 770, 80

\bibitem[{{Leutenegger} {et~al.}(2006){Leutenegger}, {Paerels}, {Kahn}, \&
  {Cohen}}]{Leutenegger06}
{Leutenegger}, M.~A., {Paerels}, F.~B.~S., {Kahn}, S.~M., \& {Cohen}, D.~H.
  2006, \apj, 650, 1096

\bibitem[{{Lucy}(1982)}]{Lucy82}
{Lucy}, L.~B. 1982, \apj, 255, 278

\bibitem[{{Lucy} \& {Solomon}(1970)}]{LS70}
{Lucy}, L.~B. \& {Solomon}, P.~M. 1970, \apj, 159, 879

\bibitem[{{Macfarlane} {et~al.}(1994){Macfarlane}, {Cohen}, \&
  {Wang}}]{Macfarlane94}
{Macfarlane}, J.~J., {Cohen}, D.~H., \& {Wang}, P. 1994, \aa, 437, 351

\bibitem[{{Macfarlane} {et~al.}(1993){Macfarlane}, {Waldron}, {Corcoran},
  {Wolff}, {Wang}, \& {Cassinelli}}]{Macfarlane93}
{Macfarlane}, J.~J., {Waldron}, W.~L., {Corcoran}, M.~F., {et~al.} 1993, \apj,
  419, 813

\bibitem[{{Martins} {et~al.}(2012){Martins}, {Escolano}, {Wade}, {Donati},
  {Bouret}, \& {Mimes Collaboration}}]{martins12}
{Martins}, F., {Escolano}, C., {Wade}, G.~A., {et~al.} 2012, \aap, 538, A29

\bibitem[{{Martins} {et~al.}(2015{\natexlab{a}}){Martins}, {Herv{\'e}},
  {Bouret}, {Marcolino}, {Wade}, {Neiner}, {Alecian}, {Grunhut}, \&
  {Petit}}]{martins15a}
{Martins}, F., {Herv{\'e}}, A., {Bouret}, J.-C., {et~al.} 2015{\natexlab{a}},
  \aap, 575, A34

\bibitem[{{Martins} \& {Hillier}(2012)}]{MartinsHillier12}
{Martins}, F. \& {Hillier}, D.~J. 2012, \aap, 545, A95

\bibitem[{{Martins} {et~al.}(2015{\natexlab{b}}){Martins},
  {Sim{\'o}n-D{\'{\i}}az}, {Palacios}, {Howarth}, {Georgy}, {Walborn},
  {Bouret}, \& {Barb{\'a}}}]{martins15b}
{Martins}, F., {Sim{\'o}n-D{\'{\i}}az}, S., {Palacios}, A., {et~al.}
  2015{\natexlab{b}}, \aap, 578, A109

\bibitem[{{Mihalas} \& {Hummer}(1973)}]{mihalas73}
{Mihalas}, D. \& {Hummer}, D.~G. 1973, \apj, 179, 827

\bibitem[{{Naz{\'e}} {et~al.}(2011){Naz{\'e}}, {Broos}, {Oskinova}, {Townsley},
  {Cohen}, {Corcoran}, {Evans}, {Gagn{\'e}}, {Moffat}, {Pittard}, {Rauw},
  {ud-Doula}, \& {Walborn}}]{naze11}
{Naz{\'e}}, Y., {Broos}, P.~S., {Oskinova}, L., {et~al.} 2011, \apjs, 194, 7

\bibitem[{{Naz{\'e}} {et~al.}(2012){Naz{\'e}}, {Flores}, \& {Rauw}}]{Naze12}
{Naz{\'e}}, Y., {Flores}, C.~A., \& {Rauw}, G. 2012, \aap, 538, A22

\bibitem[{{Naz{\'e}} {et~al.}(2013){Naz{\'e}}, {Oskinova}, \&
  {Gosset}}]{naze13}
{Naz{\'e}}, Y., {Oskinova}, L.~M., \& {Gosset}, E. 2013, \apj, 763, 143

\bibitem[{{Olson} \& {Castor}(1981)}]{Olcast81}
{Olson}, G. \& {Castor}, J. 1981, \apj, 244, 179

\bibitem[{{Oskinova} {et~al.}(2006){Oskinova}, {Feldmeier}, \&
  {Hamann}}]{Oskinova06}
{Oskinova}, L.~M., {Feldmeier}, A., \& {Hamann}, W.-R. 2006, \mnras, 372, 313

\bibitem[{{Oskinova} {et~al.}(2007){Oskinova}, {Hamann}, \&
  {Feldmeier}}]{oskinova07}
{Oskinova}, L.~M., {Hamann}, W.-R., \& {Feldmeier}, A. 2007, \aap, 476, 1331

\bibitem[{{Owocki}(1994)}]{Owocki94b}
{Owocki}, S.~P. 1994, \apss, 221, 3

\bibitem[{{Owocki} {et~al.}(1988){Owocki}, {Castor}, \& {Rybicki}}]{OCR88}
{Owocki}, S.~P., {Castor}, J.~I., \& {Rybicki}, G.~B. 1988, \apj, 335, 914

\bibitem[{{Owocki} \& {Cohen}(1999)}]{OwockiCohen99}
{Owocki}, S.~P. \& {Cohen}, D.~H. 1999, \apj, 520, 833

\bibitem[{{Owocki} \& {Cohen}(2006)}]{OwockiCohen06}
{Owocki}, S.~P. \& {Cohen}, D.~H. 2006, \apj, 648, 565

\bibitem[{{Owocki} \& {Rybicki}(1984)}]{ORI}
{Owocki}, S.~P. \& {Rybicki}, G.~B. 1984, \apj, 284, 337

\bibitem[{{Owocki} {et~al.}(2013){Owocki}, {Sundqvist}, {Cohen}, \&
  {Gayley}}]{OwockiSundqvist13}
{Owocki}, S.~P., {Sundqvist}, J.~O., {Cohen}, D.~H., \& {Gayley}, K.~G. 2013,
  \mnras, 429, 3379

\bibitem[{{Pauldrach} {et~al.}(2001){Pauldrach}, {Hoffmann}, \&
  {Lennon}}]{pauldrach01}
{Pauldrach}, A.~W.~A., {Hoffmann}, T.~L., \& {Lennon}, M. 2001, \aap, 375, 161

\bibitem[{{Pauldrach} {et~al.}(1990){Pauldrach}, {Kudritzki}, {Puls}, \&
  {Butler}}]{Pauldrach90}
{Pauldrach}, A.~W.~A., {Kudritzki}, R.~P., {Puls}, J., \& {Butler}, K. 1990,
  \aap, 228, 125

\bibitem[{{Pauldrach} {et~al.}(1994){Pauldrach}, {Kudritzki}, {Puls}, {Butler},
  \& {Hunsinger}}]{pauldrach94c}
{Pauldrach}, A.~W.~A., {Kudritzki}, R.~P., {Puls}, J., {Butler}, K., \&
  {Hunsinger}, J. 1994, \aap, 283, 525

\bibitem[{{Przybilla} {et~al.}(2010){Przybilla}, {Firnstein}, {Nieva},
  {Meynet}, \& {Maeder}}]{przybilla10}
{Przybilla}, N., {Firnstein}, M., {Nieva}, M.~F., {Meynet}, G., \& {Maeder}, A.
  2010, \aap, 517, A38

\bibitem[{{Puls}(2009)}]{puls09}
{Puls}, J. 2009, Communications in Asteroseismology, 158, 113

\bibitem[{{Puls} {et~al.}(1996){Puls}, {Kudritzki}, {Herrero}, {Pauldrach},
  {Haser}, {Lennon}, {Gabler}, {Voels}, {Vilchez}, {Wachter}, \&
  {Feldmeier}}]{Puls96}
{Puls}, J., {Kudritzki}, R.-P., {Herrero}, A., {et~al.} 1996, \aap, 305, 171

\bibitem[{{Puls} {et~al.}(1993){Puls}, {Owocki}, \& {Fullerton}}]{POF93}
{Puls}, J., {Owocki}, S.~P., \& {Fullerton}, A.~W. 1993, \aap, 279, 457

\bibitem[{{Puls} {et~al.}(2005){Puls}, {Urbaneja}, {Venero}, {Repolust},
  {Springmann}, {Jokuthy}, \& {Mokiem}}]{Puls05}
{Puls}, J., {Urbaneja}, M.~A., {Venero}, R., {et~al.} 2005, \aap, 435, 669

\bibitem[{{Rauw} {et~al.}(2015){Rauw}, {Herv{\'e}}, {Naz{\'e}},
  {Gonz{\'a}lez-P{\'e}rez}, {Hempelmann}, {Mittag}, {Schmitt}, {Schr{\"o}der},
  {Gosset}, {Eenens}, \& {Uuh-Sonda}}]{rauw15}
{Rauw}, G., {Herv{\'e}}, A., {Naz{\'e}}, Y., {et~al.} 2015, \aap, 580, A59

\bibitem[{{Raymond} \& {Smith}(1977)}]{Raysmith77}
{Raymond}, J. \& {Smith}, B. 1977, \apjs, 35, 419

\bibitem[{{Rivero Gonz{\'a}lez} {et~al.}(2011){Rivero Gonz{\'a}lez}, {Puls}, \&
  {Najarro}}]{rivero11}
{Rivero Gonz{\'a}lez}, J.~G., {Puls}, J., \& {Najarro}, F. 2011, \aap, 536, A58

\bibitem[{{Rivero Gonz{\'a}lez} {et~al.}(2012{\natexlab{a}}){Rivero
  Gonz{\'a}lez}, {Puls}, {Najarro}, \& {Brott}}]{rivero12}
{Rivero Gonz{\'a}lez}, J.~G., {Puls}, J., {Najarro}, F., \& {Brott}, I.
  2012{\natexlab{a}}, \aap, 537, A79

\bibitem[{{Rivero Gonz{\'a}lez} {et~al.}(2012{\natexlab{b}}){Rivero
  Gonz{\'a}lez}, {Puls}, {Najarro}, \& {Massey}}]{rivero122}
{Rivero Gonz{\'a}lez}, J.~G., {Puls}, J., {Najarro}, F., \& {Massey}, P.
  2012{\natexlab{b}}, \aap, 543, A95

\bibitem[{{Sana} {et~al.}(2006){Sana}, {Rauw}, {Naze}, {Gosset}, \&
  {Vreux}}]{Sana06}
{Sana}, H., {Rauw}, G., {Naze}, Y., {Gosset}, E., \& {Vreux}, J.-M. 2006,
  MNRAS, 372, 661

\bibitem[{{Seaton}(1958)}]{seaton58}
{Seaton}, M.~J. 1958, \mnras, 118, 504

\bibitem[{{Seward} {et~al.}(1979){Seward}, {Forman}, {Giacconi}, {Griffiths},
  {Harnden}, {Jones}, \& {Pye}}]{seward79}
{Seward}, F.~D., {Forman}, W.~R., {Giacconi}, R., {et~al.} 1979, \apjl, 234,
  L55

\bibitem[{{Simon} \& {Axford}(1966)}]{simox66}
{Simon}, M. \& {Axford}, W. 1966, Planet.Space Sci., 14, 901

\bibitem[{{Smith} {et~al.}(2001){Smith}, {Brickhouse}, {Liedahl}, \&
  {Raymond}}]{Smith01}
{Smith}, R.~K., {Brickhouse}, N.~S., {Liedahl}, D.~A., \& {Raymond}, J.~C.
  2001, \apjl, 556, L91

\bibitem[{{Snow} \& {Morton}(1976)}]{Snow76}
{Snow}, T. \& {Morton}, D. 1976, \apjs, 32, 429

\bibitem[{{Sundqvist} \& {Owocki}(2013)}]{Sundqvist13}
{Sundqvist}, J.~O. \& {Owocki}, S.~P. 2013, \mnras, 428, 1837

\bibitem[{{Sundqvist} {et~al.}(2012{\natexlab{a}}){Sundqvist}, {Owocki},
  {Cohen}, {Leutenegger}, \& {Townsend}}]{Sundqvist12b}
{Sundqvist}, J.~O., {Owocki}, S.~P., {Cohen}, D.~H., {Leutenegger}, M.~A., \&
  {Townsend}, R.~H.~D. 2012{\natexlab{a}}, \mnras, 420, 1553

\bibitem[{{Sundqvist} {et~al.}(2012{\natexlab{b}}){Sundqvist}, {Owocki}, \&
  {Puls}}]{Sundqvist12a}
{Sundqvist}, J.~O., {Owocki}, S.~P., \& {Puls}, J. 2012{\natexlab{b}}, in
  Astronomical Society of the Pacific Conference Series, Vol. 465, Proceedings
  of a Scientific Meeting in Honor of Anthony F. J. Moffat, ed. L.~{Drissen},
  C.~{Robert}, N.~{St-Louis}, \& A.~F.~J. {Moffat}, 119

\bibitem[{{Sundqvist} {et~al.}(2011){Sundqvist}, {Puls}, {Feldmeier}, \&
  {Owocki}}]{Sundqvist11}
{Sundqvist}, J.~O., {Puls}, J., {Feldmeier}, A., \& {Owocki}, S.~P. 2011, \aap,
  528, A64

\bibitem[{{Sundqvist} {et~al.}(2014){Sundqvist}, {Puls}, \&
  {Owocki}}]{Sundqvist14}
{Sundqvist}, J.~O., {Puls}, J., \& {Owocki}, S.~P. 2014, \aap, 568, A59

\bibitem[{{Trundle} {et~al.}(2004){Trundle}, {Lennon}, {Puls}, \&
  {Dufton}}]{Trundle04}
{Trundle}, C., {Lennon}, D.~J., {Puls}, J., \& {Dufton}, P.~L. 2004, \aap, 417,
  217

\bibitem[{{{\v S}urlan} {et~al.}(2013){{\v S}urlan}, {Hamann}, {Aret},
  {Kub{\'a}t}, {Oskinova}, \& {Torres}}]{Surlan13}
{{\v S}urlan}, B., {Hamann}, W.-R., {Aret}, A., {et~al.} 2013, \aap, 559, A130

\bibitem[{{Verner} \& {Yakovlev}(1995)}]{Verner95}
{Verner}, D.~A. \& {Yakovlev}, D.~G. 1995, \aaps, 109, 125

\bibitem[{{Walborn} \& {Panek}(1984)}]{Walborn84}
{Walborn}, N.~R. \& {Panek}, R.~J. 1984, \apjl, 280, L27

\bibitem[{{Waldron} \& {Cas\-si\-nel\-li}(2007)}]{WC07}
{Waldron}, W.~L. \& {Cas\-si\-nel\-li}, J.~P. 2007, \apj, 668, 456

\bibitem[{{Waldron} \& {Cas\-si\-nel\-li}(2010)}]{WC10}
{Waldron}, W.~L. \& {Cas\-si\-nel\-li}, J.~P. 2010, \aa, 711, L30

\bibitem[{{Waldron} \& {Cassinelli}(2001)}]{WC01}
{Waldron}, W.~L. \& {Cassinelli}, J.~P. 2001, \apjl, 548, L45

\bibitem[{{Zhekov} \& {Palla}(2007)}]{ZhekovPalla07}
{Zhekov}, S.~A. \& {Palla}, F. 2007, \mnras, 382, 1124

\bibitem[{{Zsarg{\'o}} {et~al.}(2008){Zsarg{\'o}}, {Hillier}, {Bouret}, {Lanz},
  {Leutenegger}, \& {Cohen}}]{Zsargo08b}
{Zsarg{\'o}}, J., {Hillier}, D.~J., {Bouret}, J.-C., {et~al.} 2008, \apjl, 685,
  L149

\end{thebibliography}

\Online
\appendix
\section{Ionization fractions of selected ions: Dependence on X-ray filling 
factor and shock temperature}
\label{diffxraysdescription}

\begin{figure*}
\resizebox{\hsize}{!}
	 {\includegraphics[angle=90]{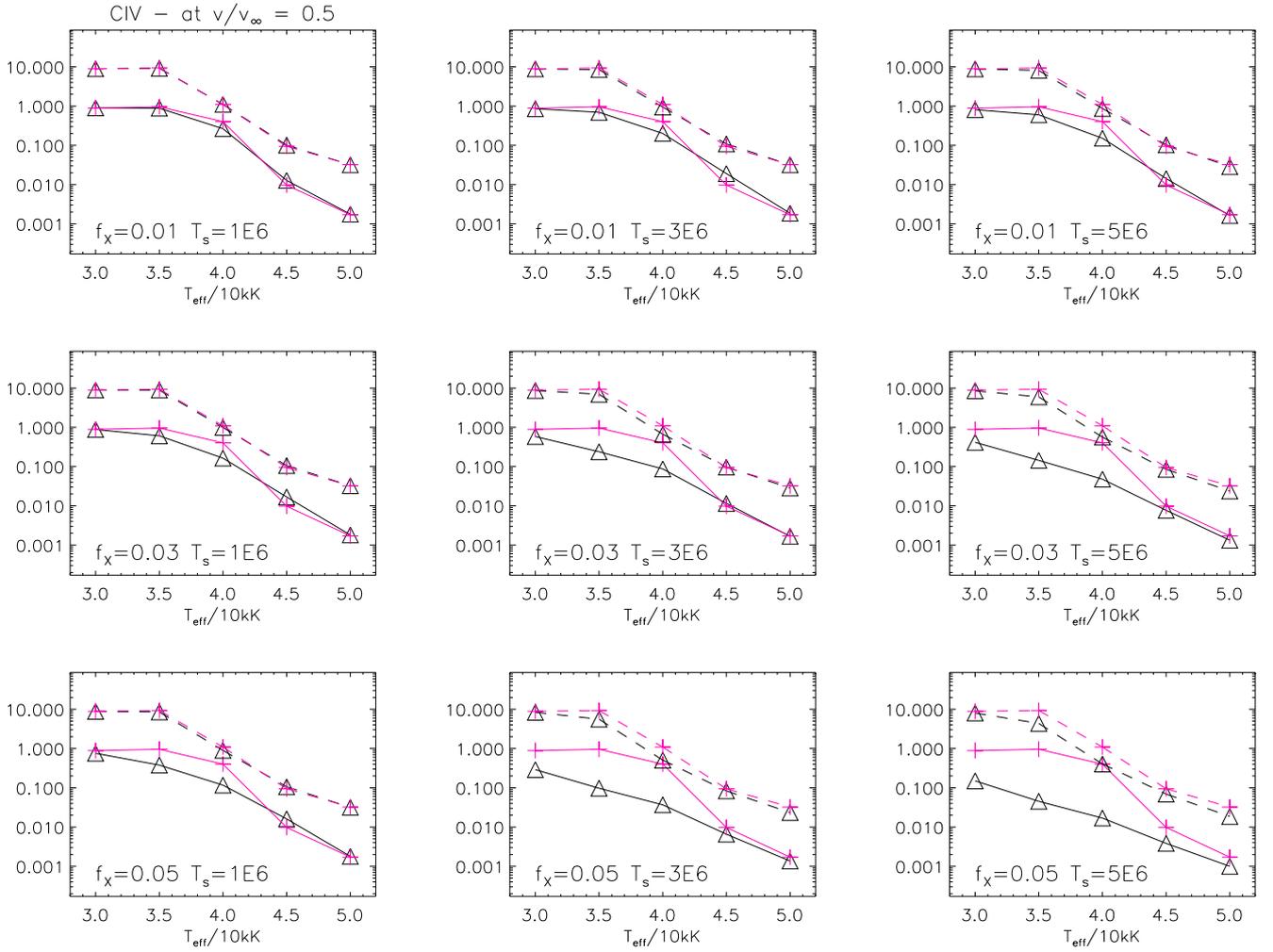}}
         \caption{Ionization fractions of \CIV\ (at $v(r) = 0.5
	 \vinf$), as a function of \Teff, and for different X-ray
	 emission parameters. Solid: supergiant models; dashed: dwarf
	 models; Black: models with shock emission; magenta: models
	 without shock emission. {\textit{For clarity, the ionization
	 fractions of dwarf models have been shifted by one dex}}.}
%        For low values of \Tshock\ or \fv\ the impact of
%	 shock radiation is barely notable. From intermediate to high
%	 X-ray parameters the depletion becomes clear for the
%	 supergiant models (the same is true for the \CIII\ parameters
%	 dependence).} 
\label{app_civ}
\end{figure*}

\begin{figure*}
\begin{minipage}{9cm}
\resizebox{\hsize}{!}
  {\includegraphics[angle=90]{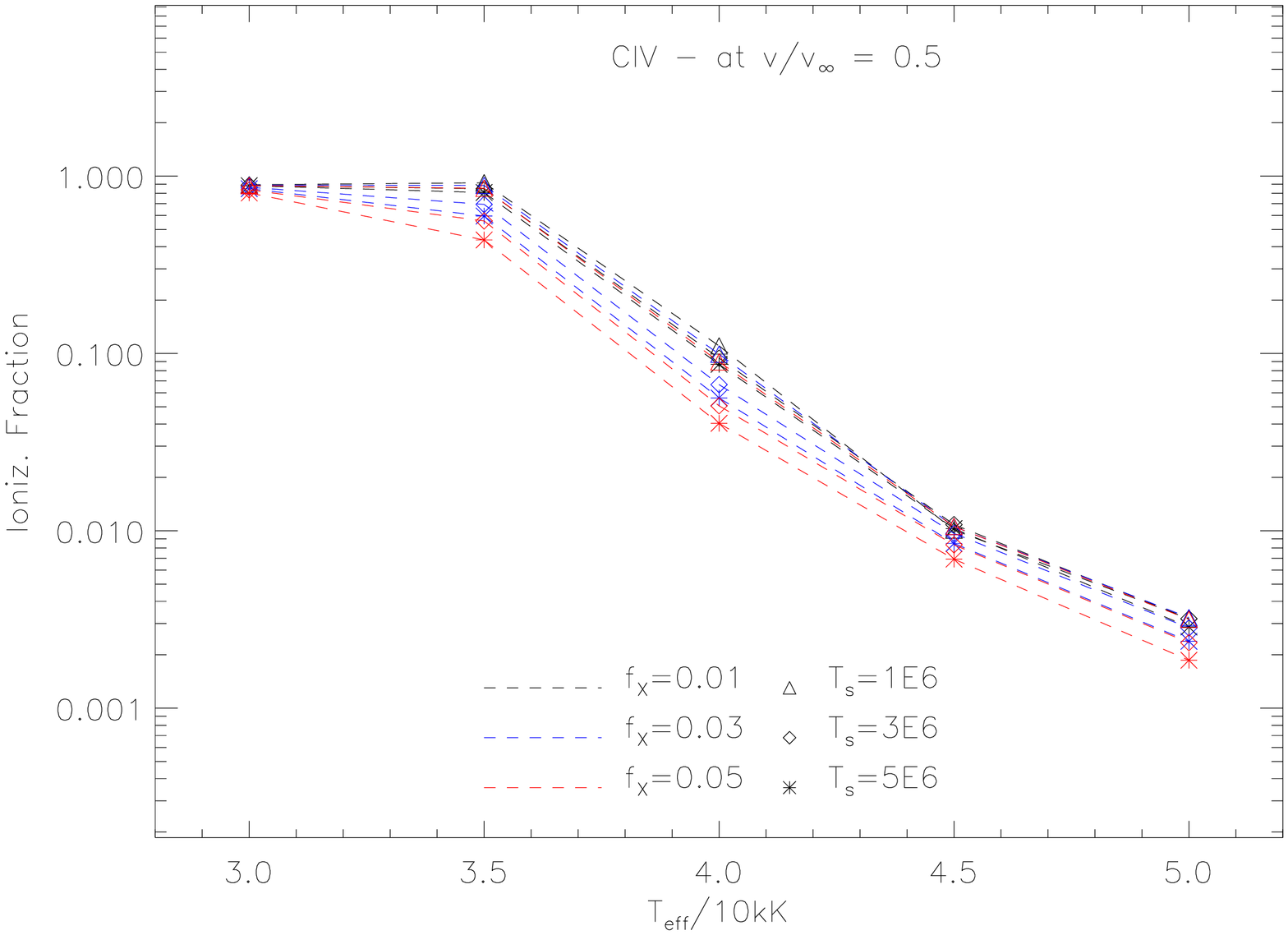}} 
\end{minipage}
\hspace{-.2cm}
\begin{minipage}{9cm}
\resizebox{\hsize}{!}
  {\includegraphics[angle=90]{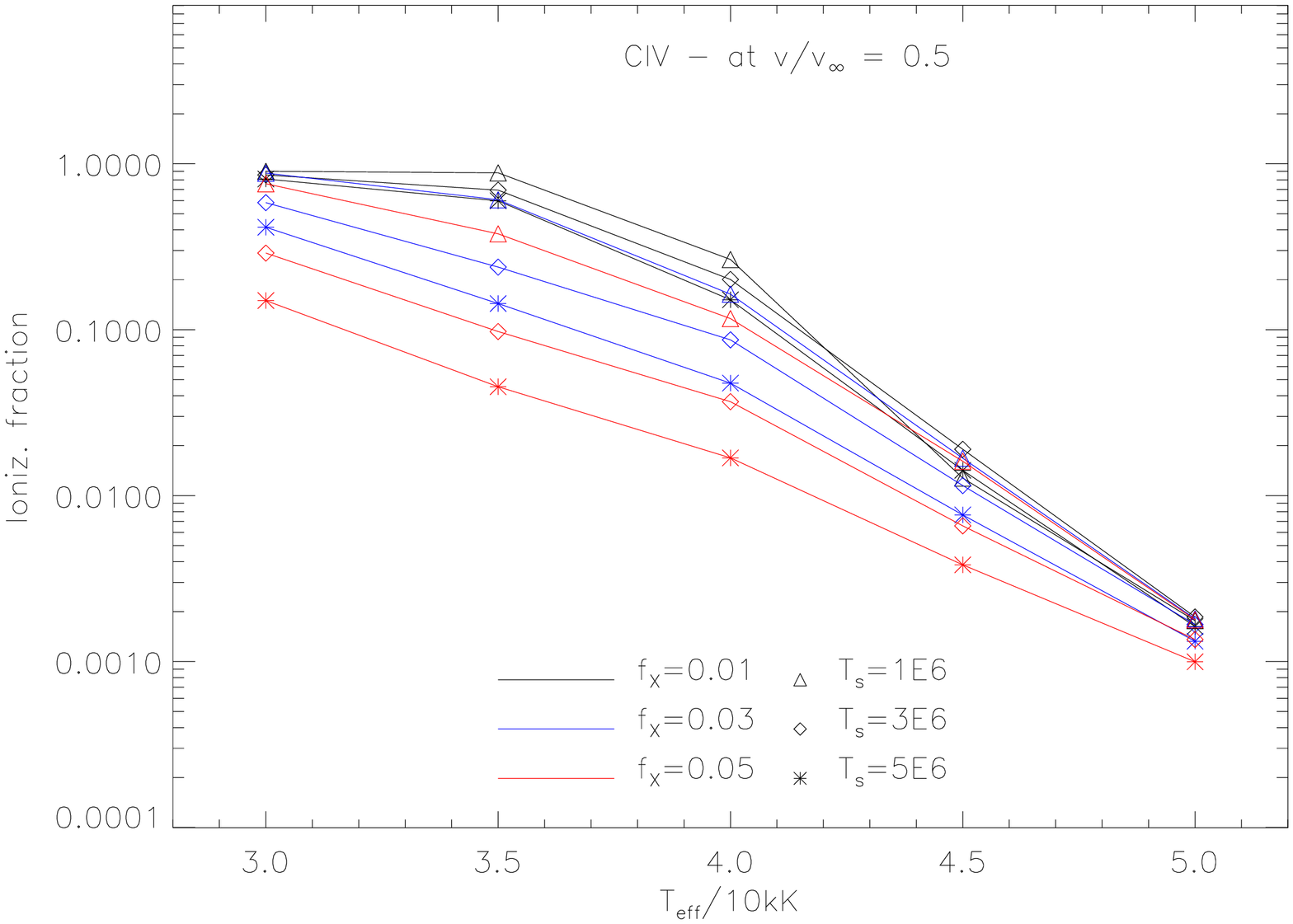}} 
\end{minipage}
\caption{{\textit{Left panel}} - As above (\CIV\ at $v(r) = 0.5 \vinf$),
	  but now for dwarf models alone and for all X-ray emission
	  parameters included in our grid. 
%	  values of \Teff\ and different X-ray description in dwarf
%	  models. Our results show in general a factor of 2 in the
%	  variance of ionization fraction for a fixed \Teff\, despite
%	  the coldest models. 
%         Therefore the shock radiation is not so influent.
          Note that the fractions have {\textit{not}} been shifted here.  
	  {\textit{Right panel}} - as left, but for supergiant models.}
%	  \Tshock\ influences the
%	  variation on occupation number for \CV\, however \fv\ plays
%	  the major role as can be seen in the distance between two
%	  curves with the same symbol. As written in
%	  Sect.~\ref{dependencefvtshock}, supergiant models with
%	  cooler than 45kK are the most affected. }
%
\label{civ_diff}
\end{figure*}

\begin{figure*}
\resizebox{\hsize}{!}
{\includegraphics[angle=90]{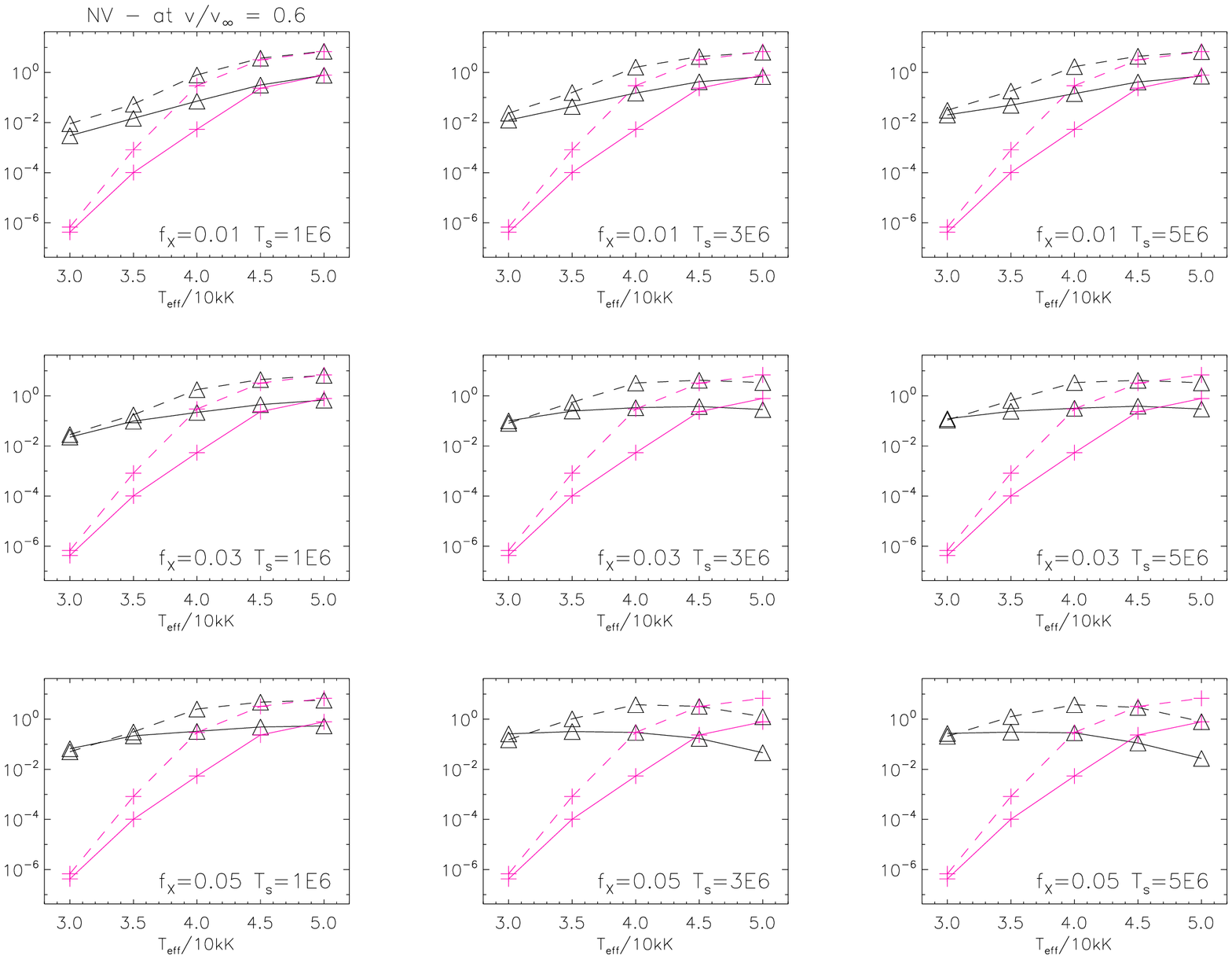}}
\caption{As Fig.~\ref{app_civ}, but for \NV\ at $v(r) = 0.6 \vinf$.}
%Ionization fraction of \NV\ for different values of temperature. The
%continuous lines are for supergiant models, while the dashed lines for
%dwarfs. {\it For clarity, the ionization fractions of dwarf models
%have been shifted by one dex}.  As pointed in section
%\ref{generaleffects}, the X-rays ionization is essential to describe
%\NV\ in cold models (both supergiants and dwarfs, except models with 
%\Teff\ = 45kK). The hottest models with moderate to high X-ray
%emission (\fv\ $\geqslant$ 0.02 and \Tshock\ $\geqslant 2E6$), 
%suggest that higher stages of ionization (\NVI) can also pass through
%changes due to shock radiation.}
%
\label{app_nv}
\end{figure*}

\begin{figure*}
\begin{minipage}{9cm}
\resizebox{\hsize}{!}
  {\includegraphics[angle=90]{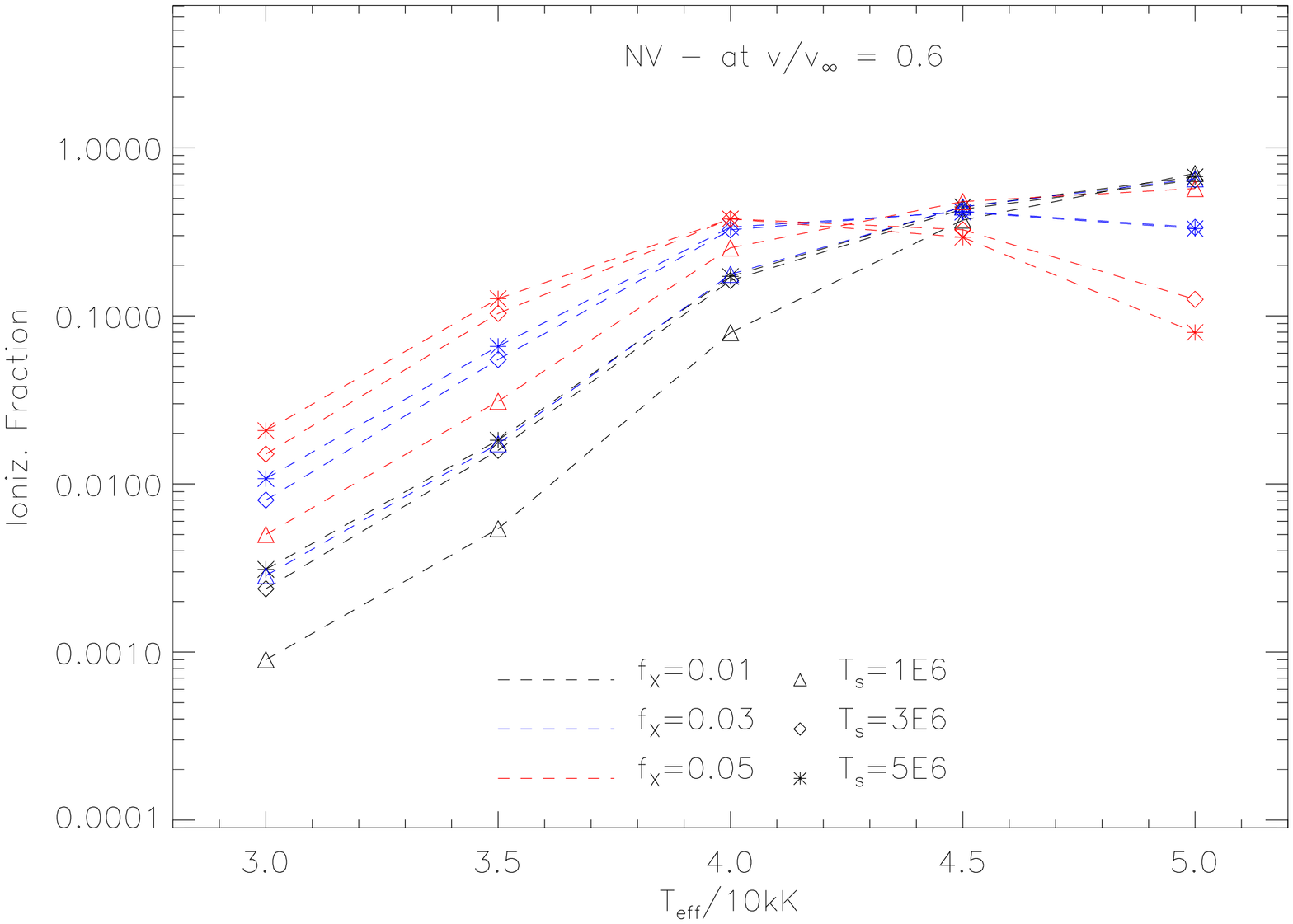}} 
\end{minipage}
\hspace{-.2cm}
\begin{minipage}{9cm}
\resizebox{\hsize}{!}
  {\includegraphics[angle=90]{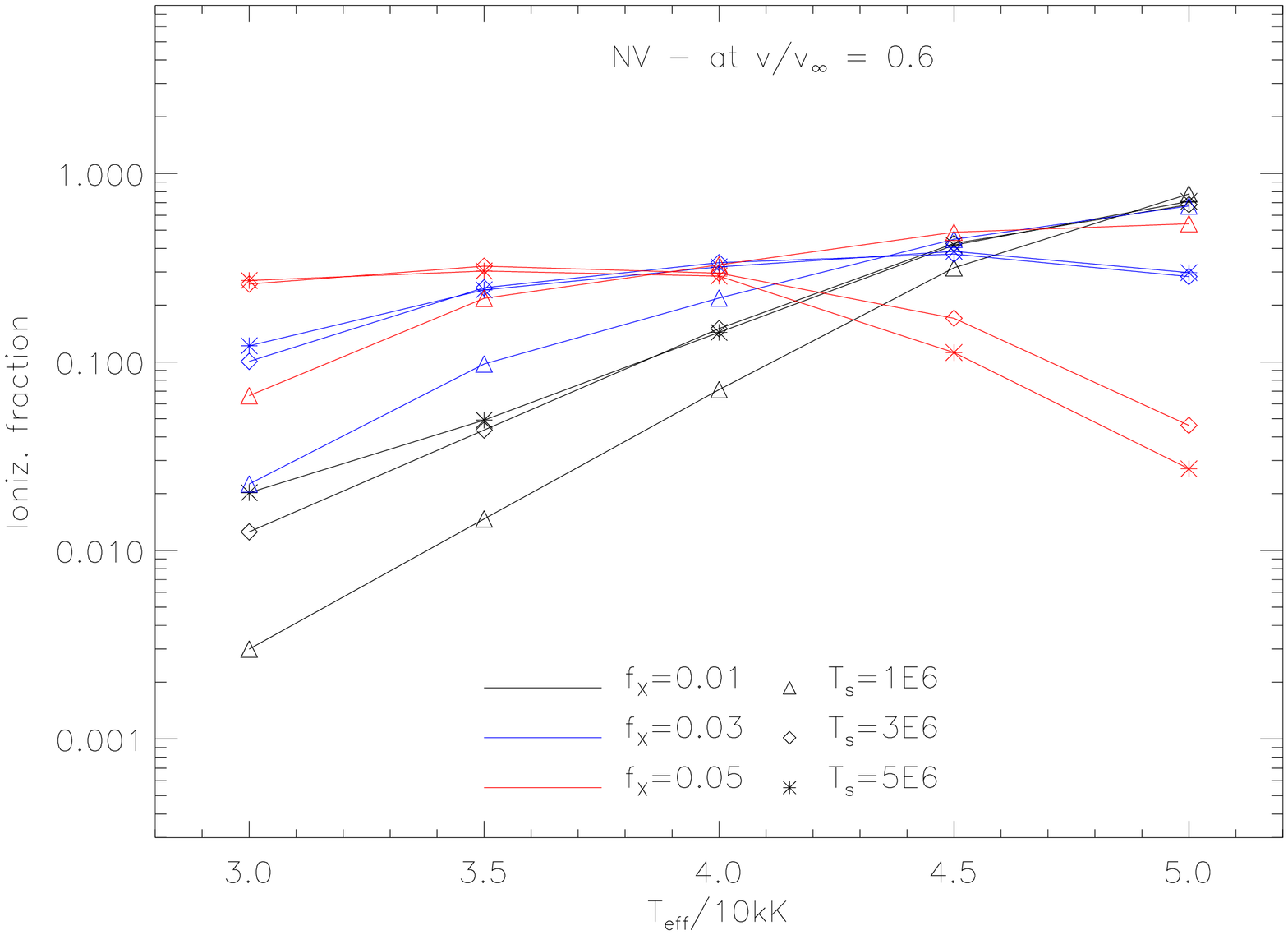}} 
\end{minipage}
\caption{As Fig.~\ref{civ_diff}, but for \NV\ ($v(r) = 0.6 \vinf$).}
\label{nv_diff}
\end{figure*}

\begin{figure*}
\resizebox{\hsize}{!}
{\includegraphics[angle=90]{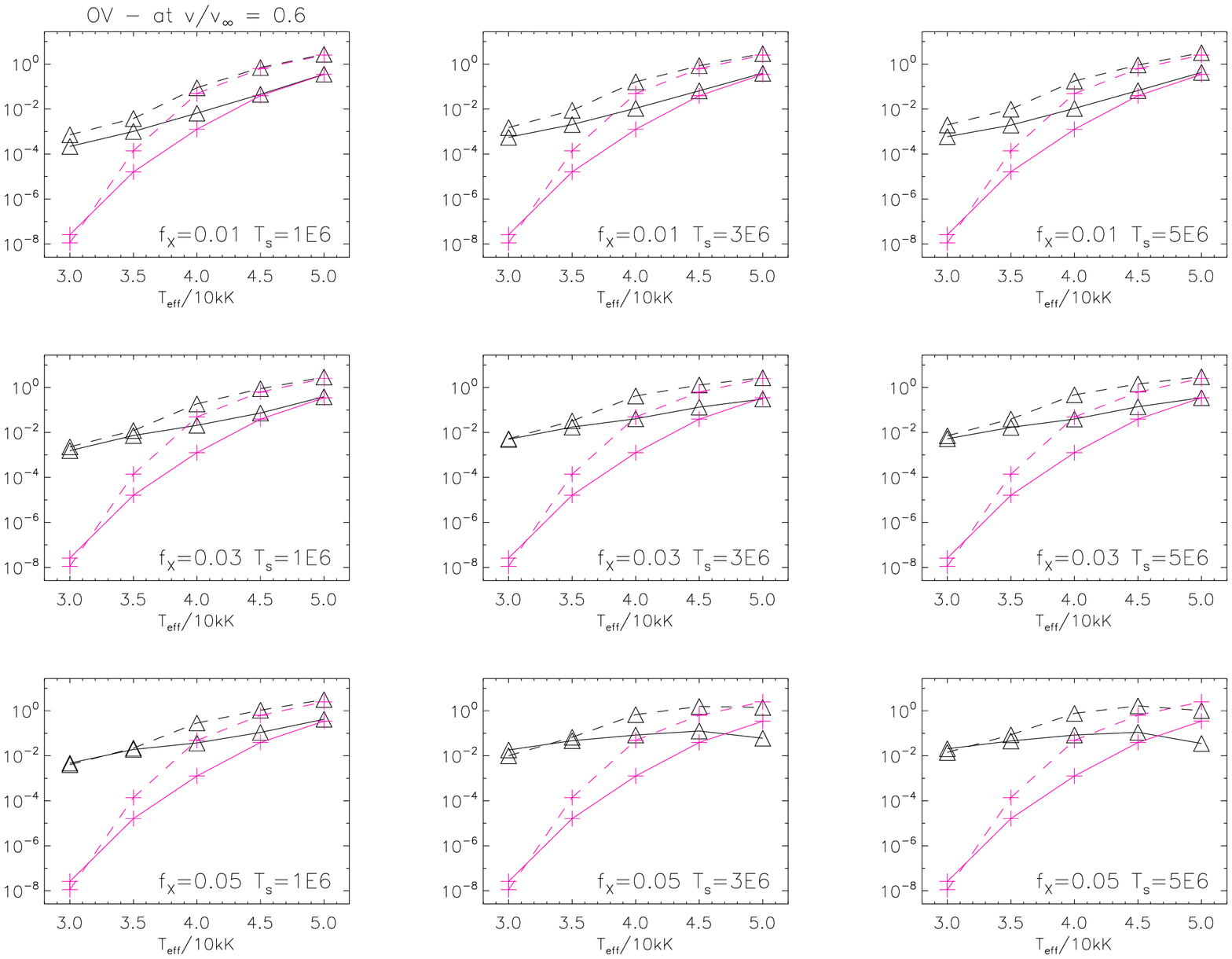}}
\caption{As Fig.~\ref{app_civ}, but for \OV\ at $v(r) = 0.6 \vinf$.}
%Ionization fraction of \OV\ for different values of temperature. The
%continuous lines are for supergiant models, while the dashed lines for
%dwarfs. {\it For clarity, the ionization fractions of dwarf models
%have been shifted by one dex}.  Shock radiation is essential for the
%description of \OV\ in cold models, even for the lowest values of \fv\
%or \Tshock. The increase of X-ray parameters changes drastically \OV\
%for cold models. When the maximum values of X-rays parameters are 
%set, a slight depletion of \OV\ is seen for the hot models. As in the
%case of \NVI, it indicates that the ionization is shifted to higher
%stages. As pointed in figure \ref{xrays43}, X-rays have a 
%considerable influence over \OVI. }
%
\label{app_ov}
\end{figure*}

\begin{figure*}
\begin{minipage}{9cm}
\resizebox{\hsize}{!}
  {\includegraphics[angle=90]{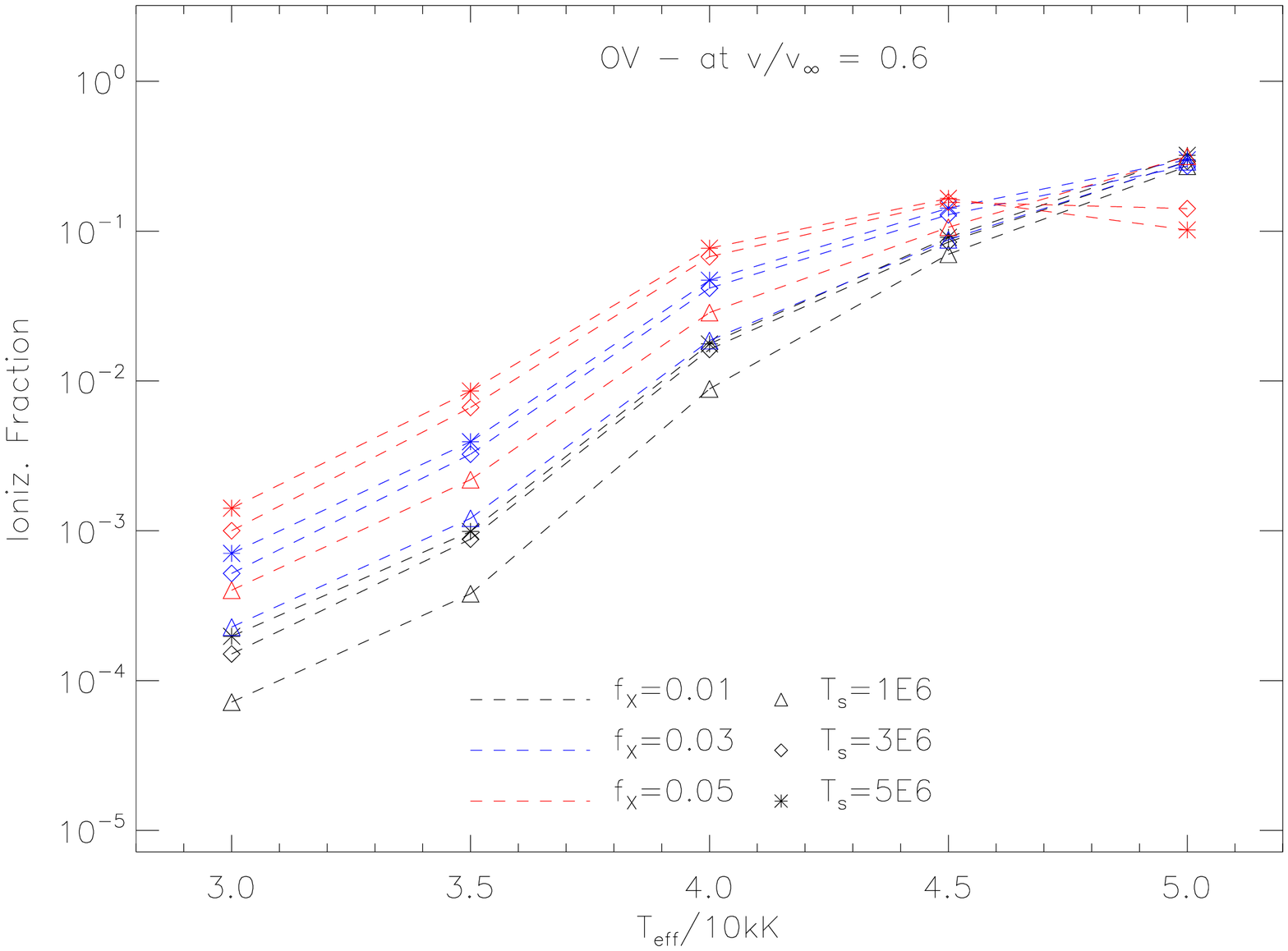}} 
\end{minipage}
\hspace{-.2cm}
\begin{minipage}{9cm}
\resizebox{\hsize}{!}
  {\includegraphics[angle=90]{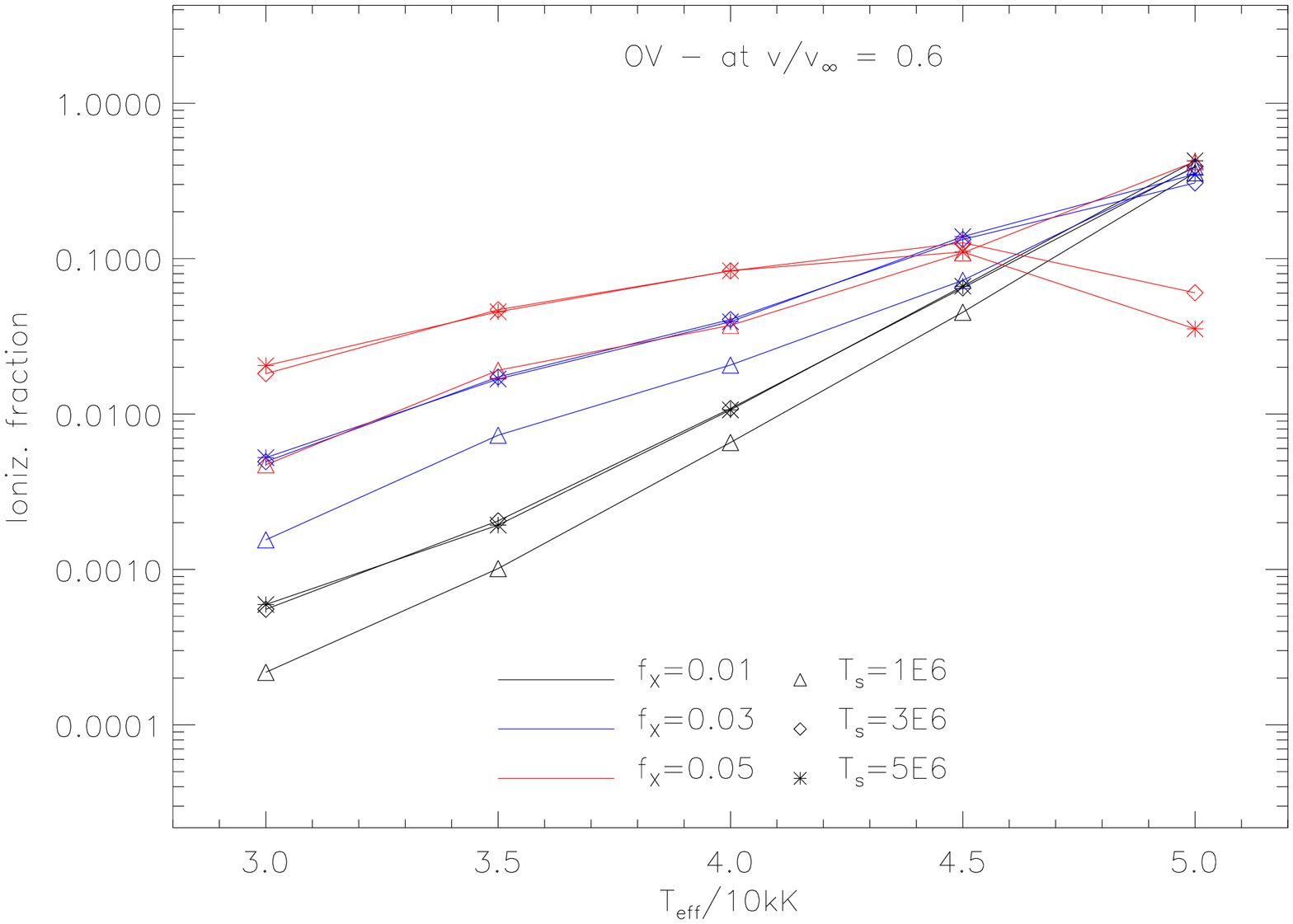}} 
\end{minipage}
\caption{As Fig.~\ref{civ_diff}, but for \OV\ ($v(r) = 0.6 \vinf$).}
\label{ov_diff}
\end{figure*}

\begin{figure*}
\resizebox{\hsize}{!}
{\includegraphics[angle=90]{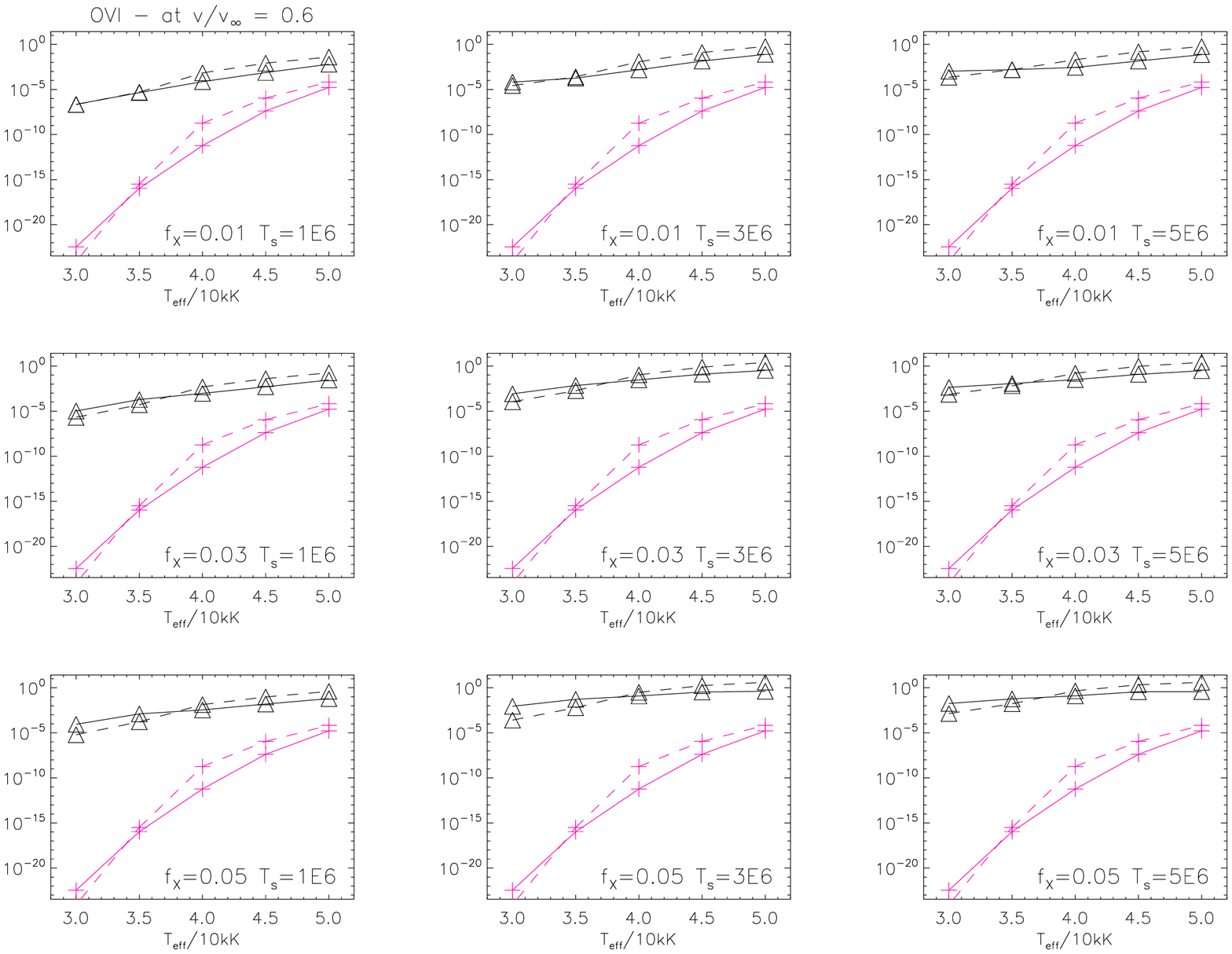}}
\caption{As Fig.~\ref{app_civ}, but for \OVI\ at $v(r) = 0.6 \vinf$.}
\label{app_ovi}
\end{figure*}

\begin{figure*}
\begin{minipage}{9cm}
\resizebox{\hsize}{!}
  {\includegraphics[angle=90]{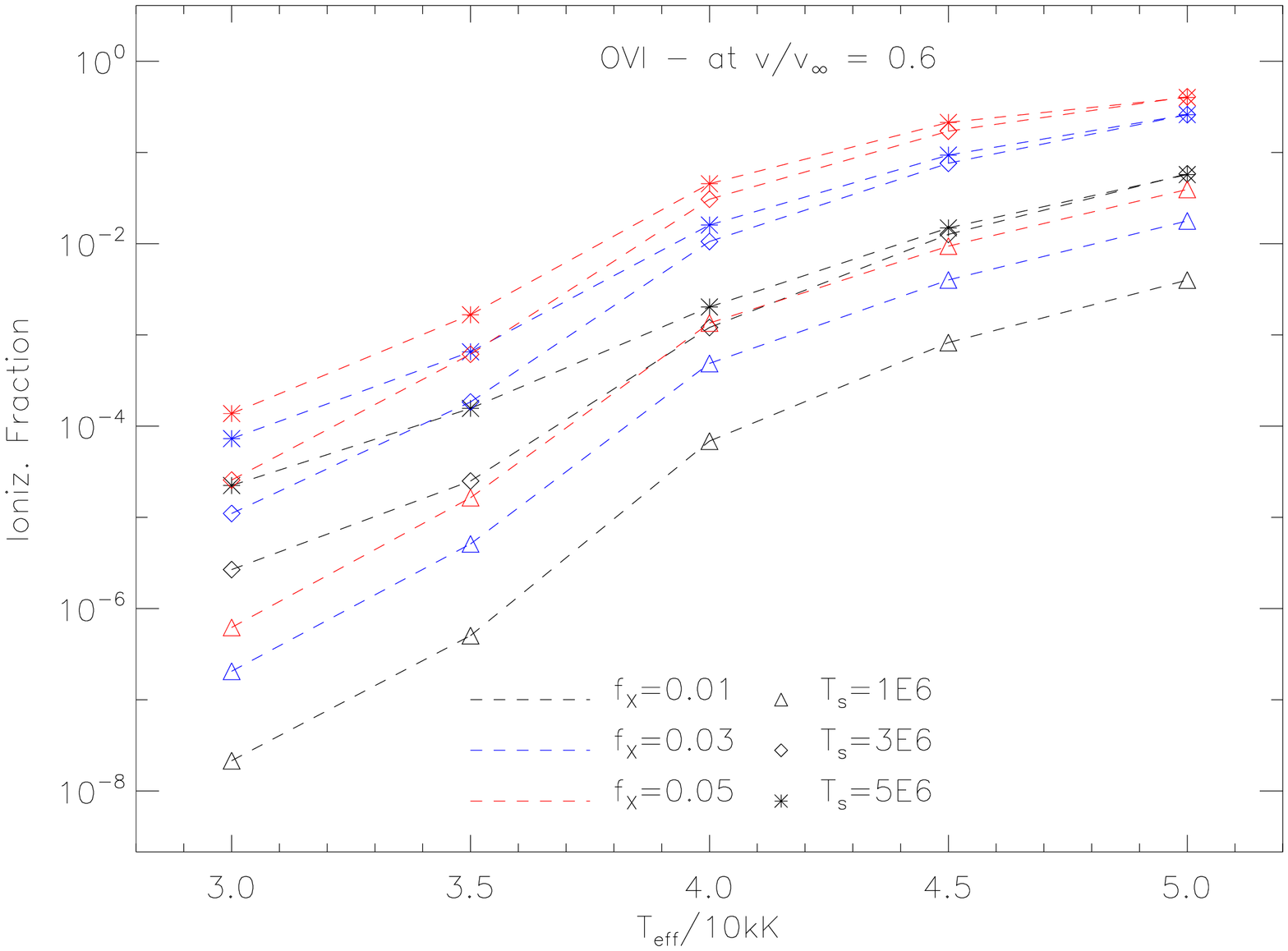}} 
\end{minipage}
\hspace{-.2cm}
\begin{minipage}{9cm}
\resizebox{\hsize}{!}
  {\includegraphics[angle=90]{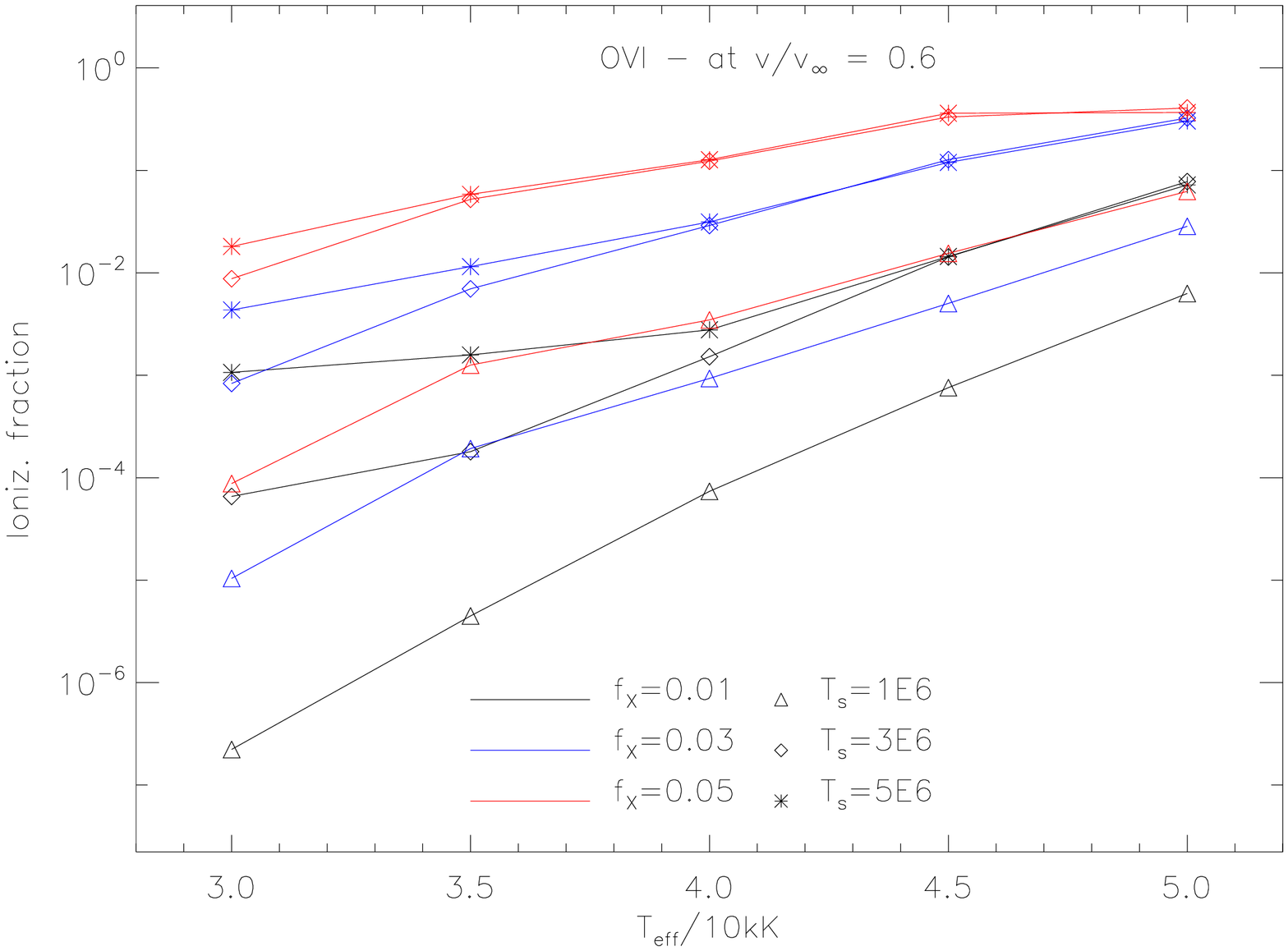}} 
\end{minipage}
\caption{As Fig.~\ref{civ_diff}, but for \OV\ ($v(r) = 0.6 \vinf$).}
\label{ovi_diff}
\end{figure*}

\begin{figure*}
\resizebox{\hsize}{!}
{\includegraphics[angle=90]{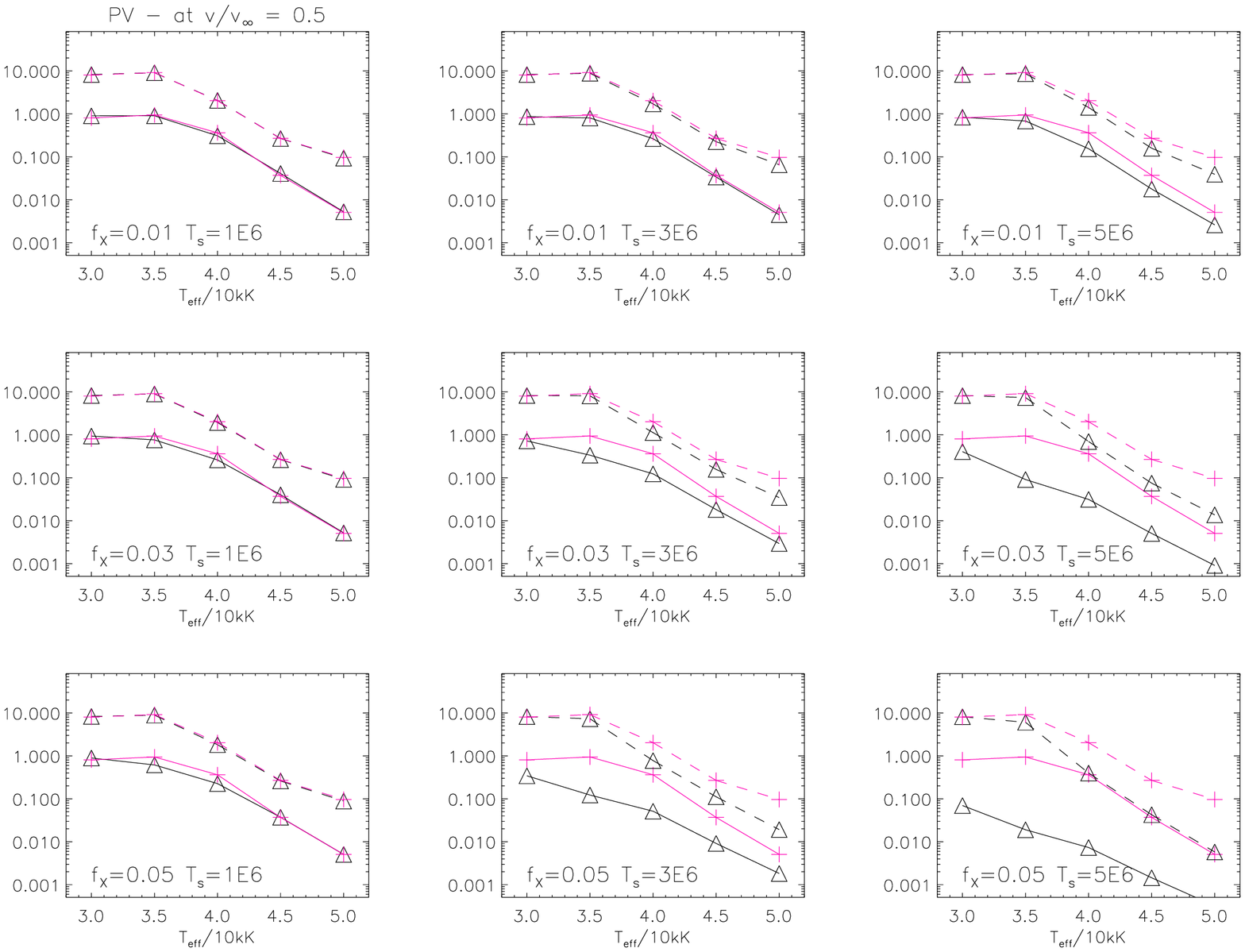}}
\caption{As Fig.~\ref{app_civ}, but for \PV\ at $v(r) = 0.5 \vinf$.}
%Ionization fraction of \PV\ for different values of temperature. The
%continuous lines are for supergiant models, while the dashed lines for
%dwarfs. {\it For clarity, the ionization fractions of dwarf models
%have been shifted by one dex}. With the lowest values of \fv\ or
%\Tshock\ almost no difference is seen between smooth and models with
%shocks, however it makes already clear that \Tshock\ has a higher
%influence in the ionization fraction. As pointed in section
%\ref{generaleffects}, with extreme X-rays parameters the depletion of
%\PV\ appears clearly for all models (both supergiants and dwarfs),
%except for D30.}
%
\label{app_pv}
\end{figure*}

\begin{figure*}
\begin{minipage}{9cm}
\resizebox{\hsize}{!}
  {\includegraphics[angle=90]{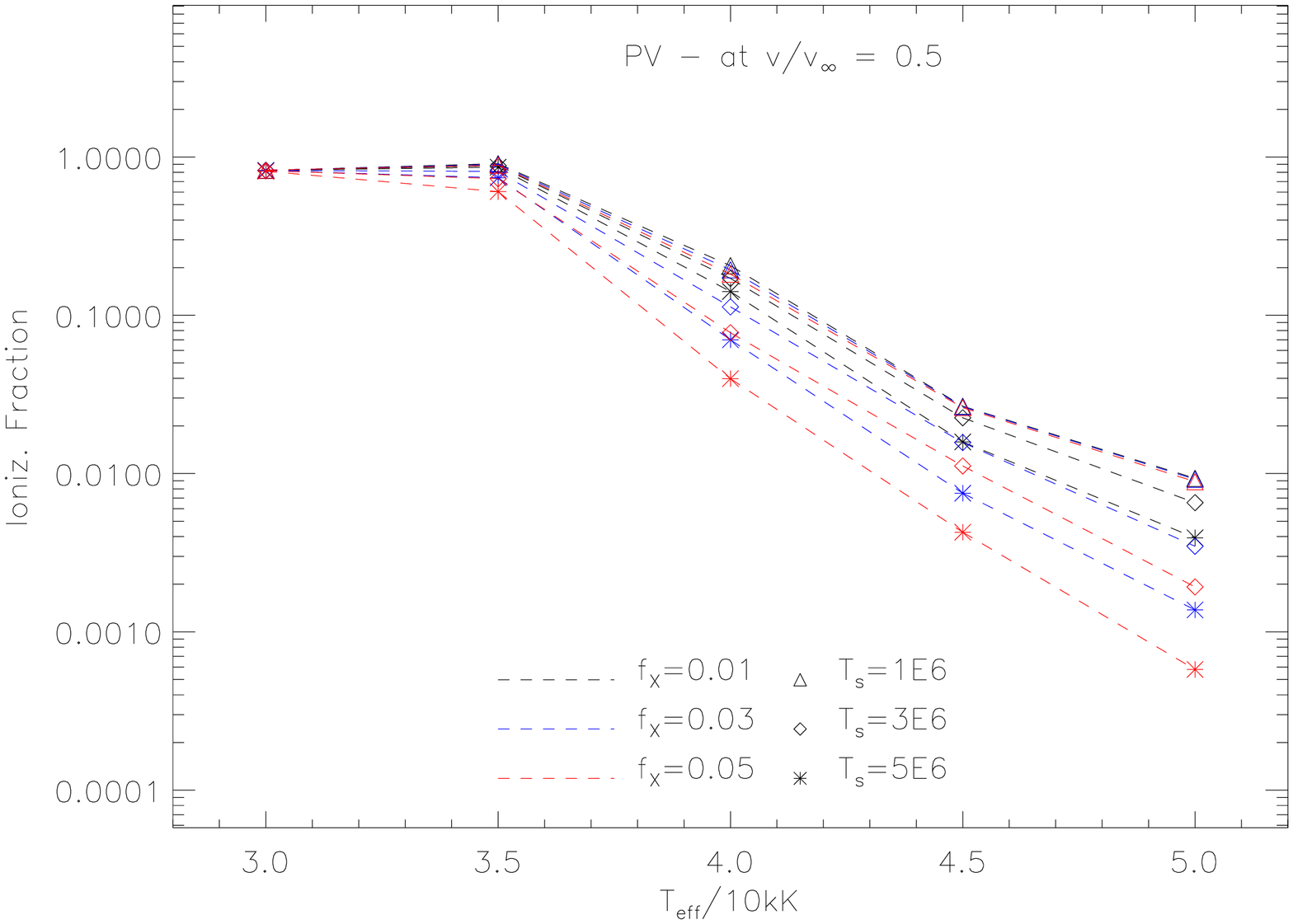}} 
\end{minipage}
\hspace{-.5cm}
\begin{minipage}{9cm}
\resizebox{\hsize}{!}
  {\includegraphics[angle=90]{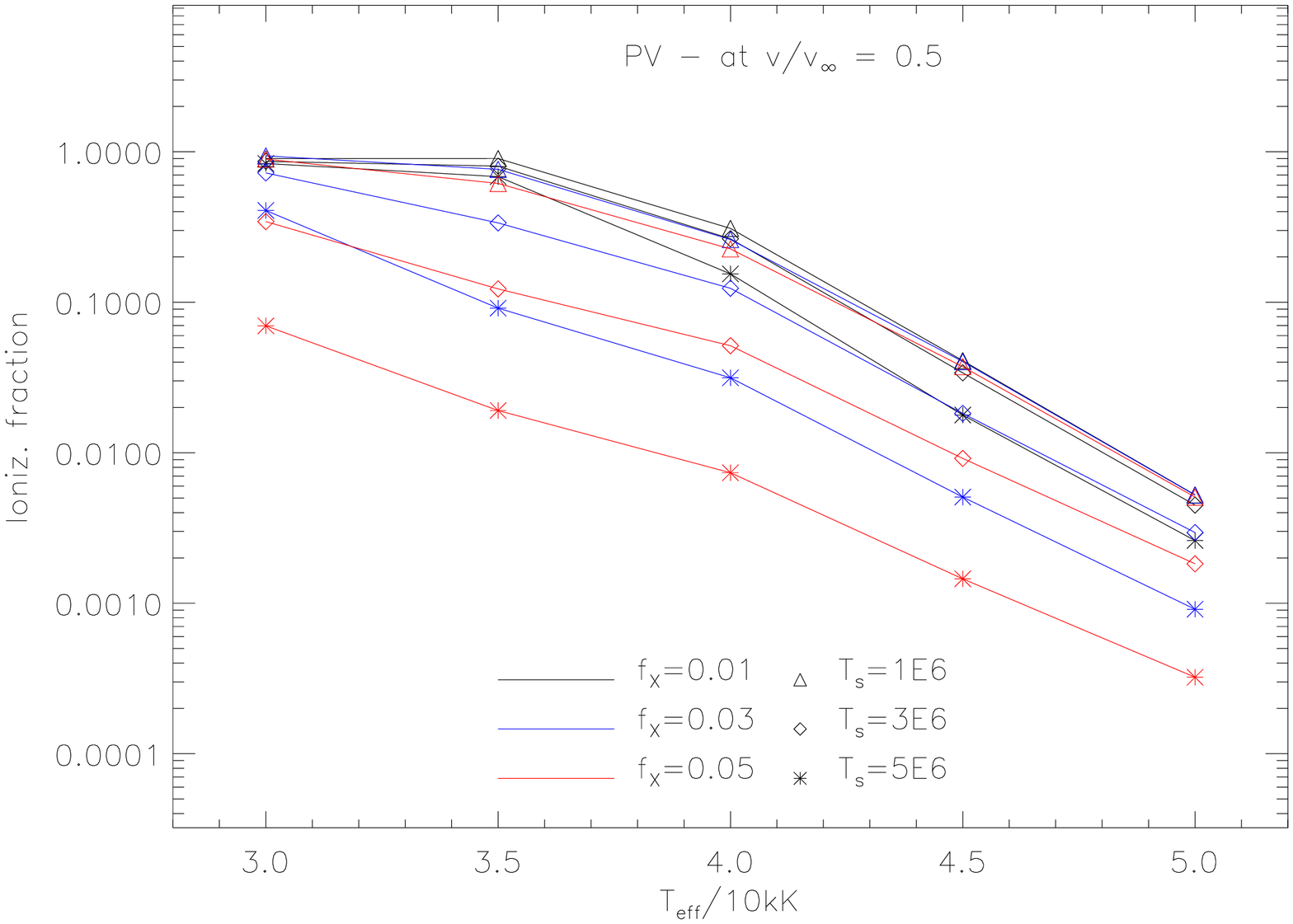}} 
\end{minipage}%
\caption{As Fig.~\ref{civ_diff}, but for \PV\ ($v(r) = 0.5 \vinf$).}
\label{pv_diff}
\end{figure*}

Figures \ref{app_civ} to \ref{pv_diff} display the reaction of \CIV,
\NV, \OV, \OVI, and \PV\ on varying the X-ray filling factors and
shock temperatures within our supergiant and dwarf models, as a
function of \Teff. For further explanation and discussion, see
Sect.~\ref{dependencefvtshock}.


\section{Comparison with WM-{\sc basic}: Ionization fractions and
UV line-profiles}
\label{comparisonwm}

\begin{figure*}[t]
\center
	 {\includegraphics[width=11cm,angle=90]{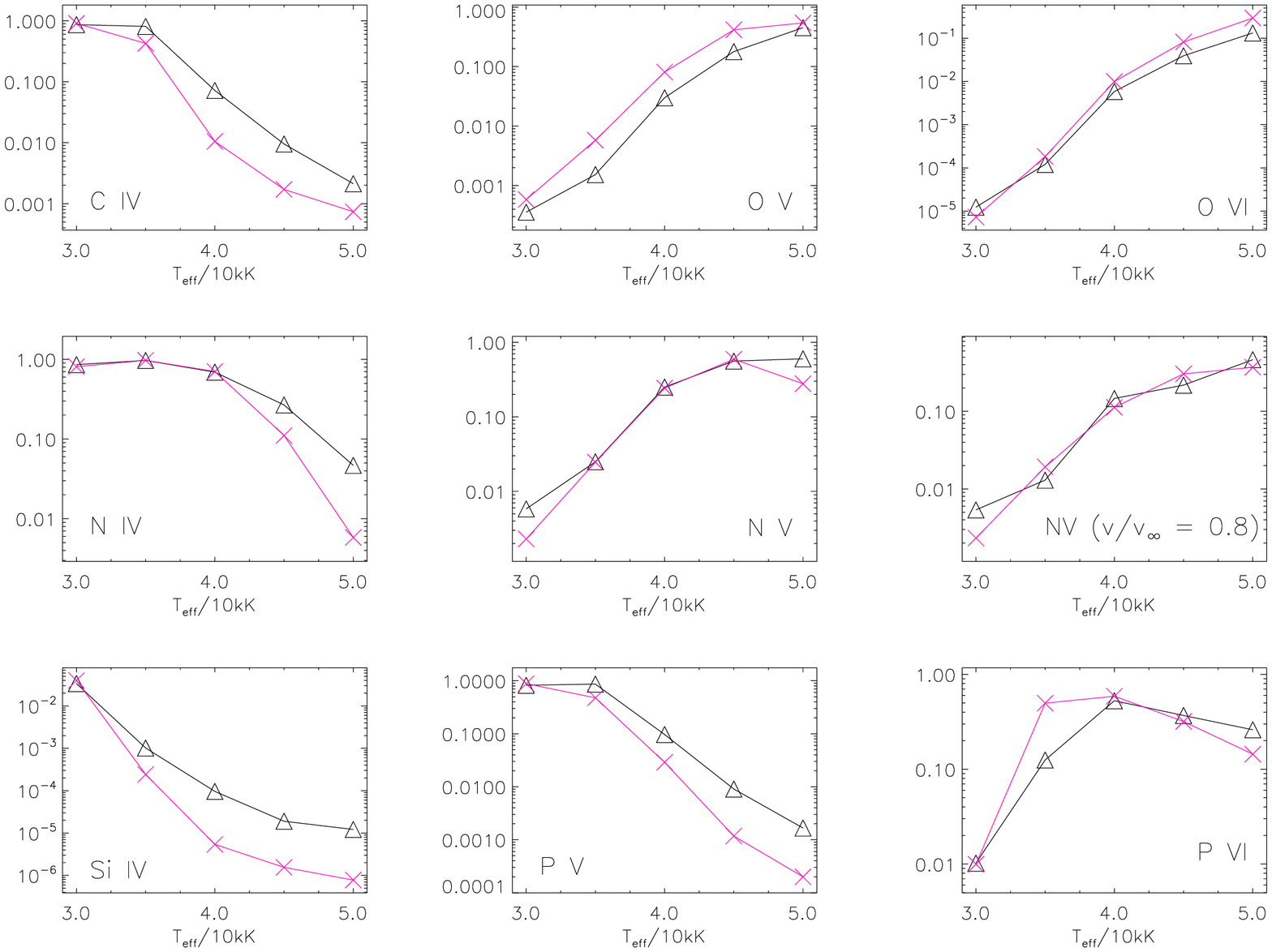}}
	 \caption{Ionization fractions of specific ions, as calculated
	 by {\sc FASTWIND} (black) and WM-{\sc basic} (magenta), for
	 our dwarf models and as a function of \Teff. If not stated
	 explicitly inside the individual panels, the fractions have
	 been evaluated at $v(r) = 0.5 \vinf$. See Sect.~\ref{comparing}.}
\label{tadziu_33_dw}
\end{figure*}

\begin{figure*}[b]
\center
{\includegraphics[width=11cm,angle=90]{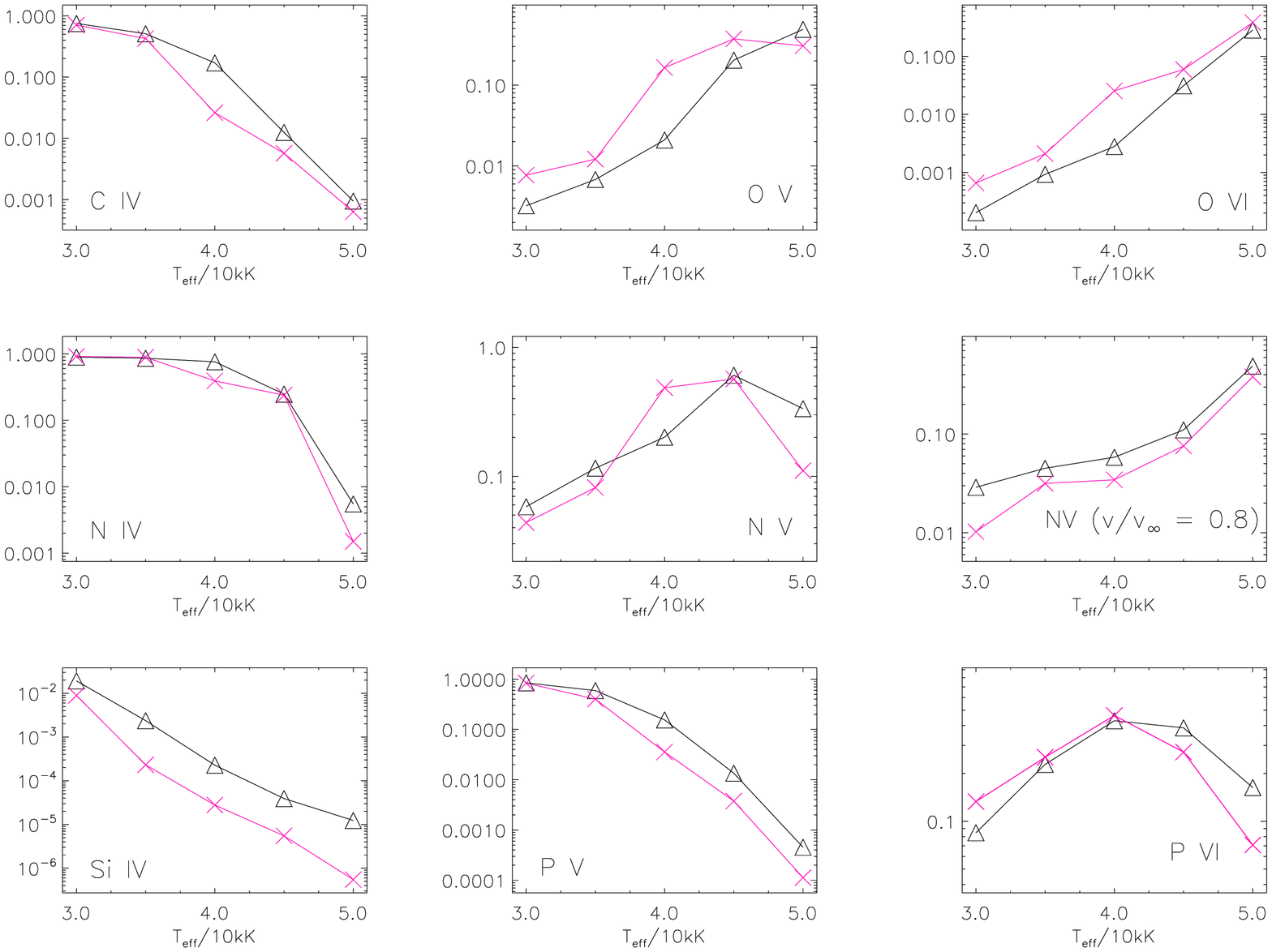}}
\caption{As  Fig.~\ref{tadziu_33_dw}, but for supergiant models.}
\label{tadziu_33_sg}
\end{figure*}

In Figs.~\ref{tadziu_33_dw} and \ref{tadziu_33_sg} we compare the 
ionization fractions of specific ions, as calculated by {\sc FASTWIND}
and WM-{\sc basic}, for dwarf and supergiant models, respectively.
Fig.~\ref{all_profiles} compares corresponding strategic UV-line
profiles for \NIV~1720, \NV~1238,1242, \OV~1371, \OVI~1031,1037, and
\PV~1117,1128. Further explanation and discussion is provided in 
Sect.~\ref{comparing}.

\begin{landscape}
\begin{figure}
\resizebox{\hsize}{!}
	 {\includegraphics[scale=14]{plots/all_profiles.ps}}
         \caption{Emergent line profiles for strategic UV lines
	 (\NIV~1720, \NV~1238,1242, \OV~1371, \OVI~1031,1037, and
	 \PV~1117,1128), as calculated by WM-{\sc basic} (green) and
	 {\sc FASTWIND} (black), for models S30 (top), D40, S40, D50,
	 and S50 (bottom). All profiles have been calculated with a
	 radially increasing microturbulence, with maximum value 
	 \vturb(max) = 0.1\vinf, and have been convolved with a
	 typical rotation velocity, \vsini\ = 100~\kms. The absorption
	 feature between the two \PV\ components is due to \SiIV~1122.
	 See Sect.~\ref{comparing}.}
\label{all_profiles}
\end{figure}
\end{landscape}

\section{Averaged mass absorption coefficients - clumped winds and 
dependence on averaging interval}
\label{mean_kappa_app}

\begin{figure*}
\begin{minipage}{10cm}
\resizebox{\hsize}{!}
{\includegraphics{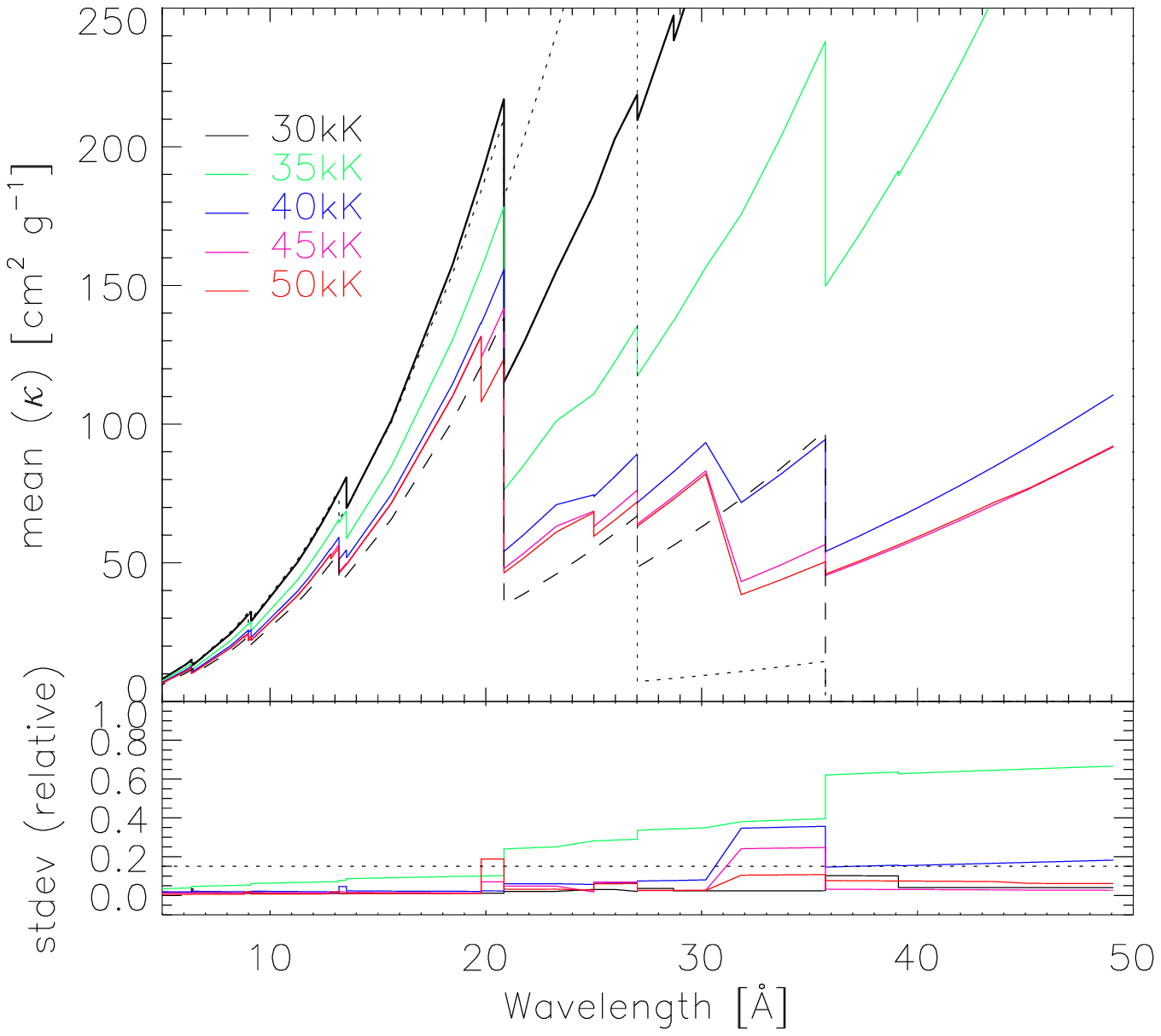}}
\end{minipage}
\hspace{-1cm}
\begin{minipage}{10cm}
\resizebox{\hsize}{!}
{\includegraphics{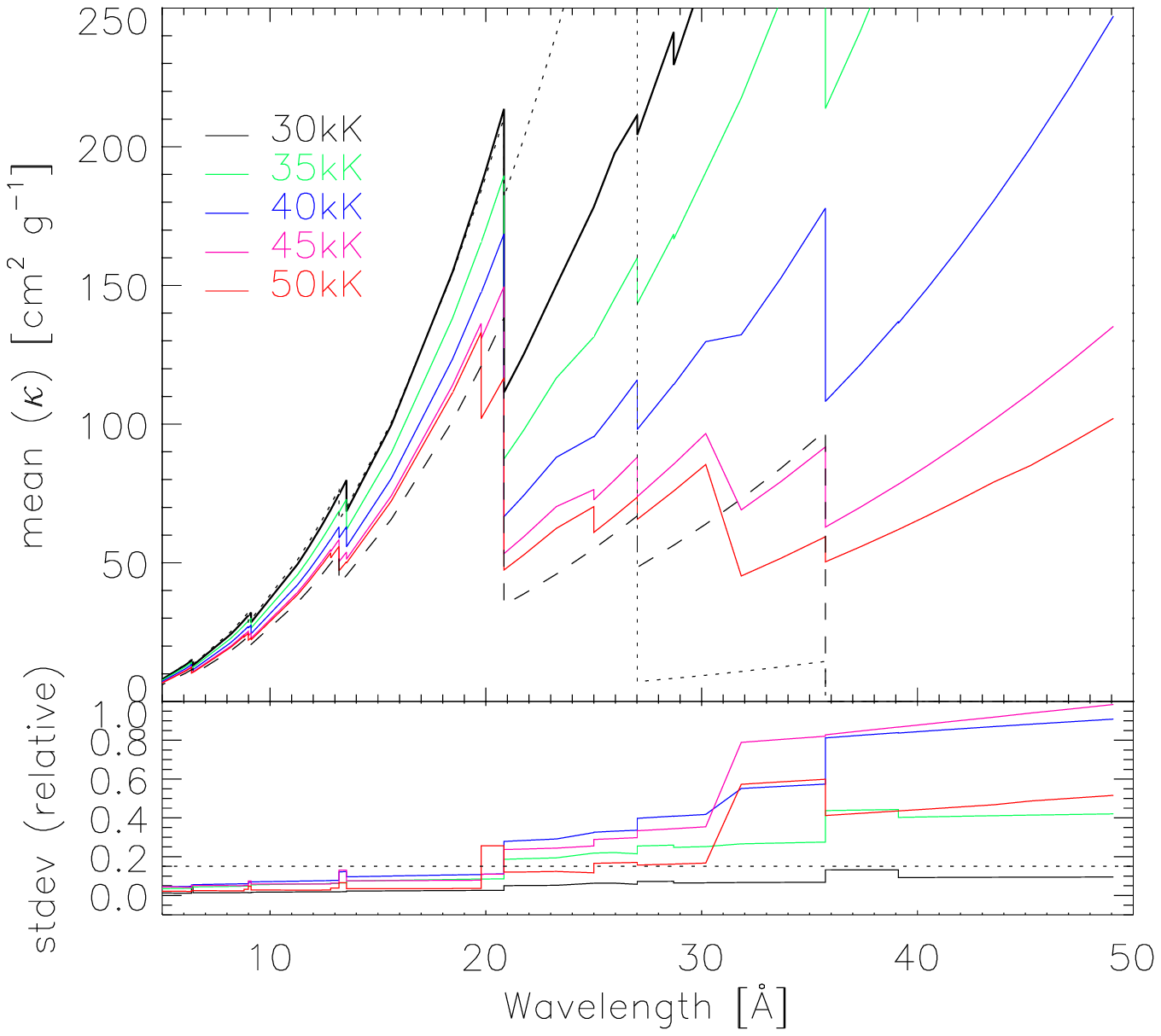}}
\end{minipage}
\caption{As Figs.~\ref{average_op_dw} and \ref{average_op_super}, but
         for clumped models with \fcl = 20 (corresponding to $f_V =
	 0.05$) and mass-loss rates reduced by a factor of $\sqrt{20}$.  
	 Left: dwarf models; right: supergiant models.}
\label{average_op_f20}
\end{figure*}

\begin{figure*}
\begin{minipage}{10cm}
\resizebox{\hsize}{!}
{\includegraphics{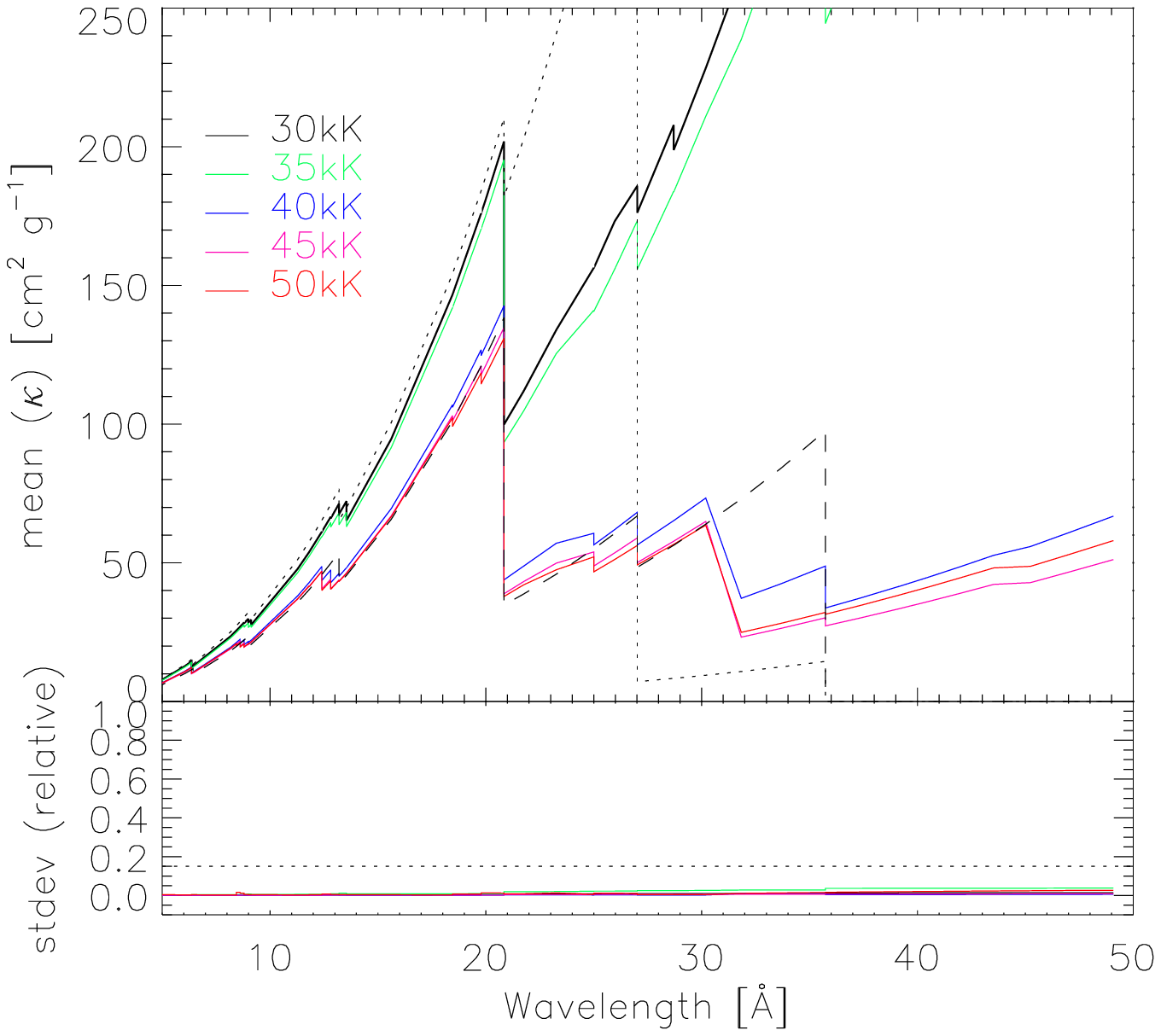}}
\end{minipage}
\hspace{-1cm}
\begin{minipage}{10cm}
\resizebox{\hsize}{!}
{\includegraphics{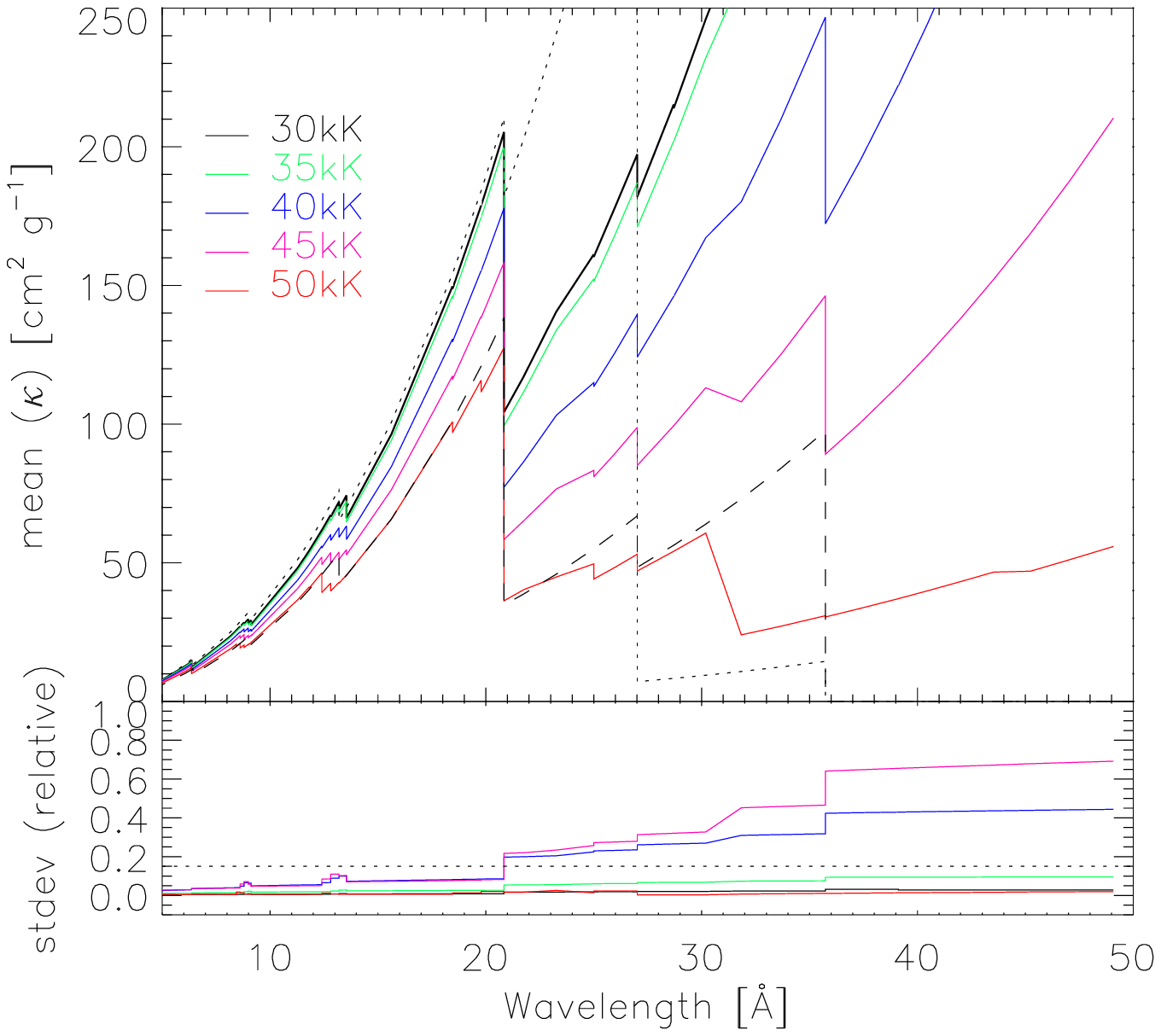}}
\end{minipage}
\caption{As Figs.~\ref{average_op_dw} and \ref{average_op_super}, but
         averaged over the interval between 10 and 110~\Rstar.
	 Left: dwarf models; right: supergiant models.}
\label{average_op_r10}
\end{figure*}

Fig.~\ref{average_op_f20} displays the density-weighted mean
(Eq.~\ref{meankappa}) of the mass absorption coefficient as a function
of wavelength, for dwarf (left) and supergiant (right) models. The
figure has a similar layout as Figs.~\ref{average_op_dw} and
\ref{average_op_super}, but has been calculated for clumped models
(\fcl\ = 20), and mass-loss rates reduced by a factor of $\sqrt{20}$.
Fig.~\ref{average_op_r10} is also analogous to
Figs.~\ref{average_op_dw} and \ref{average_op_super}, but now the
absorption coefficient has been averaged over the interval between 10
and 110~\Rstar. For details and discussion, see
Sect.~\ref{op_section}.
 
%\clearpage

\end{document}